\RequirePackage{fix-cm}
\RequirePackage{amsmath}

\documentclass{article}
\usepackage{epsfig}
\usepackage{acronym}	
\usepackage{pdflscape}
\usepackage{upgreek}
\usepackage{graphicx}
\usepackage{amssymb}   
\usepackage{textcmds}  
\usepackage{color}     
\usepackage{textcomp}
\usepackage{gensymb}   
\usepackage[]{hyperref}

\usepackage[caption=false]{subfig}

\acrodef{2MASS}[2MASS]{Two Micron All-Sky Survey}
\acrodef{AAO}[AAO]{Australian Astronomical Observatory}
\acrodef{AAT}[AAT]{Australian Astronomical Telescope}
\acrodef{AGN}[AGN]{Active Galactic Nuclei}
\acrodef{AO}[AO]{Adaptive Optics}
\acrodef{AWG}[AWG]{Arrayed Waveguide Grating}
\acrodef{DFS}[DFS]{Diverse Field Spectroscopy}
\acrodef{ELT}[ELT]{Extremely Large Telescope}
\acrodef{ESO}[ESO]{European Southern Observatory}
\acrodef{FBG}[FBG]{Fiber Bragg Gratings}
\acrodef{FM}[FM]{Few mode}
\acrodef{FMF}[FMF]{few mode fiber}
\acrodef{FPZ}[FPZ]{Free propagation zone}
\acrodef{FSR}[FSR]{free spectral range}
\acrodef{FTS}[FTS]{Fourier transform spectrograph}
\acrodef{FWHM}[FWHM]{Full Width Half Maximum}
\acrodef{HST}[HST]{Hubble Space Telescope}
\acrodef{IFU}[IFU]{Integral Field Unit}
\acrodef{IFS}[IFS]{Integral Field Spectroscopy}
\acrodef{IPS}[IPS]{Integrated Photonic Spectrograph}
\acrodef{LBT}[LBT]{Large Binocular Telescope}
\acrodef{LBTI}[LBTI]{Large Binocular Telescope Interferometer}
\acrodef{LGS}[LGS]{Laser Guide Stars}
\acrodef{LVM}[LVM]{Local Volume Mapper}
\acrodef{MCF}[MCF]{multi-core fiber}
\acrodef{MIR}[MIR]{mid-infrared}
\acrodef{MOS}[MOS]{Multi Object Spectroscopy}
\acrodef{MM}[MM]{Multi-mode}
\acrodef{MMF}[MMF]{Multi-mode fiber}
\acrodef{MOS}[MOS]{Multi Object Spectrograph}
\acrodef{NIR}[NIR]{Near Infra-Red}
\acrodef{PCF}[PCF]{photonic crystal fiber}
\acrodef{PEG}[PEG]{Phased Echelle Grating}
\acrodef{PIMMS}[PIMMS]{Photonic Integrated Multi-Mode Spectrograph}
\acrodef{PL}[PL]{Photonic Lantern}
\acrodef{PPLN}[PPLN]{periodically poled lithium niobate}
\acrodef{PSF}[PSF]{Point Spread Function}
\acrodef{SDSS}[SDSS]{Sloan Digital Sky Survey}
\acrodef{SM}[SM]{single mode}
\acrodef{SMF}[SMF]{single mode fiber}
\acrodef{SNR}[SNR]{signal-to-noise ratio}
\acrodef{ThAr}[ThAr]{Thorium Argon}
\acrodef{ULI}[ULI]{ultrafast laser inscription}
\acrodef{VLT}[VLT]{very large telescope}
\acrodef{xAO}[xAO]{extreme adaptive optics}

\begin{document}

\title{Astrophotonics: Astronomy and modern optics}

\author{S. Minardi\\  
 AMS, G\"oschwitzer-Str. 32, 07745 Jena, Germany\\ stefano@stefanominardi.eu\\       
        R. Harris    \\
        Zentrum f\"ur Astronomie der Universit\"at Heidelberg,\\ Landessternwarte Ko\"nigstuhl, Ko\"nigstuhl 12, 69117 Heidelberg\\ rharris@lsw.uni-heidelberg.de\\
        L. Labadie \\
        I. Physikalisches Institut der Universit\"at zu K\"oln\\ Z\"ulpicher Str. 77, 50937 K\"oln - Germany\\ labadie@ph1.uni-koeln.de}   


\date{Invited review article for The Astronomy and Astrophysics Review. Submitted on 9 Nov 2018}

\maketitle


\begin{abstract}

Much of the progress in Astronomy has been driven by instrumental developments, from the first telescopes to fiber fed spectrographs. In this review we describe the field of astrophotonics, a combination of photonics and astronomical instrumentation that is currently gaining importance in the development of current and future instrumentation. We begin with the science cases that have been identified as possibly benefiting from astrophotonic devices. We then discuss devices, methods and developments in the field along with the advantages they provide. We conclude by describing possible future developments in the field and their influence on astronomy.

\end{abstract}

\section{Astronomical instrumentation and science}

\begin{figure}
\epsfig{file=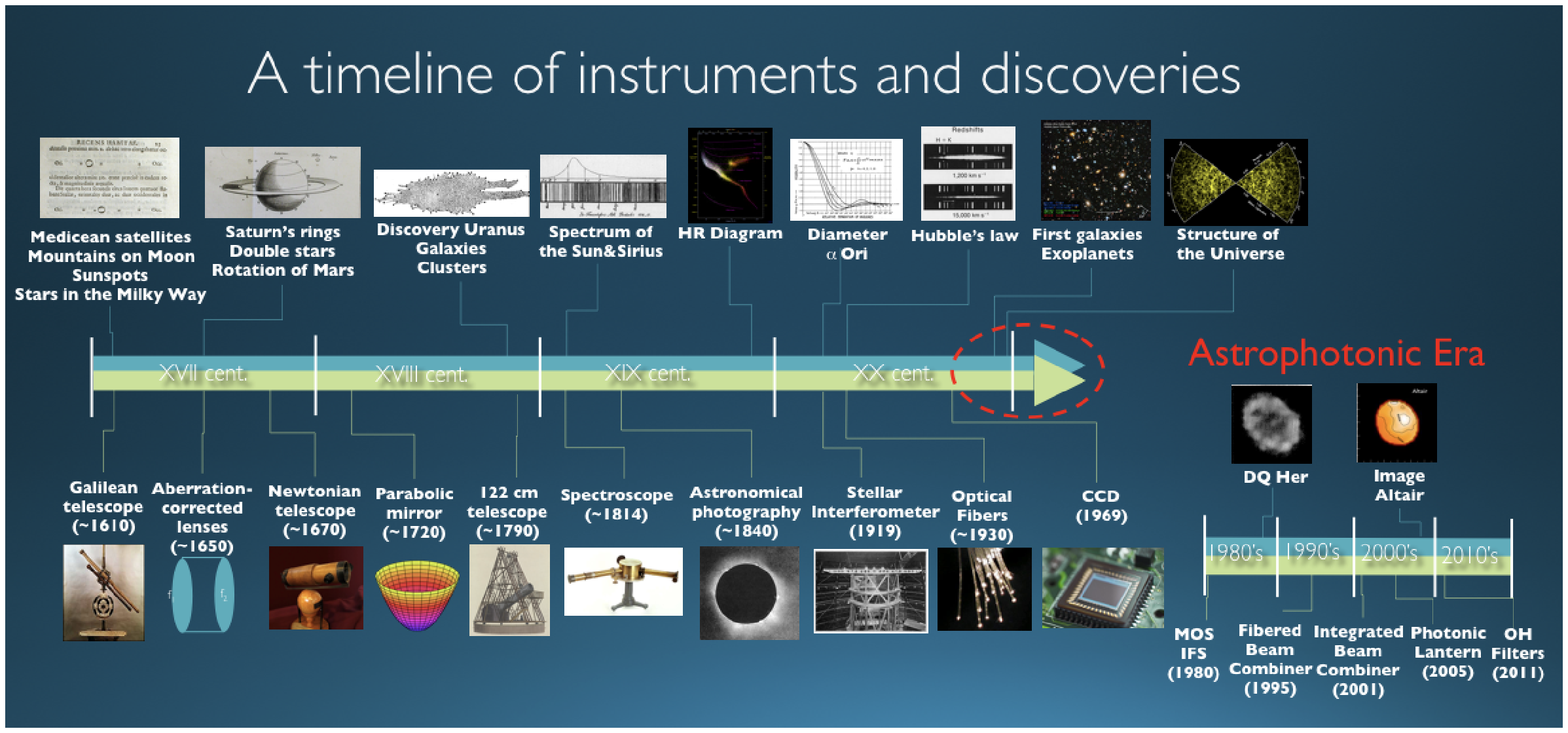, height=8 cm, angle=90}\\
\caption{\label{fig:history} An attempt to compare visually the progress of optics with the discoveries in astronomy. State-of-the-art optical instrumentation has often been the trigger for important discoveries.} 
\end{figure}

Astronomy is without doubt the empirical science which more than others relies on the analysis of electromagnetic radiation, and as such its progress has often 
been correlated to the development of optical technology.   
The most obvious example of the interaction between science and technology are the discoveries Galileo made at the beginning of the 17$^{\rm{th}}$ century, by pointing 
his telescope, then the state of the art of optical instrumentation, at the sky. 
This simple act revolutionised our understanding of the universe, and these early astronomical observations drove the technological developments that allowed science to deliver telescopes of ever increasing quality.
Before the end of the 17th century Christiaan Huyguens was able to develop an aberration corrected eyepiece allowing him to resolve the rings of 
Saturn, which were previously believed to be satellites, and detect the rotation of Mars. 
The invention of the reflecting telescope by Newton, which solved the problem of 
chromatic aberration, and the development of parabolic mirrors by Gregory ($\sim 1720$), which solved the problem of spherical aberration, opened the 
possibility to build telescopes with apertures significantly larger than the refractive ones. 
By the end of the 18th century William Herschel was observing the sky with reflecting telescopes exceeding half a meter in diameter 
(his biggest telescope had an aperture of 122 cm), enabling him to discover faint objects such as the satellites of Saturn and Uranus as well as 
to compile an extensive catalogue of 'nebulae'. 
The invention of the spectroscope by Fraunhofer (1814) and the subsequent discovery of the absorption lines in the spectrum of the Sun is another 
example of how progress in optical technology eventually changed the course of astronomy. The identification of the spectroscopic signature of 
chemical elements around 1850 gave astronomers a powerful tool to understand the structure of stars and of the universe, whose benefits we are still 
exploiting. A visual summary of the progression of optical technology and astronomical discoveries is illustrated in Fig. \ref{fig:history}, which shows qualitatively how new technologies, similarly to new paradigms, often require about a generation to bring their potential to fulfillment and open new avenues in astronomical discoveries \cite{Kuhn:1962}.

Today, the frontier in optics is represented by photonics, which is defined as the science dealing with the development of technologies which generate, transform or use tailored light states to perform a task. As the word suggests, photonic devices control and manipulate photons in a way similar to how electronic devices control and manipulate electrons. Using this definition we can include devices such as micro-optics, laser sources, optical filters, optical fibers, and optoelectronic switches, which are commonly used in applications for telecommunications and digital data storage/readout. For this article we will adopt a more restrictive definition of photonics and will deal mostly with elements requiring the wave approximation of optics to explain properly their operation.
Given the impact of optical technologies had in shaping the history of astronomy, it is not surprising that astronomers have taken an interest in photonic technologies and potential improvements they could bring to their instruments. 

The first examples of what is called now 'astrophotonics' (\textit{i.e.} the application of photonics to astronomical instrumentation \cite{BlandHawthorn:2009})
date back to the late 1970's where single optical fibers were used to connect telescope focal planes to spectrographs \cite{Hubbard1979}. Realising the potential improvements the 1980's saw further developments, when optical fibers were used to improve the productivity of spectrographs (e.g the \ac{MOS} MEDUSA \cite{Hill:1980}) or to obtain hyperspectral images of extended objects (\ac{IFS} such as ARGUS  \cite{Vanderriest:1980}). Besides representing 
a major part of all existing astrophotonic instrumentation, multi-object and integral field spectrographs have become essential tools of contemporary 
astronomy which enabled large scale surveys of the sky and kinematic studies of galaxies, both inconceivable only a few decades ago.
Given the astonishing development of telecommunication photonic and these first 
successful demonstrations of astrophotonic applications, the '2000 saw a growing interest and expectation from the astronomical community towards photonics.  
Three are the potentially appealing characteristics of photonic over conventional 'bulk optic'
instrumentation, namely \textit{(i)} the potential reduction of size and mass of the instrument, \textit{(ii)} the possibility to build highly multiplexed instruments, and \textit{(iii)} the delivery of enhanced instrumental performance.
The first advantage is clear considering the case for \textit{integrated optics}, \textit{i.e.} a technology enabling optical devices manipulating light at spatial scales of the order of the wavelength.
Complex free-space optics set-ups based on beam-splitters, folding mirrors, gratings and filters may be easily integrated on an optical fiber or a glass substrate of a few square centimetre area and 
few 100 $\upmu$m thickness. This makes the functional part of an instrument extremely compact and lightweight, a clear advantage especially for infrared instrumentation (the cryogenic enclosure is 
smaller and thus less demanding) and space missions. 
The miniaturisation and the possibility to fabricate easily many identical copies of the integrated device makes astrophotonic instrumentation appealing for highly multiplexed instruments which  
allow a broader access to unique telescope facilities and is a requirement for large scale astronomical surveys. The classical example of this feature is provided by the introduction of optical fibers 
which made multi-object spectroscopy possible (see Sec. \ref{sec:MOS_science}). Integrated optics could in the future allow for an even more aggressive multiplexing strategy levering on the inherently small size of the functional units of the instrument.  
Finally, astrophotonic devices could deliver instruments with unique performance, which could not be accomplished by other means. This is for instance the case of integrated optics 
beam combiners for stellar interferometry (Sec. \ref{sec:HAR}) which deliver a highly precise visibility measurement through the use of single mode optics, or the phase mask coronagraphy, which can 
suppress efficiently light at the diffraction limit by means of micro/nano-structured optical plates (Sec. \ref{sec:coronagraph}).

Besides these advantages, there are intrinsic and contingent limitations in contemporary photonics which limit the widespread use of photonic solutions to astronomy.
From the side of intrinsic limitations, we mention that the functionality of most of the existing photonic devices requires spatially coherent and relatively narrowband light sources, which is usually 
not the case of astronomical sources. As we will discuss throughout this review, the use of adaptive optics can solve the requirement for spatial coherence, while recent technological developments in dispersion control in integrated optics (for instance photonic crystal fibers \cite{Birks:1997}) may help extending the bandwidth of existing photonic devices.  
As for the contingent limitations, in the past 40 years integrated photonic applications have been driven mostly by the multi-billion-dollar-worth digital telecommunication industry.
As well known, digital communication protocols are relatively tolerant to noise and other spurious effects which degrade the \ac{SNR}. Digital signals can be retrieved even in the presence of relatively high noise level (\textit{e.g.} $SNR\sim3$).
Therefore specifications of telecom devices on back-reflections or intensity/polarisation contrasts are often in the 10-20 dB range, which are incompatible to the dynamic range requirements of astronomy, typically exceeding 40 dB.
Additionally, and in contrast to astronomy, light sources in telecommunication are coherent and therefore susceptible to optical amplification processes with low noise overhead. This makes the losses of telecom devices a secondary issue for their use, so that devices with 5 dB insertion losses 
(30\% transmission) are common. Such losses are clearly not acceptable for a photon starved application such an astronomical observation.   
Moreover, photonic applications developed mostly the infrared spectral region, mostly due to the existence in this spectral band of low-loss transmission bands in glasses and emission lines of laser materials.
While in principle it is possible to reproduce the photonic functionalities in other optical bands, the high development costs require a medium to large market to justify research and development. The small scope of astronomical instrumentation market is usually 
insufficient to promote such a technological shift.  

The adaptation of contemporary telecommunication technologies and the overcoming of the above mentioned limitations represent the current challenges of the astrophotonic community, which will be discussed extensively in this article. 
Nonetheless, after decades mostly spent exploring these potentialities, astrophotonics is already starting to become a favourite companion of advanced astronomical instrumentation. A proof of it is that photonic technologies will be part of the instrumentation for Extremely Large Telescopes (ELTs) already from the very beginning, as for instance laser guide stars for adaptive optics \cite{Ciliegi:2018} and optical fibers as key components of multi-object and/or integral field spectrographs\cite{Evans:2016}.\newline

\begin{figure}[ht]
\epsfig{file=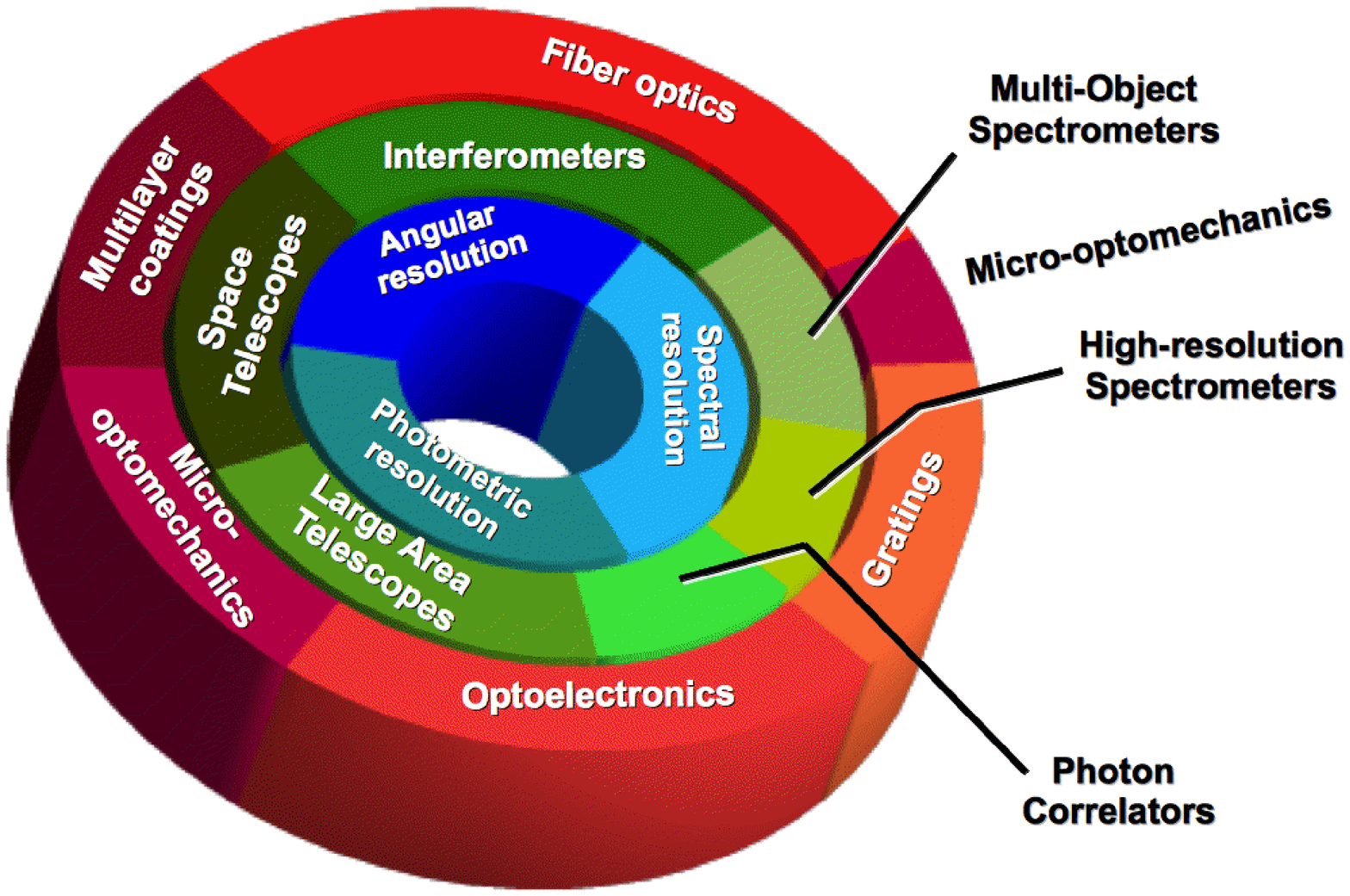, width=\textwidth}
\caption{\label{fig:astropie} Astronomical instruments (middle circle) develop in between the desired specifications derived from the astronomical target (inner circle) and the constraints imposed by the available optical technologies (outer circle).}
\end{figure}

This review seeks to give a structured and comprehensive overview of the interplay between photonics, astronomy and instrumentation, highlighting in particular the current state-of-the-art in the field as well as the directions to which the community may look at. A visualisualisation of this approach is presented in the ring-plot of Fig. \ref{fig:astropie}, where astronomical instruments (middle ring) result from the, often contrasting, specifications dictated by the astrophysics of the object under study (inner ring) and the constraints of optical physics and technologies (outer ring). Instrumentation can be therefore discussed from the perspective of the astronomer and from the one of the optical/photonic engineer.
 
As opposed to other overview works in the field (e.g. \cite{BlandHawthorn:2009,BlandHawthorn:2017}), 
this paper incorporates different branches of astrophotonics that are usually addressed separately, as for instance spectroscopy, interferometry/high-contrast, and calibration of instrumentation. Additionally, we tried to give an introduction to the physics underlying the function of photonic technologies, through an introductory chapter on optical modes and in the description of the instruments.
Section \ref{sec:astronomical_perspectives} addresses, from an astrophysical perspective, specific science cases and related requirements that can directly benefit from the photonics approach. Consistent with the ring-plot, 
Section \ref{sec:photonics_perspective} introduces the optical physics perspective by summarising the formalism of optical waveguiding, which is central to the understanding of photonic technologies. 
Section \ref{sec:astrophotonics} presents the different families of astrophotonics devices from a performance-driven point-of-view and facilitate the comparison with the theoretical background given in Section \ref{sec:photonics_perspective}. Finally Section \ref{sec:conclusions_future} concludes our paper and summarises future perspectives.

\section{Astronomical perspective}
\label{sec:astronomical_perspectives}

Imaging and spectroscopy techniques -- or the combination of them -- are the two pillars supporting any advance in observational astrophysics. These two methods branch into several sub-techniques driven by specific science cases. The potential benefit of photonics-based solutions must be appreciated under the perspective of the scientific requirements associated to these two main techniques.

\subsection{Single object spectroscopy}

Spectroscopy is a prime technique used in astronomy to investigate many properties of astrophysical objects, from chemical composition, to velocity structure, to the detection of exoplanets through the Doppler technique.  
This requires the splitting of the incoming light into its different wavelengths by means of dispersive optical elements ($e.g.$ prism, grating). 

As spectroscopy is exploited in the majority of the science cases in optical and infrared astronomy, we analyse hereafter the requirements of a subset of them. 

\paragraph{Measurement of chemical compositions and effective temperatures:}

Low-resolution spectroscopy is usually employed to extract basic information on the source (such as the effective temperature) from faint objects, for which a large dispersion would lead to a spectral flux density well below the noise level. Resolutions with $R$ up to a few ten are used in deep galactic and extra-galactic surveys, and are usually performed by means of a set of interchangeable band-pass filters rather than spectrographs.

Solid-state species and broad spectral features such as ices\cite{Pontoppidan2005}, silicates\cite{VanDenAncker1999} or PAHs\cite{VanKerckhoven2002,Taha2017} can be studied at resolution of few hundreds to few thousands in the visible and near infrared regions, where the molecular bands are active \cite{Veron-Cetty2000,VanDishoek2004,Henning2013}. 
Another use of low spectral dispersion is slit-less or long-slit spectroscopy. In slit-less spectroscopy low resolution is necessary to avoid too many overlaps between the spectra of neighbouring objects. Examples of application are found in space surveys such as GAIA, which measures the color of stars down to magnitude G$\sim$20 with a slit-less spectrograph with $R\sim20$.
Long-slit spectrographs have been used mostly in the past to deliver spatially resolved spectra of faint extended objects such as galaxies or nebulae, but this technique has been mostly replaced by integral field spectroscopic methods, where photonic technologies play a key role (see below).
Low resolution spectra from Fourier-transform spectrographs are still common in mid- and far-infrared astronomy, where until recently only noisy, single-pixel detectors where available. The availability of new low noise detector arrays in these bands has shifted instrumentation towards the more simple and robust dispersive designs.

\paragraph{Measuring motion in the Universe: from stellar physics to exoplanets and (circum)stellar physics:}
Beside investigating the chemical composition of astrophysical sources, spectroscopy is successfully and largely used for studying the radial velocities of celestial objects via the Doppler shift. 
Since the discovery of the first exoplanet around a Sun like star using the radial velocity technique\cite{Mayor:1995} the number of exoplanets discovered using this technique has grown rapidly, peaking with the discovery of a possible temperate terrestrial planet around the M dwarf Proxima Centauri\cite{Anglada2016}.  
Taking advantage of the periodic movement of an exoplanet intrinsic atmospheric line with respect to our strong telluric lines, high-resolution spectroscopy even permits to characterise the exoplanet atmosphere itself. This technique has been shown for $\tau$\,Bo\"otis\,b with the CO line using the CRIRES spectrograph at \ac{VLT} and might be implemented with the future METIS instrument on the ELT\cite{Snellen2015}.\\
For these science cases, (very) high spectral resolution with $R\sim$100\,000 is typically required if small line-of-sight velocities down to a precision of a m/s to 10 cm/s \cite{Mayor:2014} are sought according to $v$=$c\Delta\lambda$/$\lambda$. These spectrographs exploit the Echelle design where a grating is used at high diffraction orders, which are afterwards separated by a second dispersive element. To achieve a wavelength accuracy beyond the resolution limit, an extremely good stability of the instrument is required from the mechanical and thermal point-of-view as environmental disturbance may lead to drifts or jittering of the spectral lines being measured. In many instances, the possibility to set the instrumental payload in a vacuum vessel is highly valued but this requires a more demanding infrastructure depending on the size of the spectrograph. An essential tool for the observations is a stable wavelength calibration source to correct any drift of the optomechanical assembly of the spectrograph. Usually gas cells/lamps filled with heavy ions are used as calibration units, a practice which is about to be surpassed by the use of stabilized frequency combs     \cite{Steinmetz:2008}(see Sec. \ref{sec:astrocombs}). Finally, when the spectrograph is a fibered-fed unit, the seeing-induced fluctuation of the baricenter of the light distribution at the pseudo-slit (modal noise) could be a major limiting factor to achieve sub-resolving power accuracy in the wavelength measurement of stellar absorption lines. For this reason techniques reducing modal noise need to be employed (so called modal scrambling techniques, see Sec. \ref{sec:spectroscopy}).

At intermediate spectral resolutions of few thousands to ten of thousands, spectroscopy senses the gas kinematics in circumstellar environments \cite{Goto2006,Pontoppidan2008,Gustafsson2008,Millour2011} and galaxies using the techniques of spectro-astrometry \cite{Takami2001} and spectro-interferometry. Motion of gas over stellar photosphere resulting from stellar winds and mass-loss mechanisms is seen in high resolution spectra where particular shapes such as P-Cygni profiles are observed. Importantly, the magnetism of stars can be investigated using high-resolution spectro-polarimetry that reveals the splitting of atomic lines due to the Zeeman effect \cite{Artigau2014}.\\
Finally, for survey purposes accessing a large instantaneous field-of-view is a top requirement: the Radial Velocity Spectrograph (RVS) on GAIA accesses a field-of-view of 0.6\,deg$^2$ at a spectral resolution of $R\sim$11,000.

\paragraph{Solar system bodies} There are numerous types of small bodies  present in the solar system, mostly inhabiting the asteroid belt and Kuiper belt. Studies of these small objects within the solar system rely on detection and follow up capabilities, which small telescopes are ideally suited for provided the objects are bright enough. Not only can these objects tell us about the chemical make-up of the solar system, they can also give us information as to how it formed.
In addition to the small bodies in the Kuiper belt and asteroid belt comets can be used to inform theories about the formation of the solar system. In order to make useful discoveries high resolution spectroscopy ($R >$ 10,000) needs to be performed in the optical and near infrared \cite{bailyn2007renewing,drew2010report}. 

Finally, the atmospheres of solar system planets present an exciting opportunity with further study allowing us to probe the composition and dynamics of the atmospheres to greater levels than previously achieved.

\paragraph{Low mass brown dwarfs}

Brown dwarfs occupy the mass range between the largest planets and the smallest stars. Due to their size, they are unable to sustain nuclear fusion of hydrogen, though some are thought to fuse heavier elements. The first brown dwarf was discovered in 1995, though their existence was  hypothesised for many years before \cite{rebolo:1995}. Since the first discovery, the numbers and diversity in candidates has increased to hundreds, mostly due to surveys such as  the \ac{SDSS} \cite{york:2000} and \ac{2MASS} \cite{gizis:1999}. 

To increase the number of candidates will be the job of future photometric and spectroscopic surveys covering the optical and \ac{NIR}. In addition, understanding the composition and processes in their atmospheres is particularly important, and spectroscopic surveys will be required to determine parameters such as composition, temperature and density. Due to the temperature of these objects low spectral resolution ($R < 10,000)$ \ac{NIR} spectroscopy is ideal to maximise flux contrasts of these objects. 

\paragraph{Stellar variability} There are many different reasons from stellar variability, ranging from binary pairs, to stellar pulsations through to cataclysmic variables. Observations of these variable stars are extremely useful and they can be used to study many properties of the star, from determining the processes driving their variability, probing stellar evolution and properties and probing their atmospheres. In addition specific types of variable stars (e.g. RR Lyrae, Cephied variables) can be used to determine other parameters used to determine cosmological distances \cite{Freedman:2001}. In recent years the number of candidates has grown thanks to two major efforts. First through gravitational lensing surveys (e.g. MACHO \cite{Alcock:1993} and OGLE \cite{udalski:1994}) and also using wide field surveys, both for variable stars but also for exoplanets (e.g. ASAS \cite{Pojmanski:1997}, ROTSE \cite{akerlof:2000} and HATNet \cite{hartman:2011}), have provided hundreds of thousands more candidates.

Whilst time resolved photometry tells us much about these candidates, there are also advantages to performing observations with time resolved spectroscopy. Firstly to provide confirmation or clarification in the classification \cite{Downes:1996} and secondly to further probe the star.
Most recently understanding variations in stars has been helping astronomers disentangle planetary radial velocity signals from those of effects such as sun-spots on the surface of stars (e.g. \cite{oshagh:2016}).

\subsection{Integral field Spectroscopy and Multi object spectroscopy - Hyperspectral imaging}
\label{sec:MOS_science}

Whilst spectrographs observing single objects are extremely powerful, they are limited in their capabilities to observe large numbers of objects. In an era where many science cases require large statistical samples and with telescopes becoming fewer in number (due to their larger designs costing more), more pressures are put on multiplex, or the amount of objects an instrument can observe at once.
To solve this the alternative is to observe many objects at once. Spatial elements (spaxels) are observed in the focal plane and then fed into the spectrograph to produce a 3D data cube (see Fig. \ref{fig:IFS}). There are two types of these instrument, which can be loosely classed into \ac{MOS} and \ac{IFS}. \ac{MOS} allows observations of multiple objects in the field of view \cite{ellis:1988} whist \ac{IFS} allows small patches of sky to be spatially resolved by an \ac{IFU} (e.g. for observation of extended or spatially adjacent objects) \cite{allingtonsmith:2006}. To further increase the number and type of object that can be observed there is a combination of the two (using multiple \acp{IFU}) which is called \ac{DFS} \cite{Murray:2009}. There are many science cases for all of the above and some are briefly outlined below.

In a 3D-spectroscopy instrument, the detector plays a central role in determining the final cost \cite{harris:2012}, and it is therefore necessary to optimize the use of the detector pixels. Besides the spatial sampling at the telescope, it is necessary to consider the sampling element of the data cube (\textit{voxel} \cite{allingtonsmith:2006}), whose number equals the minimum required number of pixels of the detector. 
Hence it is useful to introduce a figure of merit for the instrument, known as \textit{specific information density} (SID, \cite{allingtonsmith:2006}):
\begin{equation}
\textrm{SID}=\eta\frac{N_{r}}{N_p}
\end{equation}
where $\eta$ is the throughput of the instrument $N_r$ is the number of resolution elements in the datacube and $N_p$ is the total number of pixels of the detector. 
The relation between the number of resolution elements and spaxels/voxels depends on the type of instrument and should be as close as possible to the maximum value of the SID to improve sensitivity.
The SID is maximum when the number of pixels equals the number of voxels and the throughput is unitary. However, the number of resolution elements is in general smaller than the number of voxels because of oversampling therefore, the maximum value for the SID  will be smaller than 1.  
In MOS one target is sampled by a single spaxel and 
$N_r=N_sN_{\lambda}/f_{\lambda}$, $N_{\lambda}$ being the number of samples of the optical frequency (wavelength), and $f_{\lambda}$ is the oversampling factor.
If now assume theta there are no unused pixels (condition to maximize SID) the number of voxels will be $N_p=N_s N{\lambda}$. Taking $f_{\lambda}=2$ (Niquist sampling of the wavelength), the maximum value for the SID will therefore be 1/2.
In IFS, we will have in general $N_s=N_xN_y$ spaxels (a square array of spaxels is assumed) but the actual spatial resolution elements depend on the oversampling $f_x$ and $f_y$ of the telescope point spread function (PSF) along the coordinates $x$ and $y$. Thus the number of resolution element is given by $N_r=N_sN_{\lambda}/f_xf_yf_{\lambda}$. With the Niquist sampling in the space and frequency, the maximum SID will be 1/8.

\begin{figure}
	\caption{Methods of Integral Field Spectroscopy. Taken from \cite{allingtonsmith:2006}}. The top image shows a lenslet array feeding the spectrograph, this is simple and allows for a large field of view, but requires the spectrograph to be close to the telescope. The second option, fibers, allows the spectrograph to be placed further from the telescope, usually at the cost of throughput. At the bottom, image slicers and microslicers generally allow high throughput, though also require the spectrograph to be close to the telescope.
    \label{fig:IFS}
\end{figure}

\paragraph{Galactic science} To understand how individual galaxies form, what drives their evolution and what states they tend to requires large data sets of spatially resolved galaxies. Due to this large \acp{IFS} surveys are an extremely powerful tool. There are currently lots of larger scale surveys such as MASSIVE (Mitchell Spectrograph), AMAZE (SINFONI) and KROSS (KMOS) are currently providing data on hundreds of galaxies. These are being supplemented by even bigger ones (thousands of galaxies observed) such as MaNGA, SAMI, and Califa, plus soon the \ac{LVM}.
These surveys tend to provide direct information on two components, stars and gas. 

New \ac{IFS} consisting of several hundred spaxels (e.g. MUSE at \ac{VLT}) are able to access single stars in nearby galaxies or globular clusters \cite{bacon2010muse}, making detailed kinematic and metallicity studies possible. For more distant galaxies individual stars cannot be resolved but \ac{IFS} can still deliver the average radial velocity distribution of stars, a useful information to identify the morphology of galaxies \cite{croom2012sydney}.

Studies of the gas in galaxies are equally important. They can be used to trace the metallicity and abundance ratio of the gas, telling us the processes forming stars and in the future the types of stars that will form and at what rate. In addition gas can also trace the kinematics of the galaxy. 
Combining information on both of these components in large data sets allows astronomers to answer the largest questions in galaxy formation and evolution today. These include including how physical processes evolve with time, what regulates star formation, how metals build up, what drives gas inflows and outflows, the role of the local environment and what drives strong morphological transformation.

\paragraph{Cosmology} Understanding the universe on the largest scales is one of the grandest aims of astronomy. How the universe expands, the composition of the universe and the physics governing everything are both very large questions. To solve these large sets of data are required. Obtaining these with single slits or fibers is difficult and time consuming. As such the trend is towards \ac{MOS} instruments. For these surveys the object is not typically spatially resolved and the science cases are mostly dominated by the desire to understand how galaxies and the corresponding luminosity functions evolve with time. 


For this to be known, both the composition and the redshift of the galaxy must be understood. This requires moderate spectral resolving power instruments such as VIMOS ($R$ of 200 to 2500), AAOmega ($R$ of 1000 to 8000). These instruments are used to perform survey such as zCOSMOS, VVDS, VIPERS and GAMA. These instruments and surveys have led to many interesting discoveries, such as the combined discovery of periodic variations in the density of visible matter, baryon acoustic oscillation, by the 2dF instrument \cite{peacock:2001}.
A current challenge is represented by the observation of faint galaxies with $z\simeq2$ (the so-called cosmic noon, where star formation had its peak), for which the visible spectrum of their stars appears in the near infrared. Because of chemio-luminescence generated by the formation of OH radicals in the upper earthly atmosphere, spectra of faint galaxies are difficult to acquire and evaluate. This problem has motivated the development of important astrophotonic devices which effectively suppress OH emission lines before a fibre-fed IR spectrograph (see Sec. \ref{sec:OH}).

\paragraph{Direct detection of Exoplanets} With photometric and time series observations of directly imaged exoplanets we can gain a wealth of information, constraining orbital parameters, size, temperature, surface properties and rotation rate \cite{traub2010direct}. To perform such observations is challenging, with the planet and star being separated by sub-arcsecond distances and with a star-planet contrast of $10^{6}$ - $10^{10}$. This means advanced \ac{AO} systems are needed and are often supplemented by a coronagraph to block light from the star. State-of-the-art systems often employ an additional spectrograph such as the \acp{IFU} for SPHERE \cite{Claudi:2008}, GPI \ac{IFU} \cite{Chilcote:2012} and CHARIS (SCExAO) \cite{Peters:2012}. \acp{IFU} not only allows characterisation of the planet's atmosphere \cite{bowler2010near}, but opens  up the possility of further reducing contrast by using techniques such as spectro-differential imaging \cite{racine:1999} and molecule mapping \cite{hoeijmakers2018mapping}.

\subsection{High-angular resolution}

The imaging technique is the second main pillar of modern observational astrophysics, hence the accessible level of high angular (or spatial) resolution (\textit{i.e.} beyond the limit imposed by the seeing) is a requirement shared by a large number of science cases.
In the last four decades, developments towards high-resolution imaging techniques have taken four main paths: development of large space telescopes, lucky-imaging, ground-based adaptive optics, optical long-base interferometry. The first three techniques allow to recover partially or totally the diffraction-limit of the telescopes, thus achieving angular resolutions ranging from some hundreds of mas to a few 10 mas depending on the wavelength band and the telescope aperture. Until recently, the main limit of space telescopes was represented by the limitations in weight and size of the payload of existing space carriers. This limited the aperture of the Hubble Space Telescope to 2.4 m, but represented still (at the time of launch) a remarkable improvement in resolution ($\sim$100 mas in the visible) as compared to existing ground-based facilities. Lucky imaging techniques could in principle rival or even exceed the capabilities of HST, but require a considerable data flow and post-processing overhead. Adaptive optics has seen an impressive development and diffusion in the last three decades, but (with some rare exception) is mostly  confined to less demanding infrared bands and thus to resolutions above 100 mas. The European Extremely Large Telescope with its extreme \ac{AO} system and an aperture of about 40 meters is expected to push this limit to a mere few 10 mas. 
Currently, the only technique capable of extremely high resolution at optical frequencies is long-base interferometry, which allows the synthesis of diffraction-limited apertures of several 100 meters. Again the technique is mostly possible at near- to mid-infrared with resolutions ranging from 1 to 10 mas.
These techniques have opened new opportunities in the investigation of the universe, the most significant achievements being related to the investigation of the early stages of stellar/planetary evolution and of the inner regions of galactic nuclei.
The accessibility to adaptive optics and speckle/lucky imaging in the infrared \cite{Ghez1993,Leinert1993,McCabe2006,Correia2006} complemented radial velocity surveys \cite{Mathieu1992} on young stars, confirming early studies showing that stellar multiplicity is a frequent property of solar-type stars with a binary rate larger than $\sim$50\% \cite{Duquennoy1991}. The access to milliarcsecond resolution delivered by long/base interferometry allowed furthermore to conclude that binarity was the norm rather than the exception\cite{Sana2014}. \\High resolution observations of circumstellar discs of nearby young stars with the \ac{HST} \cite{Padgett1999,Bally2000,Kalas2005,Kalas2007}, with \ac{AO}-assisted ground-based instruments at Keck and \ac{VLT} \cite{Perrin2006,Duchene2010,Garufi2017} or with long-baseline interferometers such IOTA, VLTI \cite{Monnier2002,VanBoekel2004,Renard2010,Lazareff2017} resulted in a deeper knowledge of the first stages of planetary and stellar formation, significantly contributing to the improvement of theoretical models.\\
The possibility given by imaging long-base interferometry to observe the surface of giant stars \cite{Buscher:1990} stimulated a new development in the study of convection structures in evolved stars \cite{Buscher:1990,Paladini2018,Roettenbacher2016,Ohnaka2017} or testing theoretical models of fast rotating stars \cite{Monnier:2007}.
Long baseline interferometry at Keck and VLT-I has made possible to understand the spatial properties of the circum-nuclear dust and the K-band nuclear emission in \acp{AGN} \cite{Swain2003,Burtscher2016}, while is the combination of milliarcsecond angular resolution and reverberation mapping may result in a new approach to cosmic distances estimate \cite{Hoenig2014}.\\ 
High-angular resolution techniques have been successfully used to push the limit of astrometric precision in the micro-arcsecond range. Examples are some of the major results in the Galactic Center science obtained thanks to the sharpest adaptive-optics images of the nuclear star cluster around Sagittarius A$^{*}$ \cite{Schoedel2002} and the recently started 10-$\upmu$as campaign with the GRAVITY interferometer \cite{Abuter:2017}, which may further change our understanding of strong gravity physics, while building on previous technical progress achieved in the astrometric use of long-baseline interferometers such as PTI\cite{Colavita:1999}.PRIMA \cite{Delplancke2006} and SIM \cite{Unwin2008}. \\

\subsection{High-contrast techniques}

High-contrast science may refer in the first place to the dynamic range of an image, or the amplitude between the readout noise and the saturation limit of the detector. This is particularly an important requirement for the study of stellar clusters, where a compromise between sensitivity and dynamical range needs to be found by playing on the detector (low/high) gain mode. However, with the rapid expansion in the last decades of the field of exoplanets and planet formation, high-contrast science has developed strong ties to the detection and imaging of faint targets and structures in the immediate vicinity of a bright central source. \\
Spectacular results in the field of exoplanets and disks direct imaging have been obtained with the use of coronographic instruments capable of significantly reduce the glow of the central star \cite{Oppenheimer2009}. Giant planets have been imaged within 0.5$^{\prime\prime}$ to few arcseconds from the their parent A-type stars \cite{Kalas2008,Marois2008,Lagrange2009} and offer hence excellent prospects for the future spectroscopic characterization of their atmosphere \cite{Janson2010,Bonnefoy2016}. Spatially-resolved sub-structures such as warps and spiral arms have been observed around AU\,Mic \cite{Boccaletti2015} and Herbig stars \cite{Benisty2015} using high-contrast techniques. Very high-precision V$^2$ interferometry in the near-infrared has been successfully employed to characterise the population of debris disks in the Solar neighbourhood \cite{Ertel2014}. Beside the existing experience, nulling interferometry may experience a new revival, following the DARWIN/TPF experience \cite{Fridlund2004}, for the spectroscopic evidencing of biological biomarkers in the atmosphere of Earth-like planets \cite{Kammerer2018}. The recent discovery of a terrestrial, non-transiting, planet orbiting our closest neighbour Proxima Centauri \cite{Anglada2016} will certainly further motivate rapid advances in the field of high-contrast techniques.\\
Concerning the instrumental requirements applying to high-contrast techniques, the wavefront phase needs to be controlled as -- or even more -- stringently than for imaging or classical long-baseline interferometry. Since contrasts from 10$^{-3}$ to 10$^{-9}$ need to be achieved, effects related to high ($>$90\%) Strehl ratios, control of pupil rotation, PSF centering, surface scattering, local intensity mismatches, differential polarization and long-term stability need to be addressed in order to expect small inner-working angles (IWA) \cite{Mawet2012} or deep and stable nulls \cite{Mennesson2014}. For science cases relevant to exoplanets and planet formation, the requirement of observing in the thermal infrared ($L$ to $N$, $Q$ astronomical bands) is of high relevance since the flux contrast between the central star and the planet/disk is more favourable than in the near-infrared or in the optical regimes. Considerations on the spectral richness of infrared spectra in terms of dust \& gas tracers and bio-signatures is a further motivation to operate at longer wavelengths as well. 
While a high total throughput is desirable to pick-up the faintest objects, the possibility to observe in broadband conditions is of high importance as well.

\subsection{Metrology and calibration techniques}

While not strictly related to one specific science case in astrophysics, the techniques employed for calibration and metrology in support to ground- or space-based observations are inescapable for an optimal interpretation of the science data.\\
One approach, definable as a {\it passive} calibration and likely the most used, consists in observing an astrophysical {\it standard} with exactly the same instrumental setup as the one adopted for the science target in order to identify features extrinsic to the object of interest and calibrate them out. The correction of strong telluric features in spectroscopy, the measurement of photometric standard stars for photometry and the acquisition of so-called PSF reference stars for spatially resolved imaging are the most common examples of passive calibration.\\
Beside this, it is also common to use {\it active} calibration systems when a precise knowledge of the time-dependent long-term drifts and stability/uniformity is required, which may be difficult to monitor with on-sky calibration. A good example is seen in high-resolution spectroscopy with a typical resolving power of $R \simeq$100,000 when precise radial velocity measurements are sought for the detection of planets. Since the relative spectral shift of the stellar lines must be determined to a precision on the order of m/s, meaning a precision of $\sim$1/1000 resolution element, high-stability calibration sources such as Thorium-Argon lamps or iodine cells have been employed to track instrumental drifts down to m/s. While spectral lamps have been central for the operation of the HARPS optical spectrograph, the upcoming ESPRESSO spectrograph at the \ac{VLT} will make use of spectral templates of superior stability and accuracy delivered by a laser frequency-comb \cite{Pepe2010}. Artificial sources are also employed to assess the spatial flatness of a detector response, or the quality of the sky thermal background suppression at infrared wavelengths. In wide-field imaging applications, the monitoring of static and dynamic field distortion effects need to be traced, which is typically achieved by using an artificial scene of widely and uniformly distributed point sources. Finally, the effective suppression of unwanted sky lines at optical and near-infrared wavelengths by mean of carefully tailored narrow-line filters is also highly desirable. \\
Metrology becomes critical when unavoidable drifts and flexures of a large facility or instrument hamper the ultimate expected accuracy. A Helium-Neon laser metrology is usually implemented to achieve (sub-)wavelength position accuracy at 632\,nm with an intrinsic relative precision better than 2$\cdot10^{-9}$. Some noticeable cases can be mentioned: the laser metrology system of the high-precision astrometry instrument Gravity/VLTI launches its beam from the interferometric lab up to the telescope pupil to trace back the path of the astrophysical beam \cite{Lippa2016}. Considering that the astrometric angle $\Delta\alpha$=$\Delta$OPD/$B_{p}$, the uncertainty on the length of the projected baseline needs to be sufficiently small not to dominate the overall error budget. With a laser metrology, the precision on the length of the interferometric (projected) baseline $B_{p}$ greatly surpasses what is typically delivered by pointing models; another example of critical metrology need is found with the NEAT (now THEIA) project of space-based high-precision astrometric mission \cite{Malbet2012,Boehm2017}. The small differential displacement of an Earth-hosting star with respect to the background reference stars due to the orbiting planet can only be measured if the uncertainties on the geometry of the focal plane array (FPA) are calibrated down to one part in 10$^5$. The interferometric laser metrology system illuminating the FPA can help to retrieve the ultimate position of the star PSF centroid \cite{Crouzier2016}.\\
The instrumental requirements inherent to active calibration and metrology activities are clearly driven by an excellent knowledge and understanding of the implemented hardware: this means that particular care needs to be taken to study the spectral content, stability and repeatability of a calibration lamp. Modern laser sources have made enormous progress in that sense. The characterisation of the transparency range, optical quality and modal content of any passive component involved in the calibration chain (e.g. filters, optical fibers, phase modulators) must result from dedicated lab testing and, ideally, rely on high-TRL devices. Finally, the possibility to simplify the optical design of any calibration or metrology system for stability purposed should be considered as an important requirement: in this sense, optical fibers that can efficiently transport, filter or mix light from different physical locations will play an increasingly important role in astronomical instrumentation.

\section{A very short introduction to Photonics} 
\label{sec:photonics_perspective}

In this review, we present astrophotonic instrumentation not only according to their astronomical use, but also from the perspective of the employed photonic technologies.
To this end, we use this chapter to introduce basic photonic concepts, such as waveguides and optical modes. These concepts will be used in Section \ref{sec:astrophotonics} to classify astrophotonic instruments and discuss their properties.   

Photonic components can be broadly divided into passive and active devices. Passive devices are optical elements introducing a static modification to the properties of light. In these devices, the power of light is preserved or, more often, is reduced by losses. On the contrary, active devices can dynamically modify the state of light by means of an interaction with an external agent. The output light power in active devices can be greater than its input, the gain being supplied 
by an external power source (\textit{e.g.} electricity).
As mentioned, in this review we will discuss photonic applications to astronomy based on components modifying the properties of light on spatial scales of the order of 
the wavelength of light. Under this classification we can include micro-optics, phase masks such as gratings (see Section \ref{sec:spectroscopy}) or vortex phase 
masks (see Section \ref{sec:coronagraph}) but not conventional lenses, as the former requires the structuring of an optical surface at the micro- nano-scale.

A fundamental component falling in our classification of passive photonic devices is the optical waveguide. This is an heterogeneous optical medium consisting of a region of space with high aspect ratio (the core) characterised by a refractive index higher than its surroundings (the cladding) and 
transverse dimensions comparable to the wavelength of light. In such a medium, light can be confined in the core and propagate along the long axis (or longitudinal axis, conventionally oriented to coincide with the $z$-coordinate axis) thanks to the phenomenon of total internal reflection.
The physical principle underlying waveguiding can be exemplified by simple considerations in the frame of the geometrical optics approximation. Neglecting light interference effects, we can show that light can propagate in the core of the waveguide if the external divergence angle of light (respect to the longitudinal axis) is smaller than the numerical aperture of the waveguide:

\begin{equation}
    NA=\sqrt{n_{\rm{co}}^2-n_{\rm{cl}}^2},
\end{equation} 

With $n_{\rm{co}}$ and $n_{\rm{cl}}$ being the refractive indices of the core and the cladding, respectively.
The exact modeling of waveguides with transverse core dimensions comparable with the wavelength of light require the solution of wave-equations and the introduction of the concept of spatial modes, as outlined below.  

Optical waveguides can come in the form of a glass fiber or as an element of an integrated optical circuit, \textit{i.e.} waveguides manufactured on a glass substrate. 
The state of light in waveguides can be manipulated by means of more complex passive components analogous to macroscopic devices, such as beam splitters 
(optical couplers), mirrors (Bragg gratings and micro-resonators) or phase plates (birefringent waveguides). 
 
As active devices of interest for astrophotonics, we will consider mainly lasers and phase modulators. Lasers are optical media in which the population of electrons of a 
radiative electronic transition is inverted respect to the state of thermal equilibrium. The population inversion is typically obtained by injecting power in the medium as 
optical radiation or electrical current. Light resonant to a radiative transition in an inverted medium is amplified coherently because of the occurrence of stimulated light emission. Lasing media are typically inserted in an optical resonator, where light can be amplified by many orders of magnitude by passing repeatedly in the amplifier thanks to multiple reflection in the resonator. Currently the laser resonator can be fabricated within a single optical fiber or waveguide using fiber Bragg gratings as mirrors, or manufactured in semiconductor waveguides within a laser diode.     
Photonic phase modulators are devices using an electrical signal to induce a local variation of the refractive index in a waveguide, which can advance or retard the phase of a guided optical field.
They are typically based on the electro-optical effect, by which a constant electrical field can alter the birefringence properties of the medium where the waveguide is manufactured. By integrating a phase modulator within an integrated Mach-Zehnder interferometer amplitude modulators can be realised as well. 

\subsection{Optical modes}

An important concept in photonics is the optical mode, which is a stationary state (or eigenstate) of an optical system eventually originating from its boundary conditions and the wave nature of light. 
Our everyday experience of wave phenomena can provide us a clear visualisation of the abstract concept of mode. The waves generated 
by tipping regularly on the rim of a glass filled with water or the movement of an elastic string tended between two points are all phenomena that give rise to mechanical standing waves, which can be represented mathematically as modes.
In these cases the temporal oscillation of the water or the string is modulated spatially by a sinus-like profile consisting of an integer number of half wavelengths, the periodicity of temporal and the spatial oscillations being related to each other through the speed of sound in the medium.   

Mathematically, modes are solutions of differential equations which describe the physical system under study. Optical modes are solutions of the 
electromagnetic wave propagation equation:
\begin{equation}
\nabla \times \nabla \times \vec{E} - \frac{n^2(\vec{x})}{c^2}\frac{\partial^2\vec{E}}{\partial t^2}=0,
\end{equation}
which is straightforwardly derived from the Maxwell equations. Here $n(\vec{x})$ is the spatially varying refractive index and $c$ is the speed of light.
This equation is usually cast into an eigenvalue problem assuming an harmonic 
dependence in one or more coordinates. As an example, let's consider the case of the weakly guiding optical waveguide, \textit{i.e.} a waveguide satisfying the relationship 
$\Delta n=n_\mathrm{co}-n_\mathrm{cl}\ll n_\mathrm{cl}$. In this case, we can safely assume that the electromagnetic field is transverse respect to the axis of the waveguide and the vectorial 
components of the electric field are decoupled. This allows writing the wave equation in a scalar form for the amplitude of one of the polarisation directions of the field:   
\begin{equation}
\label{eq:scalar_we}
\nabla^2 E_{x,y}- \frac{n^2(\vec{x})}{c^2}\frac{\partial^2E_{x,y}}{\partial t^2}=0.
\end{equation}
If the refractive index profile of the core is invariant along $z$,  the $z$-(longitudinal) component of the electric field has an harmonic dependence in time and z:
\begin{equation}
\label{eq:mode}
E_{x,y}(x,y,z,t)=\psi(x,y)\cdot\exp[i(\beta z - \omega t)] + c.c. .
\end{equation}  
As a consequence, the transverse profile of the field $\psi(x,y)$ obeys the following eigenvalue equation, derived by substituting Eq. (\ref{eq:mode}) in the scalar wave equation Eq.(\ref{eq:scalar_we}):
\begin{equation}
\left[\frac{\partial^2}{\partial x^2} + \frac{\partial^2}{\partial y^2}+ n(x,y)\frac{\omega^2}{c^2}\right] \psi=\beta^2\psi,
\end{equation}
where $n(x,y)$ is now the refractive index distribution in the transverse $(x,y)$ plane, which achieves its maximum in the region of the core. 
If the refractive index distribution has a peak, the eigenvalue equation has a discrete set of solutions which represent the transverse modes of a waveguide.
The eigenvalues are found by imposing everywhere  continuity in value and derivative to the field $\psi(x,y)$, a consequence of $n(x,y)$ possessing at most a finite discontinuity in value (\cite{SnyderLove}, chapter 33). 
Step index waveguides with circular cross section, \textit{i.e.}  waveguides with a cylindrical core of constant refractive index and abrupt transition to the cladding, are a particular subgroup of
waveguides which possess mode profiles in analytical form. They represent a good approximation of real optical fibres which can support one or more discrete modes, the modal behaviour being 
parametrised by the normalised frequency $V$: 
\begin{equation} 
V= \frac{2\pi}{\lambda}a\cdot NA,
\end{equation}
where $a$ is the core radius and $\lambda$ the free space wavelength of the light. For $V$ smaller than 2.405, circular waveguides can support a single mode. 
Asymptotically the number of modes supported by a step index waveguide with circular cross section can be obtained from:
\begin{equation}
\label{eq:modes_fiber}
N=\frac{V^2}{4}.
\end{equation}

Modes can also have a longitudinal attribute, as is the case of optical resonators, regions of space where light is trapped \textit{e.g.} in the volume 
between two parallel plane mirrors. The mirrors introduce a boundary condition for the electromagnetic field similar to that of a vibrating string (the optical field 
vanishes at the mirror surface) so that only a discrete set of longitudinal waves characterised by an integer number of half wavelengths are supported by 
the resonator.   

Modes can be excited by matching an external optical field to the modal field distribution at the boundaries of the photonic device. 
Examples are the excitation of modes in a fibre by illuminating its tip with a beam having the same spatial distribution of the mode or illuminating the semi-reflecting mirror of an optical resonator with 
light tuned at the frequency of its stationary waves. Modes form a complete orthonormal base for stationary fields sustained by the photonic component and thus their complex amplitude $a_
\mathrm{j}$ can be obtained by projecting the exciting field distribution $E_\mathrm{ext}(x,y)$ onto the mode profile $\psi_\mathrm{j}(x,y)$ at the interface:
\begin{equation}
a_\mathrm{j}=\sqrt{\frac{n_{co}}{2}}\left(\frac{\epsilon_0}{\mu_0}\right)^{1/4}\frac{\int_{S}\psi_\mathrm{j}(x,y) E_\mathrm{ext}^{*}dA}{\left[\int_{S}\left|\psi_\mathrm{j}(x,y)\right|^2 dA\right]^{1/2}},
\end{equation}
where $S$ represents the external surface of the modal volume. Normalisation in this case is chosen so that the power carried by the $j$-th mode is given simply by the square modulus of $a_\mathrm{j}$.

\subsection{Modes and seeing}
Highly multi-mode fibres can be described safely in the frame of the geometrical optical approximation. This gives us the possibility to use the brightness theorem \cite{BornWolf}, 
to describe the seeing limited PSF of a telescope in terms of modes. 
Moreover, as the brightness theorem is basically a formulation of the second principle of thermodynamics\cite{McMahon:1975}, we derive a useful lesson regarding 
mode transformation devices. 
The brightness theorem states that the power per unit area and solid angle (the brightness) of the image of a source of light formed by a passive optical system cannot 
exceed the brightness of the source itself. Since the brightness is related to the temperature of the source, it is clear that a violation of the brightness theorem could allow a 
perpetual motion machine to work.
In a lossless passive optical system the collected power is preserved thus the brightness theorem is equivalent to state that the etendue of the source and the image are the same, 
the etendue $\mathcal{E}$ being defined as the product of the source area $A$ by its solid angle divergence $\Omega$: 
\begin{equation}
\mathcal{E}=A\cdot \Omega.
\end{equation}
We consider now an optical system consisting of a seeing limited telescope focusing light on a multi-mode optical fibre placed in the focal plane. 
For the image of starlight in the focal plane of a telescope of diameter $D_{\rm{T}}$ and focal ratio $F_\sharp$ the etendue can be estimated as:
\begin{equation}
\mathcal{E}_\mathrm{tel}=\pi (\theta D_{\rm{T}} F_{\sharp})^2 \pi \left(\frac{1}{2F_\sharp}\right)^2=\frac{\pi^2}{4}(\theta D_{\rm{T}})^2.
\end{equation} 
Here the angle $\theta$ represents the seeing. On the other side, the etendue of light propagating in a step-index optical fibre can be estimated as the area of the core 
of radius $a$ by the square of the numerical aperture:
\begin{equation}
\mathcal{E}_\mathrm{fib}=\pi a^2 \pi NA^2=\frac{V^2}{4}\lambda^2,
\end{equation}
where we have used the definition of the normalised frequency $V$ of the waveguide Eq.(\ref{eq:modes_fiber}). In this case, $\lambda^2$ can be interpreted as the 
etendue of a single optical mode.
Because in lossless systems the etendue of light is a constant, we can write:
\begin{equation}
\frac{V^2}{4}\lambda^2 = \frac{\pi^2}{4}(\theta D_{\rm{T}})^2
\end{equation}
Recalling that the seeing angle can be roughly defined as $\lambda/r_{\rm{0}}$, $r_{\rm{0}}$ being the Fried parameter (the correlation length of the atmospheric refractive index distribution \cite{Fried:1966}) we obtain: 
\begin{equation}
\frac{V^2}{4} = \frac{\pi^2}{4}\left( \frac{D_{\rm{T}}}{r_0}\right)^2
\end{equation}
in which the expression $\frac{\pi^2}{4}\left( \frac{D_{\rm{T}}}{r_0}\right)^2$ can be interpreted as the 'modal content' of the point spread function of the telescope.

The conservation of brightness has further implications which are useful for astronomical instrumentation. In particular it shows the impossibility to have a passive, lossless 
device converting multimode light into a single mode \cite{Welford:1982}. If such a device were possible, the brightness of light confined in the single mode output should 
necessarily increase because of the reduction of the etendue, thus violating the second principle of thermodynamics.
This is the reason why multimode devices with single mode behaviour such as the photonic lantern \cite{LeonSaval:2005,Birks:2015} distribute multi-mode light to an equivalent number of single mode waveguides.

In active devices brightness is not preserved, but the corresponding reduction 
of entropy is compensated by the necessity to extract information from the system.
This is for instance the case of an adaptive optics system coupled to a single 
mode fibre, which uses the information of the wavefront sensor to correct the 
wavefront and concentrates the 
power in a single mode of the fibre.
The reduction in entropy associated to the correction of the aberrations is 
largely compensated by the entropy increase of the wavefront measurement 
operation, which requires the 
absorption of light on a detector.  

\subsection{Spectroscopy}
\label{subsec:spectroscopy}
In a simplified dispersive spectrograph, the light enters through a slit that selects a small region of the field, the light is collimated, a prism/grating or other dispersive element disperses the light, and a camera lens focuses the light onto a detection surface. 
One of the most important requirements for spectroscopy is the achievable spectral resolving power $R$=$\lambda$/$\Delta\lambda$, which is usually driven by a particular science case. 
It determines our ability to measure flux densities at two nearby wavelengths separated by $\Delta\lambda$.
For grating spectroscopy for example the spectral resolving power is calculated by

\begin{equation}
	R = \frac{m \rho \lambda W}{\theta D_{\rm{T}}}.
   \label{eqn:spec_resolving}
\end{equation}

where $m$ is the diffraction grating order, $\rho$ is the rulings on the grating, $\lambda$ is the operating wavelength, $W$ is the number of illumined rulings, $\theta$ is the angular seeing and $D_{\rm{T}}$ is the telescope diameter.

This means that in systems where diffraction is dominant ( $\theta D_{\rm{T}} = \lambda$) the spectral resolving power is limited by the properties of the diffractive element. However in non diffraction limited cases ( $\theta D_{\rm{T}} > \lambda$), $R$ will be constrained by the implemented dispersion element (\textit{e.g.} grating, prism), the diameter of the telescope feeding it, and the slit size itself. This dependence is well established in classical instrumentation and further information can be found in spectroscopy textbooks (e.g. \cite{Schroeder1987}).

\subsection{Modal noise}
\label{subsec:modal_noise}
As mentioned in the previous sections, multi-mode fibers contain different modes depending on their properties. Depending on the incident electric field, wavelength, fiber stresses and strains different modes will be excited. With a constantly moving telescope and changing atmosphere this translates to an ever changing output illumination pattern from the fiber. In an astronomical spectrograph this phenomena is known as modal noise and results in a variation in the measured position of a certain wavelength.

Whether this variation is important greatly depends on the science case, where wavelength precision is not required this is not a problem, however for highly precise measurements, e.g. exoplanet detection, this can seriously influence the accuracy of results \cite{Rawson:1980}. There are various ways of controlling this, for instance agitating the fiber \cite{baudrand2001modal} and this works sufficiently for the current generation of spectrographs, which aim for around 1 m/s precision and work in the visble with many modes. However, this problem is particularly pronounced as the number of the modes within the fiber is reduced, statistically increasing the relative uncertainty due to this movement. The next generation of insturments, which will aim for higher precision and work in the infra-red will be particularly affected.

It must also be noted, that as the number of modes reduces, eventually to one (not including polarisation), this then eliminates this effect. However, it has been noted that polarisation then becomes a dominant effect \cite{Halverson:2015}.

\section{Astrophotonics}
\label{sec:astrophotonics}

\begin{figure}
	\epsfig{file=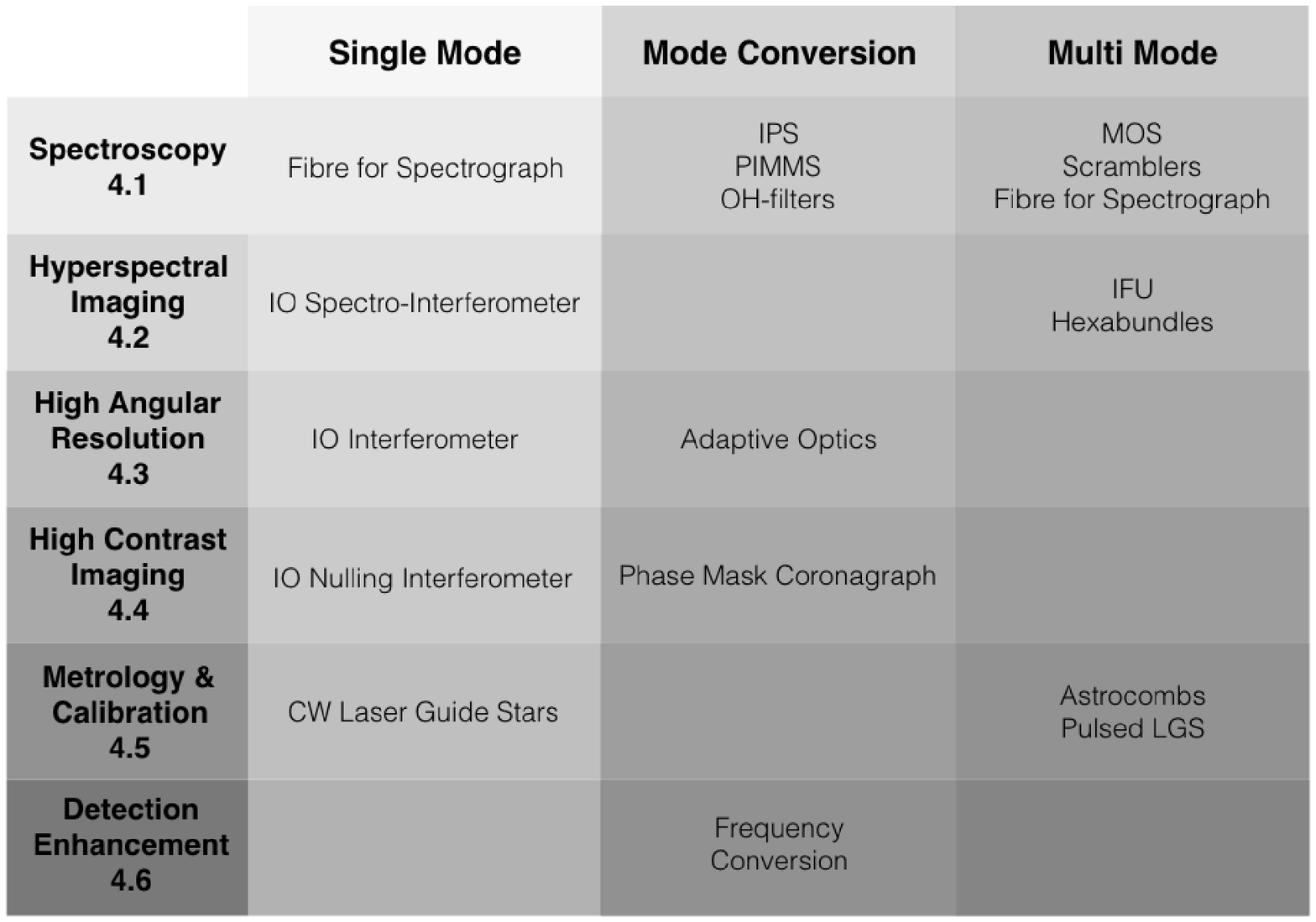, width=\textwidth}
    \caption{\label{fig:table} Astrophotonic instruments can be classified according to the scientific purpose (rows) and the modal behaviour (columns) of their key photonic component.}
\end{figure}

In this section we review photonic devices developed so far for astronomical instrumentation and classify them by astronomical technique and photonic function.
As indicated in Figure \ref{fig:table} 1, we distinguish the 6 main astronomical techniques (namely spectroscopy, high angular resolution, hyperspectral imaging, high contrast imaging and metrology/calibration, detection enhancement) whose scope and requirements have been discussed in Section 2. 
The photonic functionality distinguishes 3 categories depending on the modal content of the key photonic component of the instrument, \textit{i.e.} single mode, multi-mode and mode transformation 
devices.
Under single-mode devices we include continuous wave lasers operating at a single longitudinal mode, or devices based on single-mode optical waveguides. 
Devices like laser guide stars and integrated optics beam combiners for interferometry are therefore found in this category.
Multimode devices are pulsed laser sources, which emit on a multitude of longitudinal modes, or devices based on multi-mode optical fibers.
Laser frequency combs or 3D spectroscopy instruments (MOS or IFS) are typical representative of this category.
In between the single mode and the multimode categories we can find mode transformation devices, which operate a modification of the modal distribution between input and output.
These are for instance passive devices like the photonic lantern and diffraction
gratings (MM to MM),  phase masks (SM to SM) or active devices like the deformable mirrors in an adaptive optical system (MM to SM).
The table in Figure \ref{fig:table} gives an overview of the astrophotonic instruments according to the dual classification described above.
In the following pages the sub-sections are devoted to the afore mentioned astronomical techniques, while paragraphs distinguish the various photonic functionalities underlying the instruments.

\subsection{Spectroscopy}
\label{sec:spectroscopy}

\paragraph{Single Mode Spectrographs} \label{sec:sm_spec}
The mismatch between the diffraction-limited resolution of the telescope $\lambda/D_T$ and the seeing $\theta$ reduces the spectral resolving power of the spectrograph, which can be restored only by making the grating (and hence instrument) larger or reducing the size of the slit (which can cause a loss in light unless techniques such as image slicing is used). The smallest spectrograph is then in general the one for which $\lambda/D_T=\theta$, the so-called diffraction-limited spectrograph.
From the point-of-view of astronomical instrumentation, a truly diffraction-limited spectrograph could offer, besides size reduction, two further main advantages. Firstly, improved stability through the elimination of modal noise in fibres (see Sec. 3.4), and secondly a further reduction of cost, thanks to either mass produced components for the photonic market or smaller conventional components. 
Truly diffraction limited spectrographs could be obtained by feeding the spectrograph with one or multiple \acp{SMF} acting as spatial filters (a single mode spectrograph). 

Whilst the idea of using spectrographs fed by \acp{SMF} is not new to astronomy,  coupling starlight to them is notoriously inefficient, due to atmospheric turbulence \cite{Shaklan:1988}. A sophisticated \ac{AO} system is therefore still mandatory to reduce coupling losses. Before truly \ac{SM} coupling was considered there were attempts using \acp{FMF} in 1998 \cite{Ge:1998}, though with a light loss still unacceptable for astronomical spectroscopy. This restriction means most current spectrographs are fed using \ac{MM} fibres. In recent years developments in \ac{xAO} systems now routinely allow high Strehl ratios and hence increaingly efficient coupling into \acp{SMF}. This is leading to the first generation of \ac{SMF} fed instrumentation, with examples of coupling tests on SCExAO \cite{jovanovic:2017} and full instruments in Minerva-red \cite{Blake:2015} and ilocater \cite{crepp:2016}.

\paragraph{Single mode reformatters and spectrographs}
To overcome the requirement of using \ac{xAO} while preserving a high throughput, single mode spectrographs fed by a \ac{PL} \cite{LeonSaval:2005,Birks:2015} have been proposed (\ac{PIMMS} \cite{BlandHawthorn:2010}). \acp{PL} are tapered optical fibres which adiabatically transform a \ac{MM} optical fibre  into a collection of \acp{SMF}. Light injected at the \ac{MM} end of the device is distributed without losses in the output \acp{SMF} provided their number is at least equal to the number of modes supported in the \ac{MM} end. A lossless transition to fewer fibres is in fact prohibited by the brightness theorem (see Sec. 3.2). \acp{PL} come in the form of fibers \cite{LeonSaval:2005} or integrated 3D waveguides \cite{Thomson:2012,spaleniak:2013} and can provide single mode functionality to \acp{MMF}. This quality, originally exploited for the development of astronomical notch filters (see below), allows in principle efficient coupling from seeing-limited telescopes, while allowing light dispersion with a compact single mode spectrograph. The \ac{PIMMS} concept has two possible implementations, either the \ac{SMF} ends are rearranged into a linear array acting as the pseudoslit of a spectrograph (also known as reformatting of light), or they are fed to integrated spectrographs (see next paragraph). 
Further elaborations of the \ac{PIMMS} concept have been conceived or developed. The simplest is the photonic Tiger concept (named for the fibre fed Tiger \ac{IFU} \cite{LeonSaval:2012} where no reformatting is used and a \ac{MCF} feeds the entrance slit to the spectrograph. This means the spectra need to be dispersed such that none of the inputs overlap. Alternatives have been put forward, to use \ac{ULI} to write a 3D integrated optical component reformatting the \ac{MCF} into a long slit. These are either separate waveguides \cite{Thomson:2012,spaleniak:2013} or joining the waveguides together to form a long slit \cite{maclachlan:2016}. The advantage of this method is similar to conventional slit techniques, maximising use of the detector. 
In addition fibre reformatters, taking the \ac{PSF} from the telescope and reformatting into a slit have now been proposed and tested in the lab, showing high levels of throughput \cite{yerolatsitis:2017}, these are efficient, but in their current form limited to low mode counts due to the complexity of the devices. Throughput in low-mode fibers can be improved by improving the PSF of the telescope with a low order AO system \cite{harris:2015}. A preliminary numerical investigation of the trade-off between telescope beam quality and throughput is presented in \cite{Diab:2018}.
Some of these devices have been tested on sky either with a spectrograph (e.g. the TIGER spectrograph \cite{LeonSaval:2012}), or without (e.g. the Photonic dicer \cite{harris:2015} and the hybrid reformatter \cite{maclachlan:2016_hybrid}).

\paragraph{Integrated Photonic Spectrographs} 
\label{sec:IPS}
As astronomical instruments are constantly growing in size, techniques allowing the reduction of their size and complexity are very popular. In telecommunications this has been an aim for many years, resulting in the delivery of mature devices on centimetre size scales. 

The idea of using these technologies in astronomy was suggested as early as the mid 90s \cite{Watson:1995,Watson:1997} and was initially based upon the \ac{AWG}. The \ac{AWG} works in a similar way to a conventional spectrograph, though embedded within a glass chip, which reduces size and alignment complexity. A sketch of the device is illustrated in Figure \ref{fig:AWG}. The initial single mode waveguide can be considered like the slit, or feeding fibre to the spectrograph. The beam diffracts in the first free propagation region, which acts as a collimator thanks to the confocal curved surfaces of its edges (\cite{Goodman}, Chapter 4). The expanded beam is sampled by an array of waveguides which, thanks to their curved paths, add a constant incremental phase difference to the light propagating in neighbouring waveguides. The output of the array of waveguides is connected to a second free propagation region, this time acting as a camera lens. Similarly to a conventional spectrograph, spectral lines are focused at the curved exit surface of the free propagation region only if the optical path difference from the focus to any pair of outputs of the output array is an integer multiple of the wavelength. In telecommunication applications, the single mode fibres at the output of the second free propagation region  are placed at regular intervals  to collect the light of separate  wavelength-demultiplexed communication channels. Since the incremental optical path difference introduced by the array of waveguides usually corresponds to several optical wavelengths, the \ac{AWG} operates at high diffraction order ($\sim25$) like a conventional Echelle grating. As a consequence, the \ac{FSR} of the \ac{AWG} is very small, and their use for astronomy requires the introduction of a cross-disperser at its output to separate the diffraction orders.

At the time of \cite{Watson:1995} the AWG technology was not considered sufficiently developed to be used for astronomy. During the 2000s, the technology was reconsidered by Bland-Hawthorn\&Horton \cite{BlandHawthorn:2006} who compared the \ac{AWG} and \ac{PEG} and concluded that the technology was now sufficiently developed for use in astronomy.

\begin{figure}
\caption{A Schematic of an astronomical \acf{AWG} reproduced from \cite{Douglass:2018}. Here the input is single mode waveguide, which is fed by light from the telescope. This light passes through the input free propagation zone and enters the waveguide array where a phase difference is added (analogous to a grating). The light is recombined in the output free propagation zone and the spectrum is sampled either at the output of the chip or cross dispersed.}
\label{fig:AWG}
\end{figure}

The first on sky test of a modified commercial \ac{AWG} was performed at the \ac{AAT}, formerly the Anglo-Australian Telescope in Australia \cite{cvetojevic:2009}. For this test the device was coupled to the telescope using a single, \ac{SMF}. This led to low coupling efficiency with the seeing limited \ac{AAT}, though high enough to pick up the atmospheric emission lines. As mentioned, due to the designed low \ac{FSR} ($\approx$ 57 nm) of the device, the IRIS 2 \cite{Tinney:2004} spectrograph was used to cross disperse the output in order to obtain a useful \ac{FSR} for astronomy. This first prototype displayed a spectral resolving power of around \textit{R} $\approx$ 2100 in the lab, which was suitable for low resolution applications (see Sec. 2.1).  

In the next set of experiments various improvements were made \cite{Cvetojevic:2012}. These included 1) optimising production of the devices for efficiency. removing the tapers of the input waveguides in the \ac{FPZ} (which permitted an increased spectral resolving power, and 3) improving the coupling efficiency with the telescope using a \ac{PL}. A single optimized AWG was then coupled to the IRIS2 spectrometer for cross dispersion.

Recently more efficient devices have focussed on improving coupling, either through using a \ac{SMF} with the extreme \ac{AO} system SCExAO \cite{Jovanovic:2015,Jovanovic:2017b} or with a \ac{PL} input \cite{Cvetojevic:2017}. However, the experiments using the \ac{PL} were shown to induce modal noise due to a modal mismatch between the \ac{MMF} and \ac{PL} used \cite{Cvetojevic:2017}. This shows the importance of correctly designing \ac{IPS} systems and photonic systems in general. 

There have also been advances in developing custom \acp{AWG} for astronomy. These have been successfully manufactured for low spectral resolving power \acp{AWG} (e.g. \cite{Gatkine:2016,Gatkine:2017}). Whilst high spectral resolving power (of order \textit{R} = 60,000) \ac{AWG} \cite{Stoll:2017} has been designed, these are more difficult to manufacture and currently none are in use.

Another category of integrated spectrometers proposed for astrophotonic applications is based on the measurement of the temporal coherence of light in a waveguide. The so called Standing Wave Integrated Fourier-Transform Spectrometer (SWIFTS \cite{LeCoarer:2007}) consists in a sensing waveguide coated with a regular array of gold nano-antennas, used to sample an optical standing wave trapped in the waveguide. In order to generate the standing wave, the outputs of a 2-way waveguide splitter are connected to both ends of the sensing waveguide.  Alternatively, light is injected at one end of the sensing waveguide, which is ended by a reflecting mirror. The standing wave is then formed near the mirror interface (Lippmann configuration). The nano-antennas are excited by the evanescent field of the guided mode and radiate light, which can be collected by an optical microscope. As a consequence, only a sub-Nyquist sampling of the standing wave is possible. This implies that the SWIFTS can be used at very high resolving powers (100000 or more) on very small bandwidths. A possibility to achieve Nyquist sampling is to split the incoming light on an array of SWIFTS in Lippmann configuration, with the end mirror cut at an angle allowing to introduce an incremental phase shift for each SWIFTS, which can be used to densify the sampling of the standing wave. Even though they were initially proposed for astrophotonic applications \cite{LeCoarer:2007,Kern:2009} (see also Sec. \ref{sec:spectro_interferometry}), SWIFTS have intrinsically a low sensitivity, since the out-coupling efficiency of the nano-antennas should be low enough to avoid depleting the standing wave within a few optical cycles. Nonetheless their extremely small footprint and the possibility to paste the detector directly on the chip surface makes them attractive and sensitivity can be traded off for an increased multiplexing of the instruments.

\paragraph{OH suppression} \label{sec:OH} The constantly varying OH lines in the night sky are a large problem for ground based observations of faint objects in the \ac{NIR}, as for instance galaxies of the 'cosmic noon' which possess a redshift of $z\sim2$ \cite{BlandHawthorn:2004}. To try to reduce the 
intensity of the telluric lines, the use of a tailored mask covering them in the focal plane of a high resolution spectrograph has been attempted with a certain success \cite{Iwamuro:1994}. However, focal plane masks cannot remove the Lorentzian tails of the PSF of the spectrograph, which scatter light in the interline spectrum and adds a background to the target spectrum. Though small, this effect is 
a significant contributor to noise in the spectra of the faintest objects.

\begin{figure}
	\caption 
	{a) The top shows a schematic of a Fibre Bragg grating. Refractive index variations in the core of the fibre act as a Bragg reflector and reflect periodic wavelengths (c) back in the fibre whilst transmitting other wavelengths (d). Adapted from \cite{BlandHawthorn:2011}. b) Schematic of a ring resonator, adapted from \cite{Ellis:2017}. Light from the 'input' waveguide is evanescently coupled into the ring and on resonance light is coupled to the 'drop' waveuide, off resonance passes back to the 'through'.} 
\end{figure} 

To achieve ultimate suppression of OH lines, filtering should be introduced before the spectrograph. Following this approach, multi-notch filters \cite{Offer:1998} and volume holograms \cite{Blais-Ouellette:2004} were proposed, but the developments did not reach the level of prototype test on sky due to technical limitations of the fabrication methods. A more promising approach involves the use of Fiber Bragg Gratings (FBG) in SMF \cite{BlandHawthorn:2004,BlandHawthorn:2011}. 
A \ac{FBG} is a periodic refractive index variation within the core of a fibre. Because of the boundary conditions at the interface of media with different refractive index, the FBG acts as a periodic collection of semi-reflecting mirrors with tiny reflectivity. If the period of the grating is equal to an integer multiple of half wavelengths of the light, all these reflected waves can interfere constructively and the fibre acts as a narrow-band reflecting mirror (see Fig. \ref{fig:FBG}). As discussed, coupling starlight in SMF is very inefficient, reason why MMF are preferred. The reflection line of a FBG written in MMF splits however in several lines due to the slightly different effective wavelength of each guided mode which smear the reflection spectrum and reduce the overall rejection ratio of the filter. In order to get a single-mode-like FBGs a MMF photonic lanterns were invented \cite{LeonSaval:2005}. 
After conceptual development the devices were trialled on sky with the GNOSIS spectrograph. The \ac{AAT} fed a 7 element \ac{IFU}, of which one of these was picked off by a \ac{MMF}. This fed a 1 x 19 core \ac{PL}, turning the \ac{MMF} into 19 \ac{SMF}. Each of these fibres contained a collection \acp{FBG} tuned to reflect 145 OH lines in H-band and were then fed to a reverse \ac{PL}, turning the \acp{SMF}s into a \ac{MMF} which fed the IRIS 2 spectrograph \cite{Trinh:2013}. Though this initial experiment was only used to examine the sky background and no objects were observed, they found that the OH suppression worked well.

Because of the time consuming nature of writing \acp{FBG} into \acp{SMF} other methods of manufacturing them have been proposed, these include; writing the gratings directly into a \ac{MCF} \cite{Lindley2014}, though currently this suffers from imperfections in the gratings due to focussing effects; and using ultrafast laser inscription \ac{ULI} to reformat the waveguides into a straight line, avoiding focussing effects and write the gratings there \cite{brown:2012,spaleniak:2014}. 

Whilst no full instruments exist, one future instrument (PRAXIS) is currently in the build phase and should be commissioned in the summer of 2018 \cite{horton:2013,Ellis:2016}.

Performing OH suppression using a telecom add-drop filter based on multiple ring resonators has also been suggested (see Fig. \ref{fig:ring_res}). These devices evancescently couple light from the input waveguide (the bus waveguide) into a ring. If an integer number of wavelengths match the perimeter of the ring resonator (the resonance condition), light drops from the bus waveguide and is ‘stored’ in the resonator. A second waveguide placed opposite to the first one is eventually used to dump the light stored in the resonator. For light off-resonance, no stationary field can be sustained in the ring and light is completely transmitted through the first waveguide into the spectrograph \cite{ellis:2011,Ellis:2017}. 
In order to properly suppress the OH night sky emission multiple rings are required and as such no on-sky demonstrators exist.

\paragraph{Incomplete scrambling and modal noise}

The majority of high resolution, high stability spectrographs (with a few exceptions) are now fibre fed. This has two main advantages, the first is that as fibres confine the light,  this allows the spectrograph to optically independent from the telescope. This allows the final spectrograph to be placed remotely in a controlled environment, making stabilisation easier. The second advantage is that fibres have better scrambling properties than other types of spectrographs (e.g. \cite{Heacox1988}). This means that as light travels along the fiber it is randomised, removing the imprint of the initial conditions of the light at the telescope, this helps with stability in the final spectrograph as it is less sensitive to atmospheric conditions, telescope movement etc. Despite being better, this scrambling is imperfect however, and some residual imprint will be left at the spectrograph. To attempt to remove the residual imprint, two main techniques have been trialled. Fibre scramblers \cite{heacox1986application,Heacox1988}, tend to use double lenses to scramble both the near and far field. Alternative fibre geometries have also been trialled, such as  octagonal \cite{bouchy2013sophie+} or rectangular fibres \cite{jurgenson2016expres}, the latter of which also acts as an image slicer to increase resoliving power. Both solutions reduces or removes the problem of incomplete scrambling, particularly at lower spectral resolving powers, however at higher spectral resolving power another problem becomes apparent. Known as modal noise (see Section \ref{subsec:modal_noise}), this source of noise is due to the wave nature of light and vignetting in the spectrograph \cite{goodman:1981,lemke:2011}. To remove the modal noise in the final spectrum, the  solution is to agitate the fiber \cite{baudrand2001modal}. This 'randomises' the illumination patter of the modes over time, evening out the resulting illumination pattern and reducing modal noise. The agitation solution is limited however in what it can achieve. The next generation of high resolution spectrographs are mainly aimed at the \ac{NIR}, with the science case of detecting an earth like planet around a M-dwarf. With the longer wavelength of the light, removal of modal noise becomes more of a problem as the number of modes reduces and modal noise is more of a problem \cite{Iuzzolino2014}, with the modal noise being such a problem in the \ac{NIR} GIANO spectrograph that the fiber section was replaced with a conventional slit in GIANO-B.
It has been suggested that the \ac{PL} could be used as an efficient mode scrambler \cite{birks2012}. In particular these investigations have focused on optimising the scrambling properties of \acp{MCF}, trialling optimisation by using dissimilar core sizes \cite{Haynes2014,Haynes2018}. Results have been promising, with significant improvements over conventional fibers \cite{gris-sanchez2018}.


\subsection{3D and multi object spectroscopy}
\label{sec:MOS}

These techniques are well well established  in astronomy, and can be roughly broken down into who two subfields \ac{MOS} (multiple objects in different places) and \ac{IFS} (also known as hyperspectral imaging, spatially sampling one, or close objects). Early examples include slitless spectroscopy, which removes the slit from the spectrograph, allowing the objects to be dispersed using a prism onto a photographic plate or CCD (e.g. \cite{comte:1994}). This has the advantage that multiple objects can now be observed, but suffers from problems with crowded fields and extended objects due to overlapping spectra. This means the technique is still in use today, though is less common. Multi slit spectroscopy can solve the problems of slitless spectroscopy, using a cut mask with many smaller slits (e.g. IRIS 2 \cite{Tinney:2004} and VIMOS \cite{leFevre:2003}) This removes
 the unwanted stars, although requires accurate manufacture. Modern systems normally now rely on reconfigurable fibre optics, which allow much greater versatility, such as placement in crowded fields and rearrangement of the fibres to stop resulting spectral overlap on the detector (e.g. AAOmega \cite{Sharp:2006} and FMOS \cite{Kimura:2010}). This technique also allows the light to be brought to a spectrograph on a stable platform further from the telescope. Whilst this introduces extra optics (such as the addition of a de-rotator on the Naysmyth focus), it means the instrument can remain fixed with respect to the gravity field, increasing stability \cite{Bely:2006}. Such instruments are now allowing huge surveys on smaller research grade telescopes (e.g. BOSS \cite{smee:2013}).
\ac{IFS} first began in the 1980s \cite{Vanderriest:1980}, although it has rapidly expanded to be- come a mainstay within astronomy. It is achieved through four main methods: image slicers (e.g. GMOS \cite{Dubbeldam:2000} and NIFS \cite{McGregor:2003}, lenslet arrays (e.g. SAURON \cite{Bacon:2001}), fibre slicers (e.g. \cite{AllingtonSmith:2002}) and microslicers (e.g. \cite{Blake:2013}), these are shown and described in figure 2.4. It must be noted that although these all use novel inputs, the spectrograph behind is very similar to the long slit analogue.
\ac{DFS} is also a rapidly expanding field, with instruments such as KMOS \cite{Sharples:2004} and FLAMES \cite{Pasquini:2002} paving the way for instruments planned for the next generation of \ac{ELT}s (e.g. IRIS \cite{Larkin:2010}).
Image slicing can also be used to improve the spectral resolution of a spectrograph (by reducing the width of the slit) (e.g. \cite{Avila:2012}) and using the technologies developed for MOS and IFS allows the output from the telescope to be split into multiple replicated spectrographs (e.g. VIRUS \cite{Hill:2004}).

\paragraph{Astrophotonic Integral Field Units} 
\label{sec:IFS}
\ac{IFU}s were first developed in the 1980s. They take a contiguous spatial sample of points and allow spectra to be taken. There are many types of conventional \ac{IFU} (see \cite{allingtonsmith:2006}) all with respective advantages and disadvatanges. 

Recently multimode astrophotonic \acp{IFU} have been proposed in the form of Hexabundles. 
This involves taking multiple \acp{MMF}, removing the buffer and reducing the the cladding through etching. These fibres are then arranged into an array and fused together to form a Hexabundle \ac{IFU}. By processing the fibres in this way, the fill fraction (the percentage of light sampled) can be increased. Care must be taken in the process not to cause cross coupling \cite{bryant:2011}, leading to a degradation in the signal.

To date, Hexabundles are one of the most successful spectroscopic components for astrophotonics. Following on from their initial development, they were used in the SAMI instrument on the \ac{AAT} \cite{bryant:2015} 
which was designed to survey the kinematic 
structure of galaxies. A future instrument HECTOR is currently planned \cite{Lawrence:2012}, as an upgrade to SAMI it will use similar techniques, but with greater and also variable spatial sampling. This allows the cores of galaxies to be better sampled by the fibres.

\acp{IFU} consisting of a microlens array coupled to an array of single mode fibers have also been proposed, in order to miniaturise astronomical spectrographs. As these are single mode, they would need to be fed by a diffraction limited beam in order to achieve a high throughput. 
This means they need either a small telescope, a long wavelength or extremely good \ac{AO} correction. Currently there are various in the forms of the RHEA \cite{Feger2014}, a \ac{MCF} fed spectrograph with a microlens glued on-top. Recently using \ac{MCF}s with 3D printed microlenses on-top used as \ac{IFU}s to feed diffraction limited spectrographs has also been suggested \cite{Dietrich:2017}.

Large scale \acp{IFU} for \ac{MOS} has also been proposed for large telescopes with \acp{PSF} far from the diffraction limit. These would take the \ac{PSF} from the telescope using highly \ac{MM} \acp{PL} and convert them to \ac{FM} \acp{PL} \cite{LeonSaval:2017}. Whilst the number of fibres would be large, converting down would allow access to \ac{SM} technologies such as \acp{FBG} or ring resonators.

\paragraph{Photonic spectro-interferometry}
\label{sec:spectro_interferometry}
Spectro-interferometry is the high-angular resolution analogue of 3D spectroscopy. It consists in dispersing chromatically the interference fringes to measure the variation of their visibility
across the spectrum. If a sufficient number of different telescopes are combined, it is possible to retrieve interferometric images of the target for each wavelength\cite{Millour:2011}, 
thus creating a high-angular resolution (at the mas level) data cube as usually obtained at lower resolution with 3D spectroscopy.
In astrophotonic interferometric instruments such as PIONIER or GRAVITY the dispersion of the fringes is accomplished by interfacing the outputs of the IO component to an imaging spectrograph. 
Concepts and first experiments to accomplish the interferometric beam combination and spectral dispersion in one single chip have been put forward in recent years.
Kern et al. 2009\cite{Kern:2009} proposed to combine an ABCD integrated beam combiner with a SWIFTS (see Section \ref{sec:IPS}) to accomplish 
high-resolution spectro-interferometry (Fig. \ref{fig:SWIFTS+ABCD}). 
The design consisted in splitting each output of the beam combiner in two straight waveguides terminated by a tilted mirror coating. The standing wave originating at the mirror surface can be retrieved by dotting the waveguide surface with a periodic array of gold nano-wires, which are excited by the evanescent field of 
the standing waves and couple the waveguide power to the free space. A camera can be pasted then to the chip so that the pixels correspond to the 
nano-wires, thus avoiding the use of relay optics to collect the radiated light. To be resolved, the nanowires should be placed at distances much greater than the wavelength of the exciting light, thus realising a sub-Niquist sampling of the evanescent wave. However the tilt angle of the mirror coated edge of the chip allows to shift spatially one standing wave respect to the other, realising \textit{de facto} a sub-nanowire-period sampling  of the standing wave.  

\begin{figure}
\caption{\label{fig:SWIFTS+ABCD} Conceptual scheme of a fully integrated spectro-interferometer combining 4 telescopes. SWIFTS devices are printed upon the output waveguides of an integrated ABCD beam combiner, which can be read by detectors pasted on the surface of the photonic chip (from \cite{Kern:2009}).}
\end{figure}

In the same article, a generalisation of the concept to a planar device combining 8 inputs channels in spectro-interferometric mode was also proposed. Differently to the above mentioned
scheme the sampling of the interference fringes would require phase modulation of the input channels or perhaps the use of redundancy in the fringe sampling to compensate for the large sampling period. None of these devices were fabricated so far.
A possible intrinsic limitation of these schemes is that the sampling of the standing wave should not deplete it significantly, thus only a small fraction of the optical power carried by the interfering waves will be effectively detected. On the other side, the complete spectro-interferometric device would be very compact so that the low sensitivity could be balanced by small dimensions and low weight.

The evanescent sampling of the field in the waveguide is also at the heart of a spectro-interferometric device integrating an array of waveguides with a diffraction grating manufactured and 
characterised very recently\cite{Martin:2017}. 
This device exploits a possibility to couple in and out light in a waveguide by means of a suitable diffraction grating. Several parallel single mode waveguides buried in glass are disposed in a non-redundant array. On top of each waveguide, a diffraction grating formed by an array of nano-void structures extract and disperse the signal carried by each waveguide. Suitable focusing optics are then used to combine in free space the spectra emitted by each waveguide in a multi-axial scheme. 
Both waveguides and the grating were manufactured by means of ultrafast laser inscription in Gallium Lanthanum Sulfide glass and were tested at a wavelength of about 1560 nm. 
The grating was designed to give a resolving power of $R$=13800 but achieved a measured resolving power of $R$=2500, allegedly due to perturbations of the array periodicity and aberrations in the collection
optics. This device accomplishes the wavefront filtering through coupling in single mode waveguides and saves the space of the dispersive elements by means of a free-space multi-axial beam combination scheme.
 
Finally, a fully integrated, active spectro-interferometer was presented by Su et al. 2017 \cite{Su:2017} for wide-field aperture synthesis purposes. The planar integrated circuit delivers the light collected in 
a few on-axis as well as off-axis points of microlenses to 3-channel wavelength demultiplexers (centered at 1540 nm, 1560 nm and 1580 nm) and then are combined pairwise (on-axis with on-axis 
input, off-axis with corresponding off-axis input) for each wavelength channel by 2x2 couplers. 
Heaters were mounted on the waveguides before the couplers to perform a fringe scan. Visibility functions of centred and laterally displaced slits were measured successfully in the laboratory.  
The prototype featured propagation losses of several Decibels, thus making the device not yet suitable for astronomical use. However, the device is compact and can be in principle scaled easily up to combine several triplet of telescopes or sub-apertures at low cost, again features which could balance the low losses in a cost/benefit analysis.

\subsection{High angular resolution}
\label{sec:HAR}
\paragraph{Long baseline interferometry}
The idea of using photonic components for long baseline stellar interferometry dates back to the beginning of the 1980's
when a first analysis of the impact of optical fibres for the interferometric connection of telescopes was discussed \cite{Froehly:1981}. Fibers appeared then as an economically viable alternative to free-space optical links, especially in the case of very long baseline ($>100$ m) interferometers. 
Short afterwards proposals for a fibred space interferometer were put forward\cite{Connes:1985} and research addressed the performance 
of optical fibre links respect to the requirements of ground- and space-based stellar interferometry \cite{Shaklan:1987,Shaklan:1988}.
Despite the high insertion losses of single mode fibres connected to a seeing-limited telescope, the absence of speckle noise and modal dispersion 
were soon recognised as the key features which could enable high precision interferometry\cite{Connes:1987,Rohloff:1991}. To improve throughput, a 
method based on M-line spectroscopy (\textit{i.e.} the technique allowing the angular separation of light carried by each mode of a multi-mode waveguide) was proposed to combine mode-wise telescopes connected by multi-mode fibres\cite{Shaklan:1992}. Despite an experimental proof of principle of the scheme was carried out successfully in the laboratory, further research focused only on single mode fibre links.     
In particular, a scalable laboratory simulator of a fibred astronomical interferometer was realised and tested \cite{Reynaud:1992}.  The demonstrator 
utilised fibre stretchers for fast adjustment of the differential optical path between telescopes induced by temperature changes and mechanical vibrations. 
The progress on laboratory test as well as the first test on sky of a fibred beam combiner (see below) motivated the proposal of the OHANA 
interferometer, aiming at connecting with fibres the large telescopes on the top of Mauna Kea \cite{Mariotti:1996,Perrin:2000}. The prototype of a stabilised kilometric fibre interferometer was tested in the laboratory and an assessment of the impact of differential chromatic dispersion effects was 
undertaken \cite{Vergnole:2005,Kotani:2005}. Finally, first fringes in K band of 107 Her were measured connecting the two Keck telescopes with fluoride fibres but without active lengths stabilization
\cite{Perrin:2006}.    
In subsequent experiments with 20 cm telescopes connected with a pair of 300-m-long silica fibres showed the possibility to obtain fringes on bright stars 
in J and H band, but pointed out the need for an active compensation system for the acoustic noise picked up by the fibres \cite{Woillez:2017}. 
More recently \cite{Lehmann:2019} an actively stabilised 200+200 m fiber interferometer operating at telecom wavelength (1550 nm) was laid over the 160 m long path outdoor between one of the CHARA telescopes and the beam-combination facility on Mt. Wilson. Fibers were protected from excessive thermal and mechanical perturbation by synthetic foam tubes. Stabilization over several minutes was possible despite the relatively short stroke of the fiber stretcher used to compensate the OPD (150 $\upmu$m). An overall r.m.s. noise level of 4 nm (corresponding to $\lambda/450$ for H-band) was achieved. The test was done in the frame of the ALOHA project, aiming at extending through photonic solutions the wavelength coverage of the CHARA array to the \ac{MIR} (see also Sec. \ref{sec:detection}).

To date, fibres have not been used for connecting telescopes beyond the reported experiments. 
In contrast, the research on fiber links motivated the development of single mode photonic beam combiners, which  lead to the realisation of 
astronomical instruments, which are currently used for science.
The first beam combiner exploiting photonic technologies was FLUOR, a K-band instrument based of fluoride glass fibre couplers (\textit{i.e.} the fiber equivalent of a conventional beam splitter) which could combine 
two telescopes and measure the interference fringes as well as the light coupled in the fibres by each telescope (photometric signal)\cite{CoudeDuForesto:1992}.
The instrument proved definitely the benefit of using the optical fibre as an effective spatial filter in the measurement of the fringe visibility with a precision of 
1\% or better \cite{CoudeDuForesto:1997}.
FLUOR was initially tested at the auxiliary telescopes of the Kitt Peak solar observatory and moved (in an upgraded form) to the IOTA facility in 
1995\cite{CoudeDuForesto:1997b}, where it delivered scientifically relevant data on star diameters. An ameliorated version of FLUOR was used as 
commissioning instrument in K-band of the VLTI \cite{Kervella:2000} and is currently available at the CHARA array for high precision visibility 
measurements \cite{CoudeDuForesto:2003}. 
The drawback of filtering the spatially incoherent focal spot of a seeing-limited telescope with a single mode fiber is a small average coupling efficiency (see Sec. 3) and a dramatic variation of the transmitted photon flux over time scales of the order of the coherence time of the atmosphere \cite{Shaklan:1992}. While the low coupling efficiency limits sensitivity, the fast flux variability negatively impacts the signal to noise ratio of individual fringe measurements, requiring a post-selection of the acquired frames \cite{Tatulli:2007b}. 
The combined use of an \ac{AO} system and a fringe tracker (\textit{i.e.} a device compensating in real time for the phase fluctuation induced by atmospheric turbulence) can solve both problems, as we have seen with the introduction of second generation instrument GRAVITY at the Very Large Telescope Interferometer in Chile. 
To date, single mode optical fibres are routinely used to deliver spatial filtering of the starlight in beam combiners which use multi-axial free space 
combination schemes such as AMBER \cite{Petrov:2007} and MIRC \cite{Monnier:2004}.

\begin{figure}
\epsfig{file=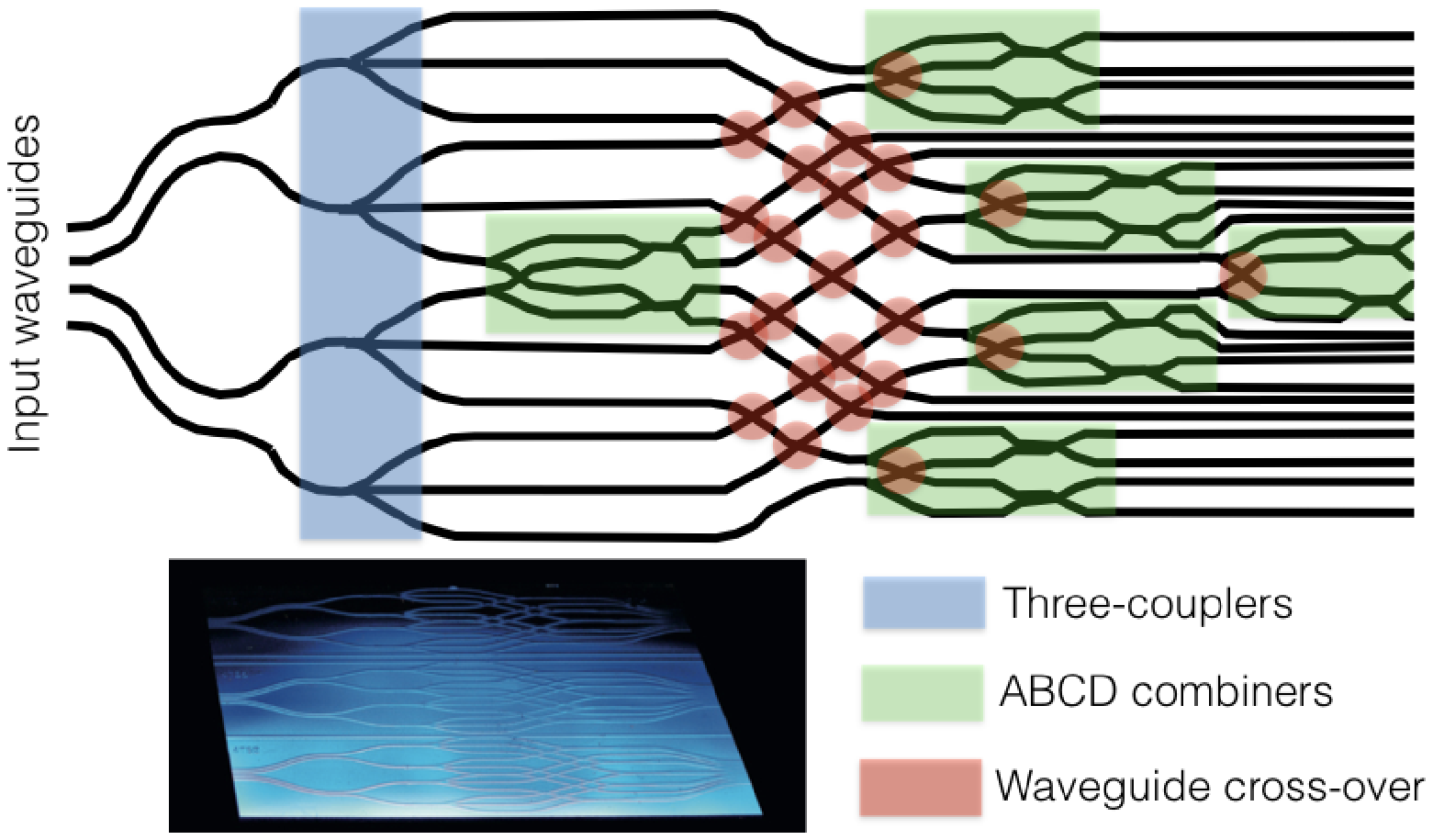, width=\textwidth}
\caption{\label{fig:PIONIER} Scheme of the integrated optics beam combiner of the PIONIER instrument, which can simultaneously retrieve the mutual coherence function of starlight collected by 4 independent telescopes (adapted from \cite{Benisty:2009}). The light from the four telescopes is coupled to the 4 input waveguides of the chip. Four three-couplers (highlighted in pale blue) divide the input light across 12 channels which are mixed pairwise by the ABCD units (highlighted in green). The four outputs of each ABCD unit deliver the 4 quadratures of one of the 6 possible pairs of input fields. Because of the planar geometry of the integrated optics chip, the combination of each possible pair of inputs is achieved only by crossing the waveguides at large angles (the so called cross-overs, marked with the red dot) which could introduce some cross-talk between the interferometric channels. Bottom on the right a photograph of a chip with three identical chips (Credit: IPAG Grenoble).}
\end{figure}

The success of the FLUOR instrument motivated research of beam combiners based on integrated optics \cite{Kern:1997}.
Besides the potential for miniaturisation of complex optical set-ups, the main advantage of integrated over fibre optics is the inherent thermomechanical 
stability of the optical paths in the former, due to the fact that waveguides are shorter and are rigidly attached to a solid substrate. 
The first 2-telescope planar integrated optics combiners for H band were manufactured with silica etching (component LETI) and ion 
exchange (component LEMO) in the late 90's \cite{Berger:1999} and later tested on sky at the IOTA interferometer \cite{Berger:2001}.
The combiners had an area of a fraction of cm$^2$ and delivered excellent visibility precision and accuracy at the 1\% level on stars up to magnitude H=2. 
After this first demonstration of the potential of integrated optics, the challenge became to scale up the integrated components to combine a larger number of telescopes
\cite{Berger:2000}, a functionality which would be rather impractical to implement with fibre coupler technology.
The IOTA facility was again used as a testbed for the first 3-telescope integrated beam combiner\cite{Berger:2003}, which enabled initial interferometric imaging 
capabilities thanks to the measurement of the closure phase.
Developments of the near-infrared combiner technology allowed to deliver by 2009 a first laboratory demonstrator of a H-band 4-telescope combiner
\cite{Benisty:2009}(Fig. \ref{fig:PIONIER}), which was afterwards used in the visitor instrument PIONIER at the VLTI\cite{LeBouquin:2011} and delivered excellent interferometric 
images of close binaries stars and protoplanetary disks. 
Today, an ameliorated component operating in K band is at the heart of the GRAVITY instrument\cite{Abuter:2017}, an \ac{AO}-assisted instrument 
capable of combining simultaneously the four 8-meter telescopes at VLTI and reach a limiting magnitude of K=17 in fringe tracking mode. 

\begin{figure}
\epsfig{file=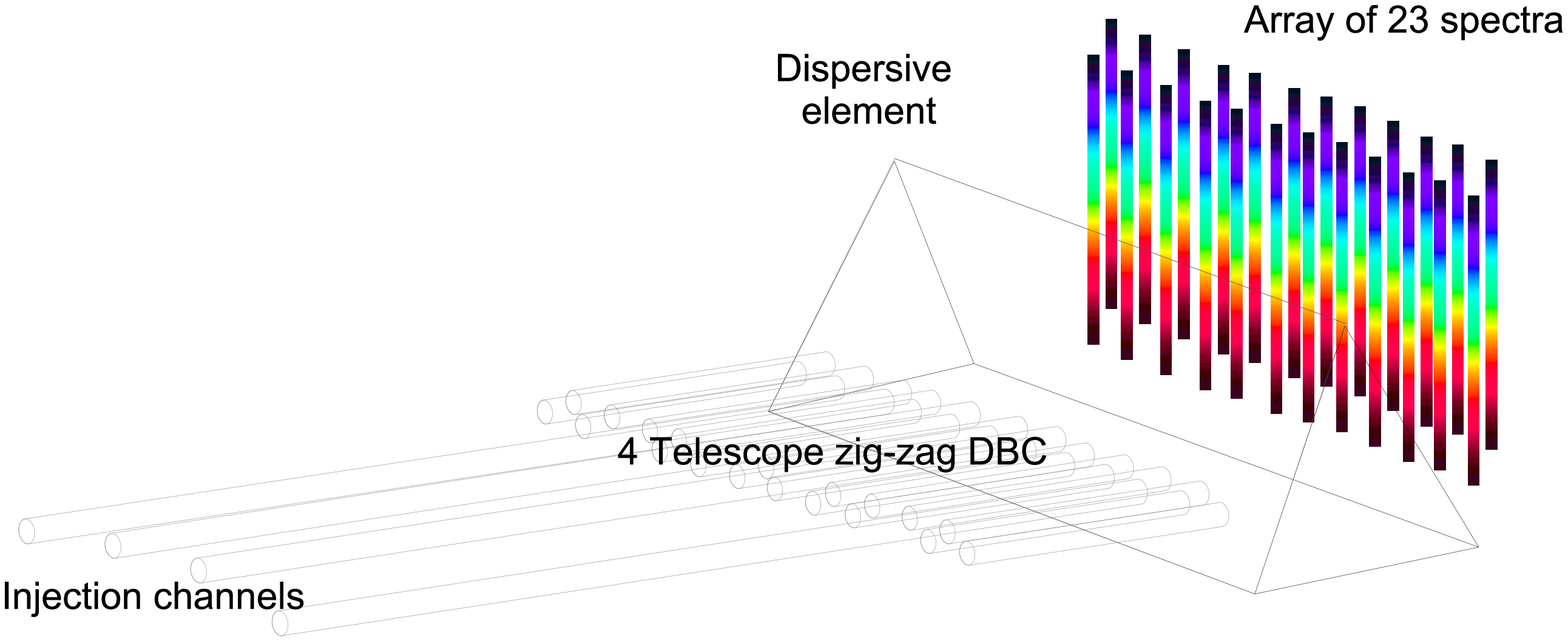, width=\textwidth}
\caption{\label{fig:DBC}Three dimensional photonics can significantly simplify the design of interferometric beam combiners. The scheme of a 4-telescope discrete beam combiner with zig-zag geometry \cite{Diener:2017}, used in spectro-interferometric mode, is illustrated in the figure. Light is injected in 4 waveguides connected to an array of 23 evanescently coupled waveguides where interferometric beam combination occurs. The waveguides are disposed on two layers mutually shifted by half a period to ease the dispersion of the light in the vertical direction. In the figure, the waveguides of the two layers are differently coloured, to help the reading. Collimation and focusing optics of the spectrograph are not included in the figure for simplicity.}
\end{figure}

Ongoing research on photonic beam combiners can be divided into three categories 1) exploration of new combination geometries, 2) increase of the number 
of combined telescopes, and 3) extension to wavelengths longer than $2.4 \upmu$m (absorption band of silica).
In the first category we mention the attempt to harness the three-dimensional (3D) structuring capabilities of \ac{ULI} \cite{Gattass:2006}) 
in various glasses. One strategy is to manufacture traditional pairwise integrated beam combiners (based on a cascade of couplers) exploiting the third 
dimension to avoid waveguide cross-overs which introduce detrimental cross-talk between the interferometric channels and are unavoidable in planar integrated optics multi-baseline beam combiners. Following this approach, three telescope beam combiners 
for mid infrared ($\lambda\sim 10 \upmu$m) were manufactured in Gallium Lanthanum Sulfide, a chalcogenide glass with transparency window extending to the 
mid infrared, and tested in the laboratory \cite{Rodenas:2012}. 
More recently L-band, so called ABCD combiners for two telescopes and with cross-over avoidance were manufactured and characterised in the same glass 
\cite{Diener:2017}.  
A more radical use of the 3D structuring of \ac{ULI} was demonstrated by the introduction of the so called discrete beam combiners (DBC \cite{Minardi:2010}), which 
exploits the propagation of light in two-dimensional, periodic arrays of evanescently coupled waveguides to combine interferometrically several beams (see Fig. \ref{fig:DBC} for an example).
Advantage of this scheme is the avoidance of losses due to bended waveguides and waveguide cross-overs, as well as the possibility to build very short devices (typically two coupling lengths).
R-band three-telescope combiners were tested in the lab with monochromatic \cite{Minardi:2012} and polychromatic light \cite{Saviauk:2013}, while a 2.5-cm-long prototype of 4-telescope combiner for L-band was reported to retrieve accurately phase and visibility of pairs of input beams \cite{Diener:2017}.

In the presence of atmospheric turbulence model free interferometric imaging requires at least the combination of triplets of telescopes to recover partial phase information from the interferograms. \cite{Rogstad:1968}. Additionally, a dense sampling of the Fourier plane of the object image should be achieved by adding many different baseline-triplets, a goal which is essentially accomplished in two different ways. Either the telescope-triplets are relocated to enable more baselines, or a large array of telescopes is combined at once. While the first approach improves the sensitivity, the second 
approach allows the rapid retrieval of the visibilities avoiding the time overheads required to relocate the telescopes. The development of multiple-telescope 
beam combiners is therefore indicated for the imaging of rapidly varying astronomical targets such as novae\cite{Chesneau:2008} or planet transits\cite{Kloppenborg:2010}.
While for the time being only up to 6 telescope may be combined at existing facilities (CHARA), plans for new facilities featuring 10 \cite{Buscher:2013} or 20 telescopes \cite{Kraus:2016} have been proposed.
CHARA is currently developing a new beam combiner in K-band which will allow to combine 3 quadruplets out of the 6-telescopes using spare components of the 
GRAVITY combiner \cite{Bummelaar:2016}. Besides the already mentioned 4-telescope combiner for mid infrared \cite{Diener:2017}, beam combiners featuring 9 input 
channels in H-band and active control of optical path have been developed for aperture masking techniques \cite{Martin:2016} (see next section).
As it may be evident from the previous text, most of the integrated optics beam combiners were developed in the near infrared band and due to the availability of the very mature silica-based technology, which has been developed for the telecommunication market.
However, the astrophysical characterisation of a target often requires the analysis of starlight over a wide spectral range, reason why ongoing research in interferometric instrumentation is focusing on devising alternative materials and technologies to extend the spectral coverage of integrated optics beam combiners.
In particular, both the booming interest towards cold targets such as debris discs and exoplanets\cite{Kraus:2012} and the difficulty to perform interferometry at short wavelength motivated research aimed at achieving beam combination in the mid infrared.
The manufacturing of integrated optics combiners for \ac{MIR} wavelengths\cite{Labadie:2008} requires the development of technological platforms allowing the 
micro-structuring of materials with transparency extending beyond the 2.4 $\upmu$m cut off of silica glass\cite{Labadie:2009}. As such, the research in the field has started to accelerate in the past few 
years, as these technology platforms began to become available, also thanks to the growing biophotonics market.
A paradigmatic technology platform is laser writing. This platform employs tightly focused, intense laser beams to modify locally the refractive index of glasses, an thus create light guiding structures. Both continuous wave lasers as well as high repetition rate femtosecond lasers have been employed to manufacture 
2x2 directional couplers for \ac{MIR} in chalcogenide 
\cite{Labadie:2011,Tepper:2017a} and Zirconium Barium Lanthanum Fluoride glasses (ZBLAN) 
\cite{Gross:2015,Tepper:2017b}.
As mentioned before, three dimensional 3-telescope \cite{Rodenas:2012} and 4-telescope \cite{Diener:2017} beam combiner prototypes were manufactured for respectively the N and the L band 
using ultrafast laser writing in Gallium Lanthanum Sulfide. 
Recently, chalcogenide planar integrated optics manufactured by conventional photolithographic techniques delivered waveguides with extremely low losses ($\sim0.3$ dB/cm \cite{Ma:2013})
and opened the perspectives for complex beam combiners for astronomical use. In this context, highly achromatic directional couplers based on multi-mode interference couplers 
(MMI \cite{Besse:1994}) were manufactured and tested \cite{KenchingtonGoldsmith2016,KenchingtonGoldsmith2017b} based on a design optimised for applications to nulling interferometry \cite{KenchingtonGoldsmith2017a}. 

\paragraph{Aperture masking}
Aperture masking is a high resolution imaging technique based on interferometric principles. In its basic form it consists in masking the pupil of the telescope with sub-apertures smaller than the 
typical correlation length of the turbulent atmosphere (the Fried parameter \cite{Fried:1966}). If the sub-apertures are chosen so that the lines connecting each possible pair (baselines) are 
unique (a so called non-redundant arrangement), the short-exposure (less than the atmospheric coherence time) intensity pattern in the focal plane of the telescope is a two-dimensional interference 
pattern from which it is possible to measure the visibility function of the astronomical target and thus retrieve its angular irradiance distribution. 
First proposed by Rhodes and Goodmann in 1973 \cite{Rhodes:1973}, the technique was initially tested on sky in the mid-1980's by Baldwin and co-workers \cite{Baldwin:1986}, who were able to 
reconstruct the image of a binary star soon afterwards \cite{Haniff:1987}.

While aperture masking can reach an angular resolution of $\lambda/2D$, $D$ being the aperture of the telescope, most of the light of the astronomical target is rejected by the mask, limiting the 
application of the method to relatively bright sources, which deliver high signal-to-noise interference patterns even with short integration times.
To overcome this limitation Perrin et al. \cite{Lacour:2007} proposed the pupil-remapping technique, in which single mode optical fibres are used to remap a densely sampled pupil plane into a non-redundant arrangement of diffraction-limited sources which are converted into an interferogram by a lens.
In this configuration, all the light collected by the telescope can be used and the redundant baselines are measured without loss of contrast. Because of the potentially large number of baselines 
measured within the pupil of the telescope, a further advantage of pupil remapping is the possibility to retrieve images with high dynamic-range and thus be able to observe nearby exoplanets.   
A first on-sky demonstration of the pupil remapping concept came with the instrument FIRST \cite{Huby:2012}. FIRST used 9 microlens-coupled, single-mode optical fibres to transform a redundant, two-dimensional array of sub-apertures of the 3 m telescope at the Lick observatory into a non-redundant linear array of beams. A fringe pattern was then imaged on a camera by focusing the beams with an anamorfic optical system. The instrument operated in the R-band and used a prism to obtain spectrally resolved visibilities ($R\sim300$). Short exposures of the fringes were possible thanks to a 
EMCCD detector. Closure phases with standard deviation well below 1 degree were obtained by averaging 100 frames (standard deviation without averaging $5^\circ$).  
FIRST was later upgraded to 
allow the combination of two separate sets of 9 sub-apertures and used to study the Capella binary system \cite{Huby:2013}.  
A miniaturisation of the pupil remapping setup was proposed and realised with the Dragonfly project\cite{Tuthill:2010}, lead by the former 
\ac{AAO} team. Instead of using optical fibres to remap redundant sub-apertures in a non-redundant array, the Dragonfly instrument used laser writing to fabricate a 
compact 2D to 1D pupil-remapper in a silica glass substrate\cite{Charles:2012}(see Fig. \ref{fig:dragonfly}). Besides reducing the size of the remapping function, the integrated optical component allowed a 
stabler operation of the instrumental transfer function. The assembled instrument was operating in the J- and H-band and tested on sky at the 4-m telescope at AAO\cite{Jovanovic:2012}. 

\begin{figure}
\caption{\label{fig:dragonfly} Scheme of the integrated optics pupil remapper at the heart of the Dragonfly instrument. Here three dimensional waveguides are used to carry light from a two dimensional arrangement of input sites to a non-redundant linear array of outputs. Output light is collimated by microlenses and combined with free-space anamorfic optic on a detector, where interference fringes are recorded (from  \cite{Jovanovic:2012}).}
\end{figure}

The best experimental fringes were obtained on Antares by combining 4 of the 8 possible sub-apertures. The used detector (an InGaAs camera) had a high noise level so that only exposures much 
longer than the coherence time of the atmosphere were recorded. This lead to a significant reduction of the measured fringe visibility. Nonetheless, a performance similar to the FIRST instrument    
(standard deviation of the closure phase $\sim5^\circ$) was obtained. 

Research in astrophotonic aperture masking is currently focusing on the development of hybrid devices allowing the combination of the sub-apertures in an integrated optics beam combiner, 
which could deliver higher sensitivity respect to multi-axial combination \cite{Minardi:2016}. A fibre-fed, 9-apertures device operating in the visible and including a Lithium Niobate phase modulator 
was recently assembled in the laboratory in the frame of the FIRST/SUBARU project\cite{Martin:2016}. This device could be used to scan fringes as well as delivering nulling capabilities to the device (see Sec. 4.4). While the proof-of-concept prototype demonstrated the functional operation of the device, losses occurring at the interfaces between different units (fibre/beam splitter, beam splitter/modulator, modulator/beam combiner, beam combiner/fibre) were believed to be mainly responsible for the high insertion losses of the device ($\sim30$ dB).

\subsection{High contrast imaging}

\begin{figure}
 \epsfig{file=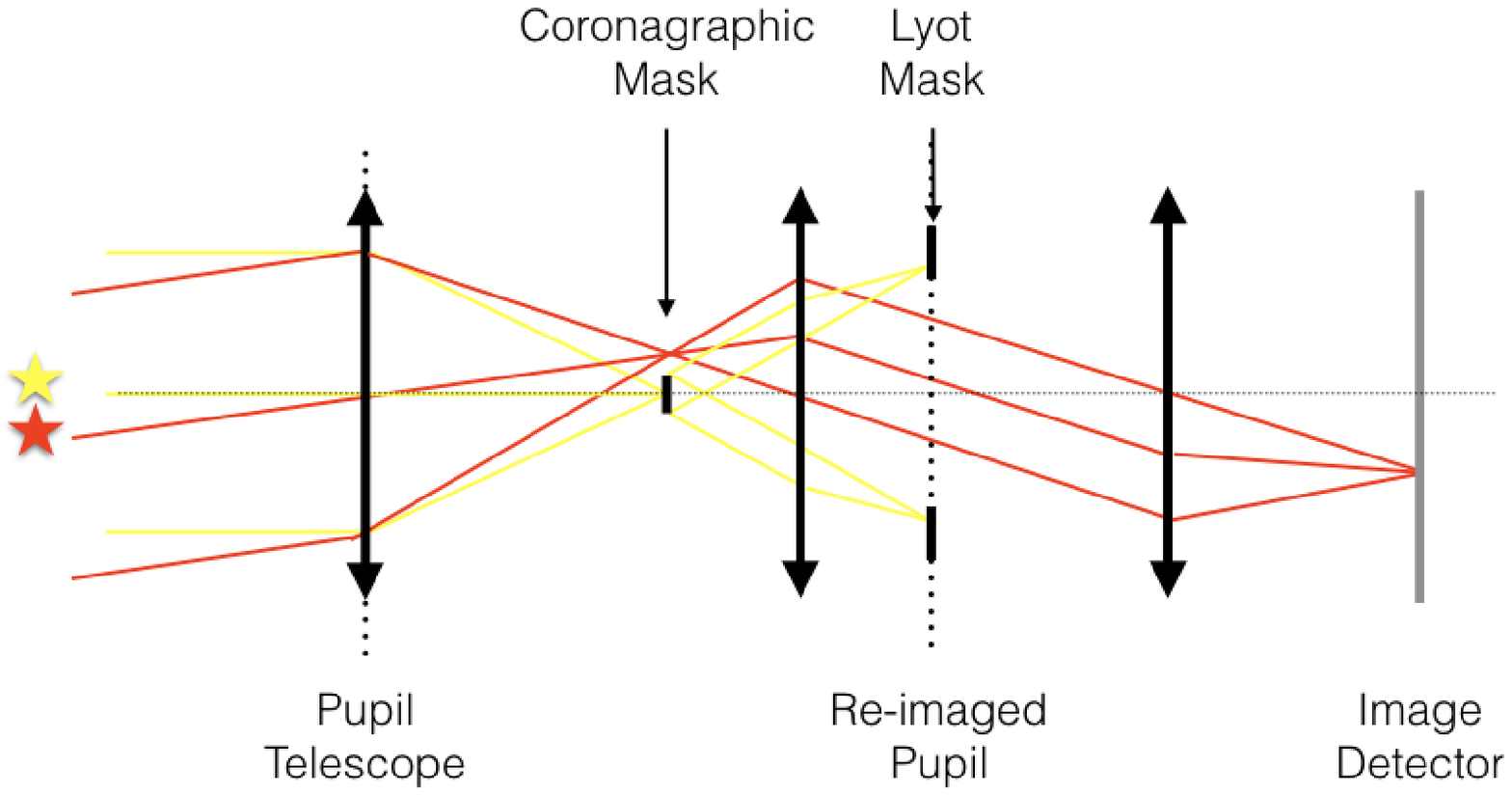, width=\textwidth}
 \caption{\label{fig:coronagraph}Scheme of the Lyot coronagraph. Light from the main star (yellow) is attenuated by the beam stop in the focal plane and the Lyot mask in the pupil plane. On the contrary, light from an off-axis source (red) propagates mostly unattenuated through the optical system and is recorded by the detector.}
 \end{figure}

\paragraph{Coronagraphic phase plates}
\label{sec:coronagraph}
 Coronagraphy owes its name to an instrument invented by Bertrand Lyot in the 1930's to create artificial solar eclipses for the observation of the 
 corona, which has a brightness several order of magnitude inferior to that of the photosphere. 
 The instrument consisted in an focal plane stop (an absorbing disk of diameter matched in size to the image of the photosphere) followed by an optical  
 spatial filter in the pupil plane (the so called Lyot filter) required to suppress light diffracted by the spider and the secondary mirror of the telescope 
 (see Fig. \ref{fig:coronagraph}). A modified version of the coronagraph was developed at the beginning of the 1980's to deliver images of faint objects surrounding stars, such 
 as low mass companions and protoplanetary disks. The first stellar coronagraph used an occulting mask in the focal plane of the telsescope followed by a
 Lyot mask (see \cite{Vilas:1987} for a description of the instrument). 
 This instrument was behind the landmark discovery of the protoplanetary disk surrounding $\beta$ Pictoris \cite{Smith:1984}.  
 Interest for stellar coronagraphy increased in the last 25 years as soon as effective \ac{AO} systems became available and the quest for direct observation
 of extrasolar planets became a mainstream goal in astrophysical research. 
 
 A particularly important result, which opened coronagraphy to the field to photonics, was the proposal of using phase plates instead of absorbing mask as field stops in the 
 coronagraphs \cite{Roddier:1997}. The rationale of the method consists in reversing the sign of a portion of the central lobe of the point spread function
of a highly corrected aperture (\textit{e.g.} a space telescope). In the pupil plane, the phase inverted portion of the PSF  interferes destructively with the starlight, which is scattered outside the pupil, where it can be eliminated by a circular aperture matching the pupil diameter. 
In practice the phase mask acts as a mode converter, transforming the PSF of the AO-corrected telescope into an annular beam.
The advantage of phase masks over traditional beam stops is that the inner working angle of the coronagraph (\textit{i.e.} the minimal distance at which a faint companion can still be detected amid the PSF of the attenuated star) is close to the diffraction limit. An experimental verification of the concept was attempted in the laboratory a few years later (\cite{Guyon:1999}), the mask being manufactured by etching a small cylinder protruding from a glass plate an 
with a height sufficient to induce a $\pi$ localised phase shift at a wavelength of 633 nm. 
The coronagraph showed a peak-to-peak attenuation of the monochromatic point spread function by a 
factor 1/16, far worse than the expected value of $10^{-4}$. 
Most of the loss of contrast was due to the round-off of the edges of the phase shifter resulting from the fabrication process. 

\begin{figure}
\epsfig{file=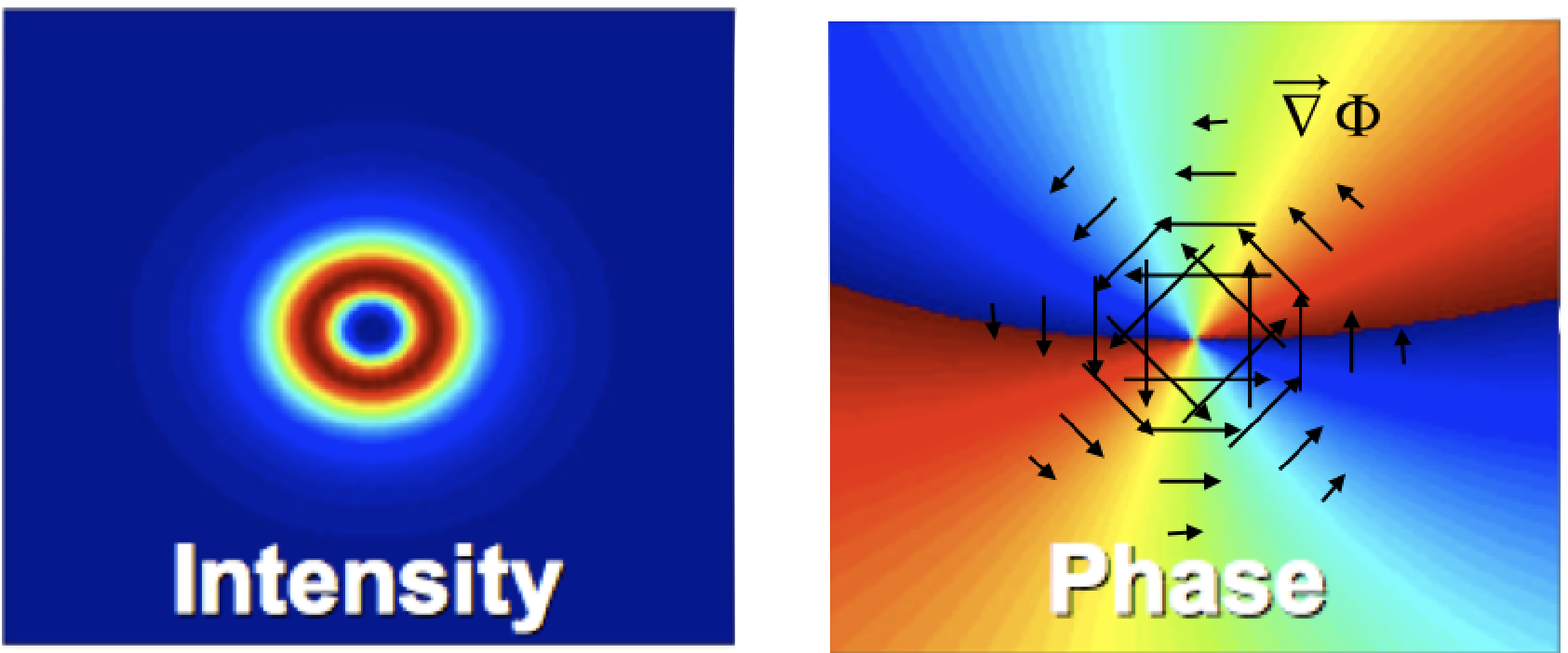, width=\textwidth}
\caption{\label{fig:vortex} Intensity (left) and phase map (right) of an optical vortex. The phase is undefined in on the axis, where the optical field vanishes. The transverse gradient of the phase (black arrows) rotates around the optical axis in analogy to the velocity field of a vortex in a liquid. The transformation of the PSF of an AO corrected telescope into a vortex is exploited to reject the light of a star and image its faint companions.}
\end{figure}

An evolution of the concept of phase mask coronagraph is the vortex coronagraph. Similar to quantum mechanical wavefunctions \cite{Berry:1984}, 
optical vortices  are electromagnetic waves nesting a phase singularity, 
\textit{i.e.} a point in the transverse plane of the optical field where the phase is undefined and the amplitude vanishes. Around the singularity the phase of 
the field grows linearly with the azimuthal angle $\theta$. Because of the continuity of the optical field, the phase difference between $\theta\rightarrow2\pi$
and $\theta\rightarrow0$ should be an integer multiple of $2\pi$ (this multiple being known as the topological charge), so that the phase ramps can take 
only discrete values of their azimuthal gradient.
The name vortex is derived by the fact that the phase gradient whirls around the central singularity in a fashion resembling the velocity field 
of a vortex in a liquid (see Fig. \ref{fig:vortex}). Alternatively, we can visualise the optical vortex as a wave with an helical wavefront, the twist of which can assume only discrete values.  
A remarkable mathematical result is that an ideal lens can transform an Airy disk nesting an optical vortex of even topological charge into a light distribution 
which is vanishing exactly within the area of the pupil \cite{Foo:2005}.
This effect can be exploited in coronagraphy by designing a phase mask with an engraved helical ramp (also described as vortex lens) and 
shaping the Lyot mask as a circular aperture. The first demonstration in the lab of a vortex coronagraph was reported in \cite{Lee:2006}, where a peak-to-
peak attenuation of 0.05 was measured. An improved phase mask was later manufactured (monochromatic attenuation in the lab up to $2\cdot10^{-3}$) 
and tested on-sky with a high Strehl, clear aperture, test telescope (diameter 1 inch) \cite{Peters:2008}. 
In the latter experiment, an average attenuation of 3\% at visible wavelength (20 nm bandwidth) was attained, most of the contrast loss being attributed to the 
chromatic dispersion of the phase mask.  

An inherent problem of phase masks is indeed that they work only at the design wavelength, as the phase delay introduced by a plate of fixed thickness 
scales as $\lambda^{-1}$. This effect can be mitigated by designing a vortex lens composed by two layers made of materials with different dispersive 
properties \cite{Swartzlander:2005,Swartzlander:2006}. The inherent complexity of the fabrication of such masks prevented however an experimental 
demonstration of the concept so far.  
A different approach was adopted by \cite{Errmann:2013}, which adapted a concept of achromatic vortex generation with ultrashort laser pulses to 
coronagraphy \cite{Bezuhanov:2004}. In this case, the vortex beam was imprinted on the PSF of a beam by an off-axis, blazed computer generated 
hologram (CGH), which unavoidably disperses broadband light. The dispersion was compensated by diffracting the vortex beam on a ordinary phase 
grating with the same periodicity of the CGH but opposite orientation of the blazing. Peak-to-peak attenuations of the PSF better than $10^{-3}$ where 
demonstrated in the laboratory for a bandwidth covering most of the R-band.
  
An alternative to conventional phase masks for vortex coronagraphy is represented by birefringent optical plates which modify locally the vectorial 
properties of light and generate polarisation vortices. 
Polarisation vortices are optical fields in which the local linear polarisation vector is oriented at an angle proportional to the azimuthal angle. 
Linearly polarised light beams can be converted into polarisation vortices by a birefringent medium which locally acts as a $\lambda/2$ retardation plate.
An early example is a glass plate coated with liquid crystals, whose slow axes are oriented and polymerised by a UV laser beam \cite{Mawet:2009}. 
If the angle of the slow axis direction is equal to the azimuthal angle, then the field in the circularly polarised base exhibits an helical wavefront of 
topological charge $+2$ and $-2$ for the left and right hand polarisation states, respectively \cite{Mawet:2005}. The far-field of these vectorial fields has 
the same light distribution as the scalar vortices of topological charge $m=|2|$ and thus can be used in coronagraphs with a Lyot mask with full aperture 
slightly smaller than the pupil.
A test on sky of the vectorial vortex coronagraph manufactured with liquid crystal technology was reported in 2010 \cite{Serabyn:2010}. 
The well corrected 1.5 meter aperture of the Hale telescope at Palomar\cite{Serabyn:2007} was used to deliver to the phase plate a beam in 
K$_\mathrm{s}$-band with Strehl ratio exceeding 90\%. A raw peak-to-peak attenuation of $\sim1/50$ was reported over a band of 14\% of the carrier, 
which was sufficient to detect three of the planets surrounding HR8799 after electronic removal of the attenuated PSF of the telescope.  
Chromatic dispersion of the retardation plate was shown in the lab to contribute significantly to the loss of contrast in the coronagraph \cite{Mawet:2010}, 
a problem that could be solved by the use of sub-wavelength gratings \cite{Mawet:2005}. 
These so called Annular Groove Phase Masks (AGPMs) consist of concentric grooves with sub-wavelength spacing. 
The sub-wavelength spacing of the grooves prevents light diffraction and acts as an effective birefringent medium \cite{BornWolf} with an azimuthally varying orientation of the slow (fast) axis. 
The shape and depth of the grooves is chosen so that the retardation between the slow and the fast axis corresponds to $\lambda/2$.
Two distinct technologies have been used to manufacture AGPMs, namely 1) the etching of grooves in diamond, and 2) auto-cloning of photonic crystal structures. 
The first technology was optimised for \ac{MIR} coronagraphy in L-band and could deliver broadband peak-to-peak attenuations at the $10^{-3}$ level in the laboratory \cite{Delacroix:2012,Delacroix:2013,VargasCatalan:2016}. Test on-sky of these phase masks were also successfully accomplished at the VLT \cite{Mawet:2013a} and the Keck telescopes \cite{Serabyn:2017}, but delivered a raw attenuation at the $10^{-2}$ level, mainly due to leakage of starlight from the central obstruction of the telescope and residual tip-tilt of the \ac{AO} system. 
Regarding the second technology, a vortex coronagraph based on photonic crystals (\textit{i.e.} sub-wavelength periodic two- or three-dimensional assemblies of materials with different refractive index) with axial symmetry was manufactured for the visible band \cite{Murakami:2013} and tested 
in the lab with circularly polarised light. A raw peak-to-peak attenuation at the $10^{-4}$ level was achieved in the lab with a clear circular pupil on a band ranging from $\lambda=543$ nm 
to $\lambda=633$ nm wavelength. Analysis pointed out that the attenuation was mainly limited by the chromatic dependence of the birefringence phase shift in the vortex plate.  
No test on sky has been reported so far for this phase mask.

A discretised version of the vortex coronagraphs is represented by the 4-quadrant and 8-octant plates. In practice, the azimuthal phase ramp is simply approximated by 0 and 
$\pi$ phase shifting regions which alternately divide the mask plane in 4 (4-quadrant mask \cite{Rouan:2000}) or 8 (8-octant mask \cite{Murakami:2013}) equal sectors. The 4-quadrant mask is 
equivalent to a vortex with topological charge $m=2$, while the 8-octant mask to a topological charge of $m=4$.
The masks work similarly to the vortex phase plates with the exception that the discovery space is limited by shadow areas extending along the interfaces between the 0 and $\pi$ 
shift sectors, where the throughput of an off-axis companion is significantly decreased. 
Silica 4-quadrant masks designed for Ks-band and manufactured by silica layer deposition techniques saw the first light at the VLT-NACO instrument in 2004 \cite{Boccaletti:2004} and delivered 
broadband attenuations of the PSF larger than $\sim10$, reaching a factor $\sim100$ with narrow band filters. The poor attenuation in broadband was mainly due to the intrinsic chromatic behaviour 
of the $\pi$-phase-shift layer, as well as the presence of the secondary mirror. 
A more achromatic behaviour was achieved in the lab with the 8-octant polarising plate which used the same photonic crystal structure mentioned above for the AGPM coronagraph. 
Tests of the 8-octant plate with monochromatic, linearly polarised light from a clear pupil revealed peak-to-peak attenuations $<10^{-5}$ at the wavelengths of $\lambda=532$ nm and 
$\lambda=633$ nm, the attenuation being apparently limited by the speckles induced by imperfections in the optics of the set-up.  
The on-sky test of an H-band 8-octant phase mask is foreseen at the SCExAO facility of the Subaru telescope in the near future \cite{Jovanovic:2015}.
 
Beyond the design and manufacturing issues, the ultimate starlight rejection of phase mask coronagraph depends on the features of the telescope and its \ac{AO} system, as well as the 
angular extension of the attenuated star \cite{Guyon:2006}.
In particular, the presence of a secondary mirror in the pupil of the telescope creates a circular pattern of light nested in the pupil of the Lyot mask, which carries a power fraction of the starlight
equal to the fraction area of the secondary \cite{Jenkins:2008}. The simplest approach to suppress this light is to include an oversized secondary stop in the Lyot mask, which however limits
the throughput of the coronagraph and does not reject completely the light scattered by the secondary mirror.    
A more advanced solution is to use pupil apodisation techniques (\textit{i.e.} transform the PSF in a fast-decreasing, ring-less light distribution such as a Gaussian beam \cite{Guyon:2003}) by means of aspherical mirrors/lenses or transmitting apodisers before the vortex phase mask.
A concept was recently proposed to apodise the pupil with 3-level stepped transmission plates which would cancel perfectly the light profile of the secondary within the pupil of a vortex coronagraph \cite{Mawet:2013b}. The SCExAO team at the Subaru telescope is currently developing suitable pupil apodising lenses to remap a pupil with secondary into a clear aperture\cite{Lozi:2009}, before focusing the starlight onto a phase mask coronagraph \cite{Jovanovic:2015}. 
A second issue on achievable contrast is related to the quality of the \ac{AO} correction. High Strehl ratios and low residual tip-tilt error are mandatory with the use of vortex phase masks with 
topological charge $m=2$, a condition somewhat relaxed by vortex plates of higher topological charge at the expense of increasing the inner working angle of the coronagraph \cite{Guyon:2006}.
Various techniques for pupil plane tip-tilt sensing and correction have been developed in recent years to keep the pointing stability of the telescope below the few mas level 
\cite{Guyon:2009,Huby:2017}.   
Qualitatively similar to a tip-tilt error, a partially resolved star will leak light in the coronagraph and reduce the contrast, an effect which affects more dramatically phase masks with smaller inner 
working angle \cite{Guyon:2006}. However, a measurement of the maximal attenuation depth with $m=2$ vortex coronagraphs could be conversely used to estimate precisely the angular 
diameter of the target star \cite{Ruane:2014}.

\paragraph{Integrated nulling interferometers}
Nulling interferometry is a high contrast technique which is the analogue of coronagraphy for  stellar interferometers.
The technique exploits the phase difference of fringes originating from on-axis sources respect to off-axis ones \cite{Bracewell:1978}.
By locking the interferometer to a dark fringe, the light of the on-axis source (\textit{e.g.} a parent star of an exoplanet system) will be rejected to the bright channel 
of the beam combiner, while the light of a dim off-axis companion (\textit{e.g.} an exoplanet) would be transmitted by the dark port.
In practice the quantity of interest in this technique is the null depth, defined as the ratio between the flux measured at a dark fringe and the corresponding bright fringe. The null depth in a planet-star system will be dependent on the separation of the two bodies (which can be retrieved varying the orientation and length of the baseline) and their relative luminosity in the wavelength band of interest. For earth-like planetary systems the null depth is expected to be as high as $\sim10^{-6}$ in the N-band.

So far, nulling interferometry has been implemented at the \ac{MIR} N-band by means of conventional optic set-ups at the Keck Interferometer \cite{Serabyn:2012} and the \ac{LBTI} \cite{Defrere:2016}. The Keck Interferometer completed a survey of the exo-zodiacal dust level of about 50 stars \cite{MillanGabet:2011,Mennesson:2014}, 
while only a few results have been published so far \cite{Defrere:2015} from the ongoing survey at LBTI\cite{Weinberger:2015}.  

Astrophotonic nulling interferometry has mainly been object of laboratory research, with the exception of the on-sky operation of a near-infrared nuller prototype at the Palomar 200" telescope, 
which exploited a single mode optical fibre as spatial filter \cite{Martin:2008,Hanot:2011}. The instrument (Palomar Fiber Nuller) consisted in two rotating apertures in the pupil plane of the  
telescope which were optimally focused by a lens on the tip of the fibre, so that when the null of the fringes is centred to the fibre axis, a negligible amount of light would be coupled in the 
fibre\cite{Haguenauer:2006}.
An important driver for nulling experiment has been the feasibility study of for the TPF-I \cite{Breckingridge:2005} and Darwin \cite{Fridlund:2004} space nulling interferometer missions, 
which were later cancelled.  
A high contrast integrated optics 2-channel beam combiner for near infrared demonstrated the possibility to reach a nulling depth of $3\cdot10^{-5}$ with monochromatic light and 
$2\cdot10^{-4}$ with polychromatic light with a bandwidth of 80 nm \cite{Weber:2004}. 
A fibre nuller laboratory experiment with \ac{MIR} laser ($\lambda=10.6 \upmu$m) beams focused on a single mode conductive waveguide demonstrated a few years later 
extinction ratio at the $6\cdot10^{-5}$ level\cite{Labadie:2007}.
The possibility to combine \ac{MIR} beams and modulate the relative phase on a single integrated optics chip was also explored. An active 2-ways Titanium indiffused Lithium Niobate beam 
combiner was designed for L-band and tested in the laboratory\cite{Hsiao:2009}. The device featured an insertion loss of 4.7 dB and an on-chip fringe scan of broadband ($\Delta\lambda=500$ nm) 
light at $\lambda$=3.4 $\upmu$m was demonstrated. However, the measurement could not asses the visibility and the nulling depth of the fringes. 
 
Renewed interest for photonic nulling interferometry rose in recent years, stimulated by the possibility to image exoplanetary systems with interferometric techniques \cite{Kraus:2016} .
In this context, we mention the a proof-of-principle experiment of an integrated optics 4-telescope nulling beam combiner \cite{Errmann:2015} based on a scheme devised by Angel\&Woolf \cite{Angel:1997}, and the aforementioned realisation of an integrated optics 2x2 multimode interference coupler in chalcogenide glass\cite{KenchingtonGoldsmith2017b}, which was designed to
achieve a deep broadband nulling level in L-band\cite{KenchingtonGoldsmith2017a}.

More recently a H-band 2x2 coupler manufactured by means of femtosecond laser writing \cite{Norris:2014} and integrated at the SCExAO testbench of the Subaru telescope on Mauna Kea was tested. The device (named GLINT\cite{Norris:2020}) was used to combine interferometrically two subapertures of the telescope separated by 5.5 meters. The phase difference between the two channels was locked in order to obtain a null in one of the two interferometric outputs of the device. Due to the residual fluctuation of the piston at each subaperture and the varying coupled flux in the two input waveguides, a statistical retrieval of the nulling depth has been used \cite{Hanot:2011}. 
A time series of the estimated null depth was used to build a statistical distribution of their values, which could be fitted by an appropriate model. To test the device, several stars with known diameter, which were partly resolved by the telescope baseline, have been measured. The fitted null depths of the stars ranged from 0.0083 to 0.18 with a precision in the order of $10^{-4}$. The estimated star diameters corresponded within 1 mas with those measured with long-base interferometry.
Upgrades of the instrument to more baselines is foreseen in the near future.

\subsection{Metrology and calibration}

\paragraph{Laser guide stars}\label{sec:LGS}
\ac{AO} systems require light from a bright star to drive the wavefront sensor, resulting in a limited sky coverage of diffraction limited imaging. 
To overcome this limitation Foy and Labeyrie proposed in 1985 to generate an artificial reference star in the sky using a laser\cite{Foy:1985}, an idea derived by 
LIDAR experiments which are used to investigate the structure of the atmosphere by recording backscattered light from a laser beam directed towards the sky.  
Two types of \ac{LGS} were discussed in this seminal paper, one relying on Rayleigh scattering by low atmospheric layers, and one utilising resonant 
scattering of sodium atoms.   

Rayleigh \ac{LGS} use powerful pulsed visible laser sources and a synchronised time gated wavefront sensor to isolate the backscattered photons from a particular layer of the 
lower atmosphere. The configuration of the deformable mirror and the exposure of the science camera should occur within the coherence time of the atmosphere.
The first experimental demonstration of this type of \ac{LGS} was indeed achieved at the end of the '80s by the US Air Force working on the development of 
the space shield Star Wars program \cite{Primmerman:1991}. Today, the evolution of Rayleigh LGS is used in astronomy mainly to compensate ground layer turbulence (first 10-20 Km) as the power of the scattered light decreases rapidly with the altitude. 

\ac{LGS} based on the detection of resonant scattering from sodium atoms use a laser tuned at the wavelength of the  $D_2$ sodium doublet to excite the Sodium layer at an altitude 
of about 93 Km. The exact altitude can vary by a few kilometers according to the season and atmospheric conditions. 
Sodium is found in this layer because of an equilibrium between the intake of meteoric debris and depletion by chemical reactions happening at lower 
altitudes. The laser beam is usually launched off the axis of the telescope/wavefront sensor in order to separate geometrically Rayleigh scattered light from the lower layers of the atmosphere.

The development of powerful laser sources suitable for \ac{LGS} is an important chapter of astrophotonics which has seen a rapid progress in recent years.
Here we review only the development of sources for sodium layer \ac{LGS}, as research in the field was mainly motivated by astronomical applications.  

Because powerful Sodium line solid state lasers are tricky to build (see below), the first examples of Sodium \ac{LGS} used dye solutions as lasing medium. These lasers are tunable over the fluorescence band of the dye molecules, which are excited by flashlamps or other lasers tuned at their absorption band.  A dye laser usually consists in a low power master oscillator which seeds one or more power amplification stages. The selection of the longitudinal mode is accomplished in the master oscillator by an etalon. 
A solution of Rodhamine 6G in ethilen glycol pumped by 4x10W frequency doubled, continuous wave Nd:YAG lasers ($\lambda_0=$532 nm) was the lasing 
medium employed by PARSEC, an early sodium \ac{LGS} at the \ac{ESO} VLT which yielded about 12 W of continuous radiation at 589 nm\cite{Bonaccini:2002}. 
 
Research on solid state laser sources progressed as well, the basic technological solution being the exploitation of the capability of Neodinium-doped garnate (Nd:YAG) to deliver laser transitions at $\lambda_1=$1064 nm and $\lambda_2=$1319 nm. Radiation at a wavelength of 589 nm can be generated by mixing in a nonlinear crystal two pump lasers oscillating at $\lambda_{1}=$1064 nm and $\lambda_{2}=$1319 nm, a process known as sum frequency generator. 
Jeys and co-workers first explored the possibility to mix two Nd:YAG Q-switched lasers to create a sodium line source for astronomical applications\cite{Jeys:1989}. Intracavity etalons with a tuneability of about 0.5 nm were used in the lasers to select the lasing wavelength, while sum frequency was achieved by a single pass in a 5-cm-long lithium-niobate crystal. The source yielded up to 395 mW of radiation at 589.159 nm with a pulse repetition rate of 1 kHz. Because of the relatively low power, the source was used only in LIDAR applications. 
Since this first experiment, considerable progress has been made to develop powerful sources for sodium \ac{LGS} based on Nd:YAG sum frequency, both in the pulsed and continuum operation regime. 
A 20W continuum source based on Nd:YAG injection-locked lasers at $\lambda_1$ and $\lambda_2$ feeding a doubly resonant, sum-frequency cavity was demonstrated by Bienfang et al. 
\cite{Bienfang:2003}. The final engineered source (Frequency-Addition Source of Optical Radiation - FASOR) achieved 50 W of output power at the sodium line and was eventually installed 
at the 3.5 m telescope of the Starfire Optical Range facility of the US Airforce, where it was used for adaptive optical imaging of satellites.
Multiwatt, pulsed Nd:YAG sources are currently employed at the Subaru\cite{Saito:2010}, Palomar \cite{Hankla:2006}, Keck I and Gemini South observatories.
In these cases two synchronised mode-locked Nd:YAG lasers are used and mixed in single pass in a periodically poled Mg-O-doped stoichiometric lithium tantalate (PPMg-O:SLT) crystal
\cite{Saito:2007}. Synchronisation of the lasers is achieved by a phase shift of the radio-frequency drivers of the loss modulators inside the cavities. Output powers of 6.8 W were reported as 
well as a power stability of 2.2\% over 8 hours.  

\begin{figure}
\caption{\label{fig:LGS} Active photonic devices such as lasers are becoming a working instrument for astronomers. Laser guide stars allow to extend the use of \acf{AO} to targets which are too faint to use natural guide stars to lock the \ac{AO} system. The picture shows an 87x87 arcsecond portion of the cluster NGC288 taken with help of Gemini South’s GeMS/GSAOI 589 nm laser guide star facility. The image is taken at 1.65 microns (H band) and features an average full-width at half-maximum resolution of slightly below 0.080 arcsecond. The insets on the right show a comparison of the same close up of the cluster taken with LGS-AO (top), with natural guide star AO (middle), and seeing-limited observations (bottom). (Credit: Gemini Observatory).}
\end{figure}

The complexity of the Sodium laser systems requires trained personnel and controlled environments for their correct operation. The need for a more rugged and turn-key sodium line source motivated \ac{ESO} 
and the company TOPTICA to develop a fibre-based sodium line \ac{LGS}. Fibre laser can deliver high power with excellent beam profile and require no alignment of the laser cavity, which is formed by spliced integrated 
components. The fiber Sodium line laser used a 70 W Yb-doped fibre laser operating at 1020 nm to amplify through Raman process a $\lambda_0=$1178 nm seed from a diode laser in a 100-m-long nonlinear fibre \cite{Feng:2008,Feng:2009}. Radiation at 589 nm was achieved by frequency doubling in a phase-locked singly resonant cavity containing a lithium triborate crystal \cite{Feng:2009}. 
Output powers of 50 W at the sodium line wavelength were achieved by combining interferometrically three Raman fibre lasers at 1178 nm\cite{Taylor:2010}. The system engineered by TOPTICA includes several servo systems enabling a power stability better than 2\% over several hours and a fully maintenance-free operation \cite{Arsenault:2012}. Four units of the laser sources are 
currently forming the laser guide star asterism of the \ac{LGS} facility at the UT4 VLT\cite{Bonaccini:2004}, 
which feeds the GRAAL/HAWK-I \cite{Paufique:2012} and the GALACSI/MUSE 
\cite{Stroebele:2012,Laurent:2010} instruments. 

\paragraph{Astrocombs} 
\label{sec:astrocombs}

As we have discussed in the introduction, the key for the future challenges of high resolution spectroscopy lies in the capability to calibrate accurately the spectrograph.
This is usually achieved by means of standard lamps exciting the emission lines of heavy elements such as Thorium and Argon \cite{Baranne:1996}, or by filtering starlight or a white light source with 
a gas cell filled with iodine vapour \cite{Butler:1996}.
While the short time accuracy of the position of the lines of \ac{ThAr} lamps is at the few cm/s level \cite{Wilken:2012}, their uneven frequency and brightness distribution does not allow for a uniform
calibration accuracy across the spectral range of the spectrograph. Moreover, ageing of the emission lamps reduces the accuracy on a time scale of several hundred hours 
\cite{Mayor:2009}. An additional limitation of atomic references is the unavailability of satisfactory standards for mid infrared bands \cite{Seemann:2014}. 

In the past decade, interest has been mounting for the realisation of artificial light calibrator sources with ideal characteristics such as even frequency spacing, sub-resolution linewidth and uniform 
intensity across the spectral window of the spectrograph. Sources with these characteristics are the so called optical frequency combs, also known as astrocombs.
Two main approaches have been followed to generate such reference sources, namely the Fabry-Perot etalon (passive device) and the laser frequency comb (active device).
Both approaches are based on the property of optical resonators to sustain a multitude of longitudinal modes (resonances) in between the reflecting surfaces of the resonator. The allowed 
propagation modes shape the transmission spectrum of the resonator into a comb.  
The resonance frequencies of an achromatic resonator $\nu_\mathrm{m}$ are periodic and depend only on two parameters, the \ac{FSR} (i.e. the periodicity of the resonances, FSR), and the offset 
frequency $\nu_0$:
\begin{equation}
\nu_\mathrm{m}=\nu_0+m\cdot \mathrm{FSR},
\end{equation}
$m$ being an integer number identifying the resonance. The \ac{FSR} is the inverse of the round-trip time of light in the resonator, which for a cavity of length $L$ composed by two parallel mirrors filled 
with a medium of refractive index $n$ corresponds to:
\begin{equation}
FSR=\frac{c}{2nL}
\end{equation}
Another important parameter of cavities is the linewidth, which can be expressed in terms of the finesse (FSR over Full Width Half Maximum linewidth)\cite{Hecht:2012}. 

The simplest implementation of the optical frequency comb source is a Fabry-Perot etalon with a vacuum filling and illuminated by a collimated broadband light source. Key for the accuracy of the comb line position is to control the spacing between the mirrors with nanometric precision over time. This is typically achieved by high precision temperature control of the environment of the etalon \cite{Wildi:2010}, or by 
active tuning of the mirror gap monitored by a reference laser \cite{Gurevich:2014}. 
Fiber pigtailed, vacuum-spaced, planar dielectric mirrors were used for the Fabry-Perot calibrator of HARPS\cite{Wildi:2010} and reported a calibration stability comparable to the \ac{ThAr} lamps on 
short ($<20$ cm/s) as well as long observation periods ($<1$ m/s)\cite{Wildi:2011}. Authors reported that possible slight misalignments of the fiber-to-fiber imaging system may be responsible for the 
observed calibration drift over long periods. 
Fabry-Perot etalons consisting of a short single mode fibre patch with spliced mirrors and single mode pigtails were proposed as possible alternative to bulk optics etalons inherently insensitive
to the alignment of the illumination source \cite{Halverson:2012,Gurevich:2014}. 
Differently from vacuum-spaced bulk-optics etalons, fibre Fabry-Perot (FFP) resonators suffer of dispersion (i.e. the comb is not exactly periodic) and are more sensitive to temperature variations, requiring a temperature stabilisation at the fraction of 
mK level in order to keep the wavelength drift of the resonances below 1 m/s\cite{Halverson:2012}. Precision temperature tuning of the FFP by monitoring the gap with a Rubidium referenced 
laser allowed the stabilisation of the FFP resonances below the 20 cm/s level averaging over an integration time of only 30s\cite{Gurevich:2014}. 
A portable calibrator for low to mid resolution, near infrared spectrographs was engineered starting from an integrated optics micro-ring resonator with drop off channel\cite{Lee:2012}. 
The micro-ring resonator featured a \ac{FSR} of 200 GHz and included a resistive thermal device on its surface for accurate control of the temperature. A fibered amplified spontaneous 
emission broadband source was butt-coupled to the input channel of the microring resonator, so that a bright line spectrum was dropped off at the output channel. 
A line wavelength stability of the order of 1 pm at 1550 nm was achieved over a period of 24 hours, which corresponds to a RV precision of about 200 m/s. 

The advent of laser frequency comb technology as ultimate time/frequency standards \cite{Udem:2002} stimulated research on their application as accurate calibrators for high-resolution 
astronomical spectroscopy. Laser frequency combs are ultrashort pulse (20-300 fs) laser sources where modulated resonator losses enable the locking of the longitudinal modes and the generation of an exactly periodic frequency comb. The key difference to a standard ultrashort laser pulse is that frequency combs use a external nonlinear interferometer referenced to an atomic clock to control the drift of the offset frequency through a feedback on the cavity length. The advantage of this scheme over conventional Fabry-Perot calibrators is the higher brilliance of the spectral lines and the stability of their position which is set by the precision of the 
atomic clock ($\Delta\nu/\nu\sim10^{-11}$ for a standard Rubidium clock). Two types of femtosecond laser sources have been used so far for astrocombs: fibre laser and solid state laser. 
The former has the advantage of being a self-aligned, turn-key system, but require usually long patches of gain fibres, reducing the free-spectral range to a few 100 MHz. This line separation is too 
small to be resolved even by the spectrograph with the highest resolution ($\Delta\nu\sim$1 GHz), so that the spectral lines are selected by several actively stabilised Fabry-Perot cavities with free 
spectral range of the order of 10-20 GHz. A prototype of a fibre based, infrared comb was installed at the German Vacuum Tower Solar Telescope and delivered a calibration 
precision at the 9 m/s level ($\Delta\nu/\nu\sim3\cdot10^{-8}$). The known drifts of the spectrograph were responsible for the measured precision, which, though remarkable, exceeded the theoretical precision of the comb \cite{Steinmetz:2008}. An improved, frequency doubled version of the 
astrocomb was later installed at HARPS and was used to complete a new atlas of the solar spectrum reflected by the Moon with significantly improved accuracy \cite{Molaro:2013}.
In the same measurement campaign, the astrocomb demonstrated a short-time repeatability of the HARPS calibration at the 2.5 cm/s level ($\Delta\nu/\nu\sim8\cdot10^{-11}$) \cite{Wilken:2012}). 
Radial velocity measurements of HD75289 were also taken both with ThAr lamp and astro-comb calibration which agreed within an accuracy of 2.5 m/s, most probably limited by the ThAr lamp 
accuracy\cite{Wilken:2012}.   
Solid state, mode-locked lasers offer more compact cavities which can have FSR of a few GHz, relaxing the requirement for the Fabry-Perot cavity required to select an astrocomb with 10s of GHz
spacing, but introducing more degrees of freedom for the periodic alignment of the laser cavity. A stabilised frequency-doubled Ti:Sapphire laser with a free-spectral range of 1 GHz was used in 
combination with a Fabry-Perot etalon to obtain a 20-50 GHz-spaced astro-comb operating in the blue region of the visible spectrum ($\lambda=420$ nm) and tested at the TRES spectrograph at 
the Fred Lawrence Whipple Observatory\cite{Benedick:2010}. The frequency comb allowed for an accurate calibration at the 1 m/s level ($\Delta\nu/\nu\sim3\cdot10^{-9}$) of the spectrograph which evidenced drifts of a few 100 m/s over days \cite{Phillips:2012}.  
An upgraded version of the comb has been recently deployed at the HARPS-N spectrograph of the Telescopio Nazionale Galileo on Canary island and demonstrated a short time calibration 
uncertainty at the 2 cm/s level \cite{Glenday:2015}.

Besides the conventional architecture of laser frequency combs based on mode-locked femtosecond laser sources, alternative schemes for the generation of trains of ultrafast laser 
pulses with GHz repetition rates were explored. Aim of these schemes is to reduce the complexity and cost of the astrocombs based on mode-locked lasers.  
A frequency comb with tuneable FSR was designed and built by generating a train of near-infrared optical solitons in a nonlinear fibre seeded by the beating of two mutually detuned, continuous 
wave lasers \cite{Zajnulina:2015}. Laboratory tests showed the possibility to obtain combs with sub-THz FSR, suitable for the calibration of low- to mid-resolution spectrographs.
No stability characterisation was carried out in the published work.
More recently a near-infrared astrocomb with 12 GHz FSR was deployed at NASA IRTF telescope and Keck II\cite{Yi:2016}.   
The source uses electrooptical modulators in phase and amplitude to generate a 12 GHz train of 2 ps pulses from a continuous wave laser source, 
referenced to narrow molecular absorption lines of acetylene or cyanide. The radio frequency driving signal of the modulators is referenced to a Rubidium atomic clock, allowing an accurate 
control of the comb FSR at the $\Delta\nu/\nu\sim 10^{-11}$ level. After amplification in Erbium doped fibre, the pulse train spectrum was broadened in a highly nonlinear fibre allowing the comb to stretch from 1400 nm to 1700 nm wavelength. The intrinsic accuracy of the comb was measured in the lab to be below 60 cm/s ($\Delta\nu/\nu=3\cdot10^{-9}$), which was degraded to 
an estimated level of $\sim1.5$ m/s after taking into account the uncertainties related to the spectrograph operation.  

Efforts to miniaturizing frequency combs for astronomical instrumentation have been recently achieved a milestone by delivering a calibrator based on integrated microcombs at the Keck II telescope \cite{Shu:2019}.
Microcombs are frequency combs excited in a microring resonator by means of an optical nonlinear process known as four-wave-mixing. The resonator is pumped with a monochromatic laser at frequency $\nu_0$ tuned near the resonance frequency of the microring. Inside the resonator a very strong optical field is stored which modifies the refractive index of the resonator and enables the split of two photons at $\nu_0$ in a signal and idler photons following the principle of energy conservation:
\begin{equation}
\nu_{\pm}= \nu_0 \pm \Delta\nu    
\end{equation}
Each new line in the spectrum can be considered as a pump field, so that a cascade of equally spaced frequencies can be generated \cite{Herr:2012}. In low dispersion resonators a wide comb can be generated and phase locked, generating a single optical pulse circulating in the resonator \cite{Herr:2013}. Because of the small dimension of the microring resonators, the frequency spacing of the generated comb is in the range of a few 10 GHz, making them attractive for the calibration of astronomical spectrographs.
In the reported work \cite{Shu:2019}, a 70 nm wide comb centered at a wavelength of 1550 nm was generated in a silicon microdisc resonator excited by a suspended fiber taper. The comb was referenced against a hydrogen cyanide gas cell (absorption line at 1559.9 nm) with a precision better than 1 MHz, since the spectrum was not wide enough to apply the usual self-referencing method. This provided a radial velocity precision of about 1 m/s. The comb was then extended to cover almost completely the H-band by passing the generated pulses into a patch of nonlinear fiber and projected into the beamline of the NIRSPEC R=25000 infrared spectrograph at the Keck II telescope. The drift of the spectrometer could be determined within a precision of a few m/s, mostly limited by the geometry of the sepctrograph and detector. The study allowed to identify a roadmap for the upgrade of the spectrograph aimed at limiting its thermo-mechanical drifts and improving its resolving power by means of adaptive optics on the incoming stellar beam. 
Even tough the whole microcomb setup required about 1.3 m of a standard instrumental rack, there is a potential for integrating microdics and pump sources within a chip of a few square centimeters footprint.

\paragraph{Wavefront sensing and adaptive optics}
Using fiber bundles as tools for guiding telescopes has been around for many years, for example the fiber guiding mode using the FLAMES \ac{IFU} at the \ac{VLT}. This instrument uses four Fiducial Acquisition Bundles (FACBs) to record positions of reference stars and then pass any offset to the instrument guiding system. With \ac{SM} astrophotonics gaining popularity, accurate guiding and compensation for vibrations in the telescope system is becoming more urgent. In 2017, \cite{Dietrich:2017} proposed using a \ac{MCF} fed with 3D printed microlenses to compensate for tip-tilt vibrations, to increase coupling into a \ac{SM} fed instruments. This concept has been further developed by \cite{hottinger2018}, taking the \ac{MCF} and replacing it with a fiber bundle for increased sensing and linearity. Outside astronomy there are also similar concepts, including using a \ac{PCF} for wavefront sensing \cite{DosSantos:2015}.
Another avenue being actively pursued is using a \ac{PL} at the focal plane to monitor the \ac{PSF} of the telescope. As the light separates from the \ac{MM} core into many \acp{SM} any irregularities in the wavefront result in different modes being excited. By monitoring the output of the \ac{PL} \cite{Corrigan:2016} were able to detect tip and tilt in the \ac{PSF} using a simulated 9 mode \ac{PL}. Recently, \cite{norris2020all} have extended this concept, showing that they can detect higher order modes using a 19 core \ac{PL}, opening up the possiblity of using a \ac{PL} for full photonic wavefront sensing for applications such as exoplanet science.

In addition to using astrophotonics for wavefront sensing, using active waveguides instead of a deformable mirror is also under investigation. \cite{Miller:2013} proposed that using on-chip cascaded Mach-Zehnder interferometers in conjunction with a \ac{PL} could be used as an \ac{AO} system (see Figure \ref{fig:active-combiner}). This would take an input \ac{PSF} and use micro phase shifters on the chip to constructively interfering the individual modes. Sensed correctly, this would either enhance, or potentially replace an \ac{AO} system, in a fully integrated chip.

\begin{figure}
\caption{\label{fig:active-combiner} One approach for coupling an arbitrary input beam to a single output beam, so called \acf{AO} on a chip. In the figure, P1 – P4 are controllable phase shifters, MA1 – MC1 are controllable MZ interferometers, and DA1 – DC1 are detectors used to give the signals for feedback loops. The dummy phase shifters are optional and could be included for equality of path lengths and/or loss. Figure reproduced from \cite{Miller:2013}.}
\end{figure}

\subsection{Detection enhancement through frequency conversion}
\label{sec:detection}
A remarkable research topic bearing potentially important consequences for the future of infrared astronomy is related to the nonlinear optical conversion of light frequencies 
in \ac{PPLN} crystals \cite{Albota:2004,Langrock:2005}.
It is well known that infrared detectors have quantum efficiencies well below their visible counterparts, especially in the \ac{MIR} bands. Moreover, as we have seen, photonic components for the \ac{MIR} are currently under development. A solution to these problems could come from nonlinear frequency conversion.

Nonlinear frequency conversion is a process exploiting the nonlinear polarization response of non-centrosymmetric crystals. In such crystals the polarization vector has a term depending on the square of the electric field (nonlinear polarization). If the driving electric field has two frequency components $\omega_1$ and $\omega_2$, then the nonlinear polarization will contain a term oscillating at the sum-frequency $\omega_3=\omega_1+\omega_2$ which gives rise to a radiating field (frequency up-conversion process).
The sum frequency field can build up coherently as it propagates in the medium provided the \textit{phase-matching} condition is fulfilled,\textit{i.e.} the propagation vector at frequency 
$\omega_3$ should equal the sum of the wavevectors of the two initial fields: $\mathbf{k}(\omega_3)=\mathbf{k}(\omega_1)+\mathbf{k}(\omega_2)$.
Usually phase matching is achieved by exploiting the birefringence of quadratic nonlinear materials, but in this case the phase matching direction may not coincide with the direction of higher nonlinear coefficient.
In periodically poled materials, phase matching is achieved by inverting periodically the orientation of the crystal, so that the phase matching direction coincides with the direction where the nonlinear coefficient of the crystal is maximum.

The first application of nonlinear frequency conversion to enhance the detection quantum efficiency of infrared photons was reported in \cite{Albota:2004}. In the experiments, a 4-cm long \ac{PPLN} crystal was operated within a singly resonant ring cavity, pumped with a 400 mW Nd:YAG laser at a wavelength of 1064 nm. Photons at the wavelength of 1.55 $\upmu$m were converted to visible light 
($\lambda=631$ nm) with an estimated efficiency of 90\% (the ratio of upconverted photons to the input in infrared photons).
The setup for the wavelength conversion was later miniaturized by Langrock et al. \cite{Langrock:2005}, who used a waveguide written in the \ac{PPLN} crystal to achieve high conversion efficiency with a pump power of a few tens of mW. The scheme could be used to convert either photons at $\lambda=1.32$ $\upmu$m or $\lambda=1.55$ $\upmu$m to photons at $\lambda=713$ nm, in order to exploit the maximal quantum efficiency of silicon, single-photon detectors. In the experiments an overall quantum detection efficiency of 40\% was reported.

While these experiments were mainly motivated by quantum optics applications, the first use of optical frequency conversion for astronomy was proposed by Brustlein et al. \cite{Brustlein:2008}. The seminal paper investigated the coherence properties of unconverted light in \ac{PPLN} waveguides. To this end, a fiber interferometer was built, featuring a wavelength converter in each arm (see Figure \ref{fig:converter_interferometer}). 
   
\begin{figure}
    \caption{\label{fig:converter_interferometer} Scheme of the experimental layout of a fiber interferometer with optical frequency conversion stage. An infrared light source consisting of two point sources is collimated and sampled by two fiber couplers which could move across the collimated beam, thus simulating an astronomical interferometer observing a binary star. The sampled light is carried by the fibers to a coupler which injets a the pump beam necessary to perform the ooptical sum frequency in the periodically poled lithium niobate waveguides (PPNL). The generated visible light from each PPNL crzstal is then coupled into a fiber suitable for visible ligt and mixed inside a 2x2 coupler before being detected. A parallel infrared interferometer is built for reference purposes (from \cite{Brustlein:2008}).} 
\end{figure}
 
By combining the up-converted beams, interference fringes were recorded with a visibility corresponding to the independent measurements taken by combining the 
 original infrared signals. The experiments showed unequivocally that the coherence properties of light are preserved by the up-conversion scheme. Further confirmation came later on with an experiment performed with a 3-channel fiber interferometer which showed that also the closure phases are preserved by the up-conversion scheme \cite{Ceus:2011}.
 
The frequency conversion scheme was later tested on sky \cite{Ceus:2012} by connecting one arm of the frequency converter to a C8 Celestron telescope used for the OHANA-Iki project \cite{Baril:2010}. The setup allowed converting a bandwidth of 3 nm centered at 1544 nm (H band) in a 4-cm-long \ac{PPLN} waveguide. Even with this narrow conversion bandwidth, 
a flux of a few tens of photons per second was recorded for observations of Betelgeuse, Antares and Pollux. The calculated overall detection quantum efficiency was estimated in $1.75\times10^{-6}$, but this included a particularly low coupling efficiency to the single mode fiber ($3.5\times10^{-3}$) and the insertion losses of the various optical elements (filters, lenses, collimators). Theoretically\cite{Shaklan:1988}, an increase of the coupling efficiency from telescope to fiber by a factor 100 can be expected, while the use of integrated optics instead of bulk optics relay and higher laser pump powers could increase further the efficiency of the photon conversion stage (overall conversion efficiency 1\%).

Despite the low overall quantum detection efficiency of the first on sky test, the result of \cite{Ceus:2012} opens a large range perspectives for the application of nonlinear optics to astronomy. It's easy to think about potential applications such as conversion of \ac{MIR} light into telecom wavelengths to exploit the availability of photonic components for signal transport and processing over large distances.
This is indeed the basic concept that the project ALOHA aims to test at the CHARA interferometric array \cite{Lehmann:2018}.
The project aims at enabling \ac{MIR} interferometric observations by converting \ac{MIR} photons to the telecom wavelengths by means of a \ac{PPLN} waveguide and transfer the up-converted light to the beam combiner using hectometric single mode optical fibers\cite{Lehmann:2019}. As the efficient conversion of photons requires long Lithium Niobate waveguides (2-4 cm), the conversion bandwidth is small and only narrow lines of the \ac{MIR} spectrum can be explored. A possibility to extend the bandwidth of the the frequency conversion is to use several non-redundant phase-modulated pump sources at slightly different wavelengths which achieve phase-matching at different portions of the \ac{MIR} spectrum \cite{Lehmann:2018b}. The phase modulation of the different pump sources allows disentangling the \ac{MIR} spectral channels by a simple Fourier-transformation of the interferometric signal of both arms of the interferometer.
The initial results of this project are very promising, however the conversion efficiency of the \ac{PPLN} conversion stage requires improvements (e.g. longer \ac{PPLN}, higher pump power, reduction of insertion losses), as the achieved overall conversion efficiency is in the order of 0.4\% \cite{Lehmann:2018b}.

\section{Conclusion and future}
\label{sec:conclusions_future}
\acresetall

In this paper we have discussed how state-of-the-art optical technologies, \textit{i.e.} photonics, are slowly being incorporated into the design and realisation of new astronomical instrumentation, with enhanced performance and which could exploit the light collection parameters of current very/large- and forthcoming extremely-large-telescopes. This process of testing advanced optical technologies to investigate the sky can be seen as a continuation of a scientifically fruitful tradition initiated by Galileo’s  attempt to point his telescope to the Moon and Jupiter. 
As discussed in Section \ref{sec:astronomical_perspectives}, a number of important scientific cases would benefit greatly from widespread incorporation of photonic technologies thanks to their intrinsic small footprint, multiplexing capability and potentially superior performance as compared to bulk optics, provided the PSF of the telescope is suitably controlled.
 
As we have seen, the small footprint offered by integrated optical components is a mayor asset for astrophotonic instrumentation. This aspect has been already exploited to deliver extremely compact beam combiners for optical astronomical interferometry, which offer enhanced stability to climatic and mechanical perturbations of the instrument (see Sec. \ref{sec:HAR}). Additionally, several groups are trying to develop miniaturised integrated photonic spectrographs, which could defeat the harsh scaling of the size of conventional spectrographs with the telescope diameter (Sec. \ref{sec:IPS}), provided many replicas of the miniaturized devices are connected to the ends of a photonic lantern \cite{Cvetojevic:2012} or a single miniaturized device is linked to an \ac{AO}-corrected telescope\cite{Jovanovic:2017b}.  
This would open a new era for spectroscopic instrumentation, allowing a remarkable reduction of weight and realisation/maintenance costs of spectrographs for extremely large telescopes. Miniaturisation can also bring tremendous benefits to space borne instrumentation, saving precious fuel or enabling more experiments on the same space probe. In this context, we mention the flight in 2017 of an integrated photonic spectrometer \cite{Betters:2012} on the Australian nanosatellite i-INSPIRE 2, and the attempt to launch in early 2018 an integrated optics interferometric beam combiner with the French PicSat \cite{Lacour:2014}. More experiment will certainly follow in the future and demonstrate the full potential of astrophotonics technologies for space astronomy.

Related to miniaturisation is also the opportunity to integrate in the instrument many small copies of optical devices delivering or processing the light signal from the sky. Under this aspect, the introduction of optical fibres in spectroscopy revolutionised astronomy allowing to multiply by up to three orders of magnitude the productivity of spectrographs (Sec. \ref{sec:MOS}) or to enable spatially and spectrally resolved images of distant galaxies (Sec. \ref{sec:IFS}). This trend will certainly continue in the future as planned large scale MOS (\textit{e.g.} 4MOST, or ELT/MOSAIC) will see the light and new astrophotonic functionalities may be added (\textit{e.g.} multiple PIMMS \cite{BlandHawthorn:2006}).

Enhanced performance is also a key element of astrophotonic instrumentation, which may develop fully in the coming years. As we mentioned in Sec. \ref{sec:HAR}, the superior performance of integrated optics beam combiners for optical interferometry has been already demonstrated and employed in instruments like GRAVITY, which promises to revolutionise our knowledge on the supermassive black-hole nested at the center of our Galaxy\cite{Abuter:2017}. Extension of the concept to \ac{MIR} wavelengths is rapidly progressing 
\cite{Diener:2017,KenchingtonGoldsmith2017b,Labadie:2018} as is the increase of the number of combined telescopes at the yet unexplored J spectral band \cite{Pedretti:2018}.
As we have seen in Section \ref{sec:OH}, photonics is the only viable existing option to enable the suppression of telluric OH emission lines before the spectrograph, a key functionality for spectroscopy of galaxies with redshifts between 2 and 3 (the cosmic noon). While proof-of-principle  experiments have been already carried out with GNOSIS, the project PRAXIS 
has moved from technology to science demonstrator, delivering first science grade observations of galaxies. Active photonics is also becoming more and more an essential tool of modern astronomy. Laser Guide Stars (Sec. \ref{sec:LGS}) will be the standard equipment in Large and Extremely Large telescopes required to extend the use of \ac{AO} to fainter objects, while laser frequency combs (Sec. \ref{sec:astrocombs}) promise to become the ultimate calibrators for mid- and high-resolution spectroscopy.
Laser light sources are becoming also common in space exploration. For instance, the ChemCam of NASA’s Curiosity Mars Rover features a powerful nanosecond pulse infrared laser to perform Laser-Induced Breakdown Spectroscopy (LIBS) of distant rocks. Light emitted by the laser plasma is delivered via optical fibre from the collection optics on a robotic arm to the spectrometer hosted in the chassis of the rover. In the future, lasers on board of deep-space probes could enable extremely accurate ranging to push our knowledge on planetary dynamics to unprecedented precision \cite{Smith:2006}, while large datasets could be beamed optically to Earth at a rate of several hundred Mbit/s \cite{BlandHawthorn:2002,Sodnik:2014,Robinson:2014} exploiting the high brilliance of laser sources.

We believe these three features will be the cornerstones of future applications of photonic technologies to astronomical instrumentation. While the choice between astrophotonic and conventional solutions has to be evaluated for each instrumental project, at present three main challenges for the deployment of the potential of photonic technologies in astronomy.
   
One key issue affecting mainly single mode astrophotonic components is the low throughput, which is mainly due to the mismatch in the modal content of uncorrected starlight and the supported modes in photonic component (see Sec. \ref{sec:photonics_perspective}). This problem is particularly severe for large telescopic apertures and short optical wavelengths. Besides using photonic lanterns to sort multimode light into several single mode outputs, the definitive solution to this problem is matching the single mode components to a telescope equipped with \ac{AO}. While diffraction limited beams are already possible at infrared wavelengths on 8-meter class telescopes \cite{Dekany:2013,Jovanovic:2014}, recent advances allow the achievement of moderate Strehl ratios in the visible band at 6-meter class telescopes \cite{Pedichini:2017}. At the same time, the development of micro-electro-mechanical-systems has consistently reduced the prices of high-order deformable mirrors, making \ac{AO} more affordable. We expect that the increasing number of telescopes equipped with moderate to extreme \ac{AO} systems will contribute to increase the competitiveness of single-mode astrophotonic instrumentation as PIMMS, while bringing to fruition the decade-long development of  phase mask coronagraphs (Sec. \ref{sec:coronagraph}).

A second aspect hindering the use of astrophotonic components is the relatively narrow spectral coverage of existing manufacturing technologies. Most of the photonic components has been optimised for the digital telecommunication market, which operates in a narrow infrared band (approximatively corresponding to the shorter half of the astronomical H-band) where silica is most transparent. The availability of commercial components outside this range is scarce and often not compliant to the requirements of astronomical use, e.g. they feature higher insertion or scattering losses compared to telecom band equivalents. Clearly, astronomers will need to rely on existing manufacturing technologies, as the optimisation of photonic production processes is beyond the possibilities of the community. However growing markets, such as biophotonics and security, currently drive photonic applications in the visible and \ac{MIR} bands, which may deliver the required platforms for the development of astrophotonic components. As we have seen in Sec. \ref{sec:HAR}, advances in manufacturing technologies for \ac{MIR} photonics are behind recent progress towards \ac{MIR} integrated optics beam combiners for stellar interferometry. 

A third aspect is the achievable dynamical range of integrated optical devices. As mentioned, these have mostly been developed for digital telecommunications where modulation contrasts of 20 dB (1:100) or less are usually sufficient to reduce bit error rates to negligible values. With these specifications, stray light originating from modal mismatches or scattering in waveguides does not affect substantially the performance of the devices. On the contrary, several astronomical science cases (such as high precision spectroscopy, coronagraphy or nulling interferometry) require contrasts exceeding 30 dB. At these levels the already mentioned sources of stray light may result unacceptable and pose a challenge for the development of astronomy-grade integrated optics.

We put forward that dealing with these challenges will be at the focal point of new astrophotonic developments in the years to come, which will certainly witness new fascinating discoveries in astronomy enabled by increasingly sophisticated instrumentation, where state-of-the-art photonics will play a crucial role.

\section*{Acknowledgements}
SM would like to thank his family for allowing him to devote many evenings to the writing of the manuscript.
RJH would like to thank Rosalie McGurk for her useful input regarding the current questions important to astronomy.

\bibliographystyle{spmpsci}      
\bibliography{references}   

\begin{thebibliography}{100}
\providecommand{\url}[1]{{#1}}
\providecommand{\urlprefix}{URL }
\expandafter\ifx\csname urlstyle\endcsname\relax
  \providecommand{\doi}[1]{DOI~\discretionary{}{}{}#1}\else
  \providecommand{\doi}{DOI~\discretionary{}{}{}\begingroup
  \urlstyle{rm}\Url}\fi

\bibitem{akerlof:2000}
Akerlof, C., Amrose, S., Balsano, R., Bloch, J., Casperson, D., Fletcher, S.,
  Gisler, G., Hills, J., Kehoe, R., Lee, B., et~al.: Rotse all-sky surveys for
  variable stars. i. test fields.
\newblock The Astronomical Journal \textbf{119}(4), 1901 (2000)

\bibitem{Albota:2004}
{Albota}, M.A., {Wong}, F.N.C.: {Efficient single-photon counting at 1.55
  {$\mu$}m by means of frequency upconversion}.
\newblock Optics Letters \textbf{29}, 1449--1451 (2004).
\newblock \doi{10.1364/OL.29.001449}

\bibitem{Alcock:1993}
{Alcock}, C., {Akerlof}, C.W., {Allsman}, R.A., {Axelrod}, T.S., {Bennett},
  D.P., {Chan}, S., {Cook}, K.H., {Freeman}, K.C., {Griest}, K., {Marshall},
  S.L., {Park}, H.S., {Perlmutter}, S., {Peterson}, B.A., {Pratt}, M.R.,
  {Quinn}, P.J., {Rodgers}, A.W., {Stubbs}, C.W., {Sutherland}, W.: {Possible
  gravitational microlensing of a star in the Large Magellanic Cloud}.
\newblock \nat \textbf{365}, 621--623 (1993).
\newblock \doi{10.1038/365621a0}

\bibitem{allingtonsmith:2006}
Allington-Smith, J.: Basic principles of integral field spectroscopy.
\newblock New Astronomy Reviews \textbf{50}(4), 244--251 (2006)

\bibitem{AllingtonSmith:2002}
Allington-Smith, J., Murray, G., Dodsworth, G., Davies, R., Miller, B.W.,
  Jorgensen, I., Hook, I., Crampton, D., Murowinski, R.: Integral field
  spectroscopy with the gemini multiobject spectrograph. i. design,
  construction, and testing.
\newblock Publications of the Astronomical Society of the Pacific
  \textbf{114}(798), 892 (2002)

\bibitem{Angel:1997}
{Angel}, J.R.P., {Woolf}, N.J.: {An Imaging Nulling Interferometer to Study
  Extrasolar Planets}.
\newblock Astroph. J. \textbf{475}, 373--379 (1997).
\newblock \doi{10.1086/303529}

\bibitem{Anglada2016}
{Anglada-Escud{\'e}}, G., {Amado}, P.J., {Barnes}, J., {Berdi{\~n}as}, Z.M.,
  {Butler}, R.P., {Coleman}, G.A.L., {de La Cueva}, I., {Dreizler}, S., {Endl},
  M., {Giesers}, B., {Jeffers}, S.V., {Jenkins}, J.S., {Jones}, H.R.A.,
  {Kiraga}, M., {K{\"u}rster}, M., {L{\'o}pez-Gonz{\'a}lez}, M.J., {Marvin},
  C.J., {Morales}, N., {Morin}, J., {Nelson}, R.P., {Ortiz}, J.L., {Ofir}, A.,
  {Paardekooper}, S.J., {Reiners}, A., {Rodr{\'{\i}}guez}, E.,
  {Rodr{\'{\i}}guez-L{\'o}pez}, C., {Sarmiento}, L.F., {Strachan}, J.P.,
  {Tsapras}, Y., {Tuomi}, M., {Zechmeister}, M.: {A terrestrial planet
  candidate in a temperate orbit around Proxima Centauri}.
\newblock \nat \textbf{536}, 437--440 (2016).
\newblock \doi{10.1038/nature19106}

\bibitem{Arsenault:2012}
Arsenault, R., Madec, P.Y., Paufique, J., Penna, P.L., Str\"obele, S., Vernet,
  E., Pirard, J.F., Hackenberg, W., Kuntschner, H., Jochum, L., Kolb, J.,
  Muller, N., Louarn, M.L., Amico, P., Hubin, N., Lizon, J.L., Ridings, R.,
  Abad, J., Fischer, G., Heinz, V., Kiekebusch, M., Argomedo, J., Conzelmann,
  R., Tordo, S., Donaldson, R., Soenke, C., Duhoux, P., Fedrigo, E., Delabre,
  B., Jost, A., Duchateau, M., Downing, M., Moreno, J., Dorn, R., Manescau, A.,
  Calia, D.B., Quattri, M., Dupuy, C., Guidolin, I., Comin, M., Guzman, R.,
  Buzzoni, B., Quentin, J., Lewis, S., Jolley, P., Kraus, M., Pfrommer, T.,
  Biasi, R., Gallieni, D., Bechet, C., Stuik, R.: Eso adaptive optics facility
  progress report.
\newblock In: Proc. SPIE, vol. 8447, p.~0J (2012)

\bibitem{Artigau2014}
{Artigau}, {\'E}., {Kouach}, D., {Donati}, J.F., {Doyon}, R., {Delfosse}, X.,
  {Baratchart}, S., {Lacombe}, M., {Moutou}, C., {Rabou}, P., {Par{\`e}s},
  L.P., {Micheau}, Y., {Thibault}, S., {Reshetov}, V.A., {Dubois}, B.,
  {Hernandez}, O., {Vall{\'e}e}, P., {Wang}, S.Y., {Dolon}, F., {Pepe}, F.A.,
  {Bouchy}, F., {Striebig}, N., {H{\'e}nault}, F., {Loop}, D., {Saddlemyer},
  L., {Barrick}, G., {Vermeulen}, T., {Dupieux}, M., {H{\'e}brard}, G.,
  {Boisse}, I., {Martioli}, E., {Alencar}, S.H.P., {do Nascimento}, J.D.,
  {Figueira}, P.: {SPIRou: the near-infrared spectropolarimeter/high-precision
  velocimeter for the Canada-France-Hawaii telescope}.
\newblock In: Ground-based and Airborne Instrumentation for Astronomy V,
  \emph{Proc. SPIE}, vol. 9147, p. 914715 (2014).
\newblock \doi{10.1117/12.2055663}

\bibitem{Avila:2012}
Avila, G., Guirao, C., Baader, T.: High efficiency inexpensive 2-slices image
  slicers.
\newblock In: SPIE Astronomical Telescopes+ Instrumentation, pp.
  84,469M--84,469M. International Society for Optics and Photonics (2012)

\bibitem{bacon2010muse}
Bacon, R., Accardo, M., Adjali, L., Anwand, H., Bauer, S., Biswas, I., Blaizot,
  J., Boudon, D., Brau-Nogue, S., Brinchmann, J., et~al.: The muse
  second-generation vlt instrument.
\newblock In: Ground-based and Airborne Instrumentation for Astronomy III, vol.
  7735, p. 773508. International Society for Optics and Photonics (2010)

\bibitem{Bacon:2001}
Bacon, R., Copin, Y., Monnet, G., Miller, B.W., Allington-Smith, J., Bureau,
  M., Marcella~Carollo, C., Davies, R.L., Emsellem, E., Kuntschner, H., et~al.:
  The sauron project--i. the panoramic integral-field spectrograph.
\newblock Monthly Notices of the Royal Astronomical Society \textbf{326}(1),
  23--35 (2001)

\bibitem{bailyn2007renewing}
Bailyn, C., Clemens, C., Johnson, J., Joseph, R., Kawaler, S., Salzer, J.,
  Thornley, M., Weintraub, D.: Renewing small telescopes for astronomical
  research (2007)

\bibitem{Baldwin:1986}
{Baldwin}, J.E., {Haniff}, C.A., {Mackay}, C.D., {Warner}, P.J.: {Closure phase
  in high-resolution optical imaging}.
\newblock Nature \textbf{320}, 595--597 (1986).
\newblock \doi{10.1038/320595a0}

\bibitem{Bally2000}
{Bally}, J., {O'Dell}, C.R., {McCaughrean}, M.J.: {Disks, Microjets, Windblown
  Bubbles, and Outflows in the Orion Nebula}.
\newblock \aj \textbf{119}, 2919--2959 (2000).
\newblock \doi{10.1086/301385}

\bibitem{Baranne:1996}
{Baranne}, A., {Queloz}, D., {Mayor}, M., {Adrianzyk}, G., {Knispel}, G.,
  {Kohler}, D., {Lacroix}, D., {Meunier}, J.P., {Rimbaud}, G., {Vin}, A.:
  {ELODIE: A spectrograph for accurate radial velocity measurements.}
\newblock Astronomy and Astrophysics, Supplement \textbf{119}, 373--390 (1996)

\bibitem{Baril:2010}
Baril, M., Lai, O., Zahariade, G., Bouchacourt, F., Perrin, G., Fedou, P.,
  Woillez, J.: 'ohana-iki: A test-bed for the 'ohana beam combiner and delay
  line at cfht.
\newblock In: Proc. SPIE, vol. 7734, pp. 7734--2C (2010)

\bibitem{baudrand2001modal}
Baudrand, J., Walker, G.A.: Modal noise in high-resolution, fiber-fed spectra:
  A study and simple cure.
\newblock Publications of the Astronomical Society of the Pacific
  \textbf{113}(785), 851 (2001)

\bibitem{Bely:2006}
Bely, P.: The Design and Construction of Large Optical Telescopes.
\newblock Astronomy and Astrophysics Library. Springer New York (2006).
\newblock \urlprefix\url{https://books.google.de/books?id=e7gGCAAAQBAJ}

\bibitem{Benedick:2010}
{Benedick}, A.J., {Chang}, G., {Birge}, J.R., {Chen}, L.J., {Glenday}, A.G.,
  {Li}, C.H., {Phillips}, D.F., {Szentgyorgyi}, A., {Korzennik}, S., {Furesz},
  G., {Walsworth}, R.L., {K{\"a}rtner}, F.X.: {Visible wavelength astro-comb}.
\newblock Opt. Exp. \textbf{18}, 19,175 (2010).
\newblock \doi{10.1364/OE.18.019175}

\bibitem{Benisty:2009}
Benisty, M., Berger, J.P., Jocou, L., Labeye, P., Malbet, F., Perraut, K.,
  Kern, P.: An integrated optics beam combiner for the second generation vlti
  instruments.
\newblock A\&A \textbf{498}, 601--613 (2009)

\bibitem{Benisty2015}
{Benisty}, M., {Juhasz}, A., {Boccaletti}, A., {Avenhaus}, H., {Milli}, J.,
  {Thalmann}, C., {Dominik}, C., {Pinilla}, P., {Buenzli}, E., {Pohl}, A.,
  {Beuzit}, J.L., {Birnstiel}, T., {de Boer}, J., {Bonnefoy}, M., {Chauvin},
  G., {Christiaens}, V., {Garufi}, A., {Grady}, C., {Henning}, T., {Huelamo},
  N., {Isella}, A., {Langlois}, M., {M{\'e}nard}, F., {Mouillet}, D.,
  {Olofsson}, J., {Pantin}, E., {Pinte}, C., {Pueyo}, L.: {Asymmetric features
  in the protoplanetary disk MWC 758}.
\newblock \aap \textbf{578}, L6 (2015).
\newblock \doi{10.1051/0004-6361/201526011}

\bibitem{Berger:2000}
Berger, J.P., Benech, P., Schanen-Duport, I., Maury, G., Malbet, F., Reynaud,
  F.: Combining up to eight telescope beams in a single chip.
\newblock In: Proc. SPIE, vol. 4006, pp. 986--995 (2000)

\bibitem{Berger:2003}
Berger, J.P., Hagenauer, P., Kern, P., Perraut, K., Malbet, F., Gluck, S.,
  Lagny, L., Schanen, I., Laurent, E., Delboulbe, A., Magnard, Y., Tatulli, E.,
  Traub, W., Carleton, N., Millan-Gabet, R., Monnier, J.D., Pedretti, E.: An
  integrated-optics 3-way beam combiner for iota.
\newblock In: Proc. SPIE, vol. 4838, pp. 1099--1106 (2003)

\bibitem{Berger:2001}
Berger, J.P., Hagenauer, P., Kern, P., Perraut, K., Malbet, F., Schanen, I.,
  Severi, M., Millan-Gabet, R., Traub, W.: Integrated optics for astronomical
  interferometry - iv. first measurements on stars.
\newblock A\&A \textbf{376}, L31--L34 (2001)

\bibitem{Berger:1999}
Berger, J.P., Rousselet-Perraut, K., Kern, P., Malbet, F., Schanen-Duport, I.,
  Reynaud, F., Hagenauer, P., Benech, P.: Integrated optics for astronomical
  interferometry - ii. first laboratory white-light interferograms.
\newblock A\&AS \textbf{139}, 173--177 (1999)

\bibitem{Berry:1984}
Berry, M.: Quantal phase factors accompanying adiabatic changes.
\newblock Proceedings of the Royal Society of London A: Mathematical, Physical
  and Engineering Sciences \textbf{392}(1802), 45--57 (1984).
\newblock \doi{10.1098/rspa.1984.0023}.
\newblock
  \urlprefix\url{http://rspa.royalsocietypublishing.org/content/392/1802/45}

\bibitem{Besse:1994}
{Besse}, P.A., {Bachmann}, M., {Melchior}, H., {Soldano}, L.B., {Smit}, M.K.:
  {Optical bandwidth and fabrication tolerances of multimode interference
  couplers}.
\newblock Journal of Lightwave Technology \textbf{12}, 1004--1009 (1994).
\newblock \doi{10.1109/50.296191}

\bibitem{Betters:2012}
Betters, C., Leon-Saval, S., Bland-Hawthorn, J.: Nanospec: A diffraction
  limited micro-spectrograph for pico- and nano-satellites.
\newblock In: Proceedings of SPIE - The International Society for Optical
  Engineering, vol. 8442 (2012)

\bibitem{Bezuhanov:2004}
{Bezuhanov}, K., {Dreischuh}, A., {Paulus}, G.G., {Sch{\"a}tzel}, M.G.,
  {Walther}, H.: {Vortices in femtosecond laser fields}.
\newblock Optics Letters \textbf{29}, 1942--1944 (2004).
\newblock \doi{10.1364/OL.29.001942}

\bibitem{Bienfang:2003}
Bienfang, J.C., Denman, C.A., Grime, B.W., Hillman, P.D., Moore, G.T., Telle,
  J.M.: 20 w of continuous-wave sodium $d_2$ resonance radiation from
  sum-frequency generation with injection-locked lasers.
\newblock Opt. Lett. \textbf{28}, 2219 (2003)

\bibitem{Birks:2015}
Birks, T.A., Gris-S{\'a}nchez, I., Yerolatsitis, S., Leon-Saval, S., Thomson,
  R.R.: The photonic lantern.
\newblock Advances in Optics and Photonics \textbf{7}(2), 107--167 (2015)

\bibitem{Birks:1997}
{Birks}, T.A., {Knight}, J.C., {Russell}, P.S.J.: {Endlessly
  single-modephotonic crystal fiber}.
\newblock Optics Letters \textbf{22}(13), 961--963 (1997).
\newblock \doi{10.1364/OL.22.000961}

\bibitem{birks2012}
{Birks}, T.A., {Mangan}, B.J., {D{\'{\i}}ez}, A., {Cruz}, J.L., {Murphy}, D.F.:
  {``Photonic lantern'' spectral filters in multi-core Fiber}.
\newblock Optics Express \textbf{20}, 13,996 (2012).
\newblock \doi{10.1364/OE.20.013996}

\bibitem{Blais-Ouellette:2004}
{Blais-Ouellette}, S., {Artigau}, {\'E}., {Havermeyer}, F., {Matthews}, K.,
  {Moser}, C., {Psaltis}, D., {Steckman}, G.J.: {Multi-notch holographic
  filters for atmospheric lines suppression}.
\newblock In: E.~{Atad-Ettedgui}, P.~{Dierickx} (eds.) Optical Fabrication,
  Metrology, and Material Advancements for Telescopes, \emph{Proc. SPIE}, vol.
  5494, pp. 554--561 (2004).
\newblock \doi{10.1117/12.552116}

\bibitem{Blake:2015}
{Blake}, C., {Johnson}, J., {Plavchan}, P., {Sliski}, D., {Wittenmyer}, R.A.,
  {Eastman}, J.D., {Barnes}, S.: {MINERVA-Red: A Census of Planets Orbiting the
  Nearest Low-mass Stars to the Sun}.
\newblock In: American Astronomical Society Meeting Abstracts \#225,
  \emph{American Astronomical Society Meeting Abstracts}, vol. 225, p. 257.32
  (2015)

\bibitem{Blake:2013}
Blake, S., Dunlop, C., Nandi, D., Sharples, R., Talbot, G., Shanks, T.,
  Donoghue, D., Galiatsatos, N., Luke, P.: New microslice technology for
  hyperspectral imaging.
\newblock Remote Sensing \textbf{5}(3), 1204--1219 (2013)

\bibitem{BlandHawthorn:2011}
Bland-Hawthorn, J., Ellis, S., Leon-Saval, S., Haynes, R., Roth, M.,
  L{\"o}hmannsr{\"o}ben, H.G., Horton, A., Cuby, J.G., Birks, T.A., Lawrence,
  J., et~al.: A complex multi-notch astronomical filter to suppress the bright
  infrared sky.
\newblock Nature communications \textbf{2}, 581 (2011)

\bibitem{BlandHawthorn:2004}
Bland-Hawthorn, J., Englund, M., Edvell, G.: New approach to atmospheric oh
  suppression using an aperiodic fibre bragg grating.
\newblock Optics Express \textbf{12}(24), 5902--5909 (2004)

\bibitem{BlandHawthorn:2002}
{Bland-Hawthorn}, J., {Harwit}, A., {Harwit}, M.: {Laser Telemetry from Space}.
\newblock Science \textbf{297} (2002).
\newblock \doi{10.1126/science.1074113}

\bibitem{BlandHawthorn:2006}
Bland-Hawthorn, J., Horton, A.: Instruments without optics: an integrated
  photonic spectrograph.
\newblock In: Society of Photo-Optical Instrumentation Engineers (SPIE)
  Conference Series, vol. 6269 (2006)

\bibitem{BlandHawthorn:2009}
{Bland-Hawthorn}, J., {Kern}, P.: {Astrophotonics: a new era for astronomical
  instruments}.
\newblock Optics Express \textbf{17}, 1880--1884 (2009).
\newblock \doi{10.1364/OE.17.001880}

\bibitem{BlandHawthorn:2010}
Bland-Hawthorn, J., Lawrence, J., Robertson, G., Campbell, S., Pope, B.,
  Betters, C., Leon-Saval, S., Birks, T., Haynes, R., Cvetojevic, N., et~al.:
  Pimms: photonic integrated multimode microspectrograph.
\newblock In: Proceedings of SPIE-The International Society for Optical
  Engineering, vol. 7735. University of Bath (2010)

\bibitem{BlandHawthorn:2017}
{Bland-Hawthorn}, J., {Leon-Saval}, S.G.: {Astrophotonics: molding the flow of
  light in astronomical instruments}.
\newblock Optics Express \textbf{25}, 15,549 (2017).
\newblock \doi{10.1364/OE.25.015549}

\bibitem{Boccaletti:2004}
{Boccaletti}, A., {Riaud}, P., {Baudoz}, P., {Baudrand}, J., {Rouan}, D.,
  {Gratadour}, D., {Lacombe}, F., {Lagrange}, A.M.: {The Four-Quadrant Phase
  Mask Coronagraph. IV. First Light at the Very Large Telescope}.
\newblock Publ. Astron. Soc. Pac. \textbf{116}, 1061--1071 (2004).
\newblock \doi{10.1086/425735}

\bibitem{Boccaletti2015}
{Boccaletti}, A., {Thalmann}, C., {Lagrange}, A.M., {Janson}, M., {Augereau},
  J.C., {Schneider}, G., {Milli}, J., {Grady}, C., {Debes}, J., {Langlois}, M.,
  {Mouillet}, D., {Henning}, T., {Dominik}, C., {Maire}, A.L., {Beuzit}, J.L.,
  {Carson}, J., {Dohlen}, K., {Engler}, N., {Feldt}, M., {Fusco}, T., {Ginski},
  C., {Girard}, J.H., {Hines}, D., {Kasper}, M., {Mawet}, D., {M{\'e}nard}, F.,
  {Meyer}, M.R., {Moutou}, C., {Olofsson}, J., {Rodigas}, T., {Sauvage}, J.F.,
  {Schlieder}, J., {Schmid}, H.M., {Turatto}, M., {Udry}, S., {Vakili}, F.,
  {Vigan}, A., {Wahhaj}, Z., {Wisniewski}, J.: {Fast-moving features in the
  debris disk around AU Microscopii}.
\newblock \nat \textbf{526}, 230--232 (2015).
\newblock \doi{10.1038/nature15705}

\bibitem{Bonaccini:2002}
Bonaccini, D., Hackenberg, W., Cullum, M., Brunetto, E., Ott, T., Quattri, M.,
  Allaert, E., Dimmler, M., Tarenghi, M., Kersteren, A.V., DiChirico, C.,
  Buzzoni, B., Gray, P., Tamai, R., Tapia, M.: Eso vlt laser guide star
  facility.
\newblock In: Proc. SPIE, vol. 4494, pp. 276--289 (2002)

\bibitem{Bonnefoy2016}
{Bonnefoy}, M., {Zurlo}, A., {Baudino}, J.L., {Lucas}, P., {Mesa}, D., {Maire},
  A.L., {Vigan}, A., {Galicher}, R., {Homeier}, D., {Marocco}, F., {Gratton},
  R., {Chauvin}, G., {Allard}, F., {Desidera}, S., {Kasper}, M., {Moutou}, C.,
  {Lagrange}, A.M., {Antichi}, J., {Baruffolo}, A., {Baudrand}, J., {Beuzit},
  J.L., {Boccaletti}, A., {Cantalloube}, F., {Carbillet}, M., {Charton}, J.,
  {Claudi}, R.U., {Costille}, A., {Dohlen}, K., {Dominik}, C., {Fantinel}, D.,
  {Feautrier}, P., {Feldt}, M., {Fusco}, T., {Gigan}, P., {Girard}, J.H.,
  {Gluck}, L., {Gry}, C., {Henning}, T., {Janson}, M., {Langlois}, M., {Madec},
  F., {Magnard}, Y., {Maurel}, D., {Mawet}, D., {Meyer}, M.R., {Milli}, J.,
  {Moeller-Nilsson}, O., {Mouillet}, D., {Pavlov}, A., {Perret}, D., {Pujet},
  P., {Quanz}, S.P., {Rochat}, S., {Rousset}, G., {Roux}, A., {Salasnich}, B.,
  {Salter}, G., {Sauvage}, J.F., {Schmid}, H.M., {Sevin}, A., {Soenke}, C.,
  {Stadler}, E., {Turatto}, M., {Udry}, S., {Vakili}, F., {Wahhaj}, Z.,
  {Wildi}, F.: {First light of the VLT planet finder SPHERE. IV. Physical and
  chemical properties of the planets around HR8799}.
\newblock \aap \textbf{587}, A58 (2016).
\newblock \doi{10.1051/0004-6361/201526906}

\bibitem{BornWolf}
Born, M., Wolf, E.: Principles of optics.
\newblock Cambridge, Cambridge (1997)

\bibitem{bouchy2013sophie+}
Bouchy, F., D{\'\i}az, R., H{\'e}brard, G., Arnold, L., Boisse, I., Delfosse,
  X., Perruchot, S., Santerne, A.: Sophie+: First results of an
  octagonal-section fiber for high-precision radial velocity measurements.
\newblock Astronomy \& Astrophysics \textbf{549}, A49 (2013)

\bibitem{LeBouquin:2011}
Bouquin, J.B.L., Berger, J.P., Lazareff, B., Zins, G., Haguenauer, P., Jocou,
  L., Kern, P., Millan-Gabet, R., Traub, W., Absil, O., Augereau, J.C.,
  Benisty, M., Blind, N., Bonfils, X., Bourget, P., Delboulbe, A., Feautrier,
  P., M.Germain, Gitton, P., Gillier, D., Kiekebusch, M., Kluska, J.,
  Knudstrup, J., Labeye, P., Lizon, J.L., Monin, J.L., Magnard, Y., Malbet, F.,
  Maurel, D., M\'enard, F., Micallef, M., Michaud, L., Montagnier, G., Morel,
  S., Moulin, T., Perraut, K., Popovic, D., Rabou, P., Rochat, S., Rojas, C.,
  Roussel, F., Roux, A., Stadler, E., Stefl, S., Tatulli, E., Ventura, N.:
  Pionier: a 4-telescope visitor instrument at vlti.
\newblock A\&A \textbf{535}, A67 (2011)

\bibitem{bowler2010near}
Bowler, B.P., Liu, M.C., Dupuy, T.J., Cushing, M.C.: Near-infrared spectroscopy
  of the extrasolar planet hr 8799 b.
\newblock The Astrophysical Journal \textbf{723}(1), 850 (2010)

\bibitem{Bracewell:1978}
{Bracewell}, R.N.: {Detecting nonsolar planets by spinning infrared
  interferometer}.
\newblock Nature \textbf{274}, 780 (1978).
\newblock \doi{10.1038/274780a0}

\bibitem{Breckingridge:2005}
{Breckinridge}, J., {Lindensmith}, C.: {The Astronomical Search for Origins}.
\newblock Optics \& Photonics News \textbf{16}, 24 (2005).
\newblock \doi{10.1364/OPN.16.2.000024}

\bibitem{brown:2012}
Brown, G., Thomson, R.R., Kar, A.K., Psaila, N.D., Bookey, H.T.: Ultrafast
  laser inscription of bragg-grating waveguides using the multiscan technique.
\newblock Optics letters \textbf{37}(4), 491--493 (2012)

\bibitem{Brustlein:2008}
{Brustlein}, S., {Del Rio}, L., {Tonello}, A., {Delage}, L., {Reynaud}, F.,
  {Herrmann}, H., {Sohler}, W.: {Laboratory Demonstration of an
  Infrared-to-Visible Up-Conversion Interferometer for Spatial Coherence
  Analysis}.
\newblock Physical Review Letters \textbf{100}(15), 153903 (2008).
\newblock \doi{10.1103/PhysRevLett.100.153903}

\bibitem{bryant:2011}
Bryant, J., O'Byrne, J., Bland-Hawthorn, J., Leon-Saval, S.: Characterization
  of hexabundles: initial results.
\newblock Monthly Notices of the Royal Astronomical Society \textbf{415}(3),
  2173--2181 (2011)

\bibitem{bryant:2015}
Bryant, J., Owers, M., Robotham, A., Croom, S., Driver, S., Drinkwater, M.,
  Lorente, N., Cortese, L., Scott, N., Colless, M., et~al.: The sami galaxy
  survey: instrument specification and target selection.
\newblock Monthly Notices of the Royal Astronomical Society \textbf{447}(3),
  2857--2879 (2015)

\bibitem{Burtscher2016}
{Burtscher}, L., {H{\"o}nig}, S., {Jaffe}, W., {Kishimoto}, M.,
  {Lopez-Gonzaga}, N., {Meisenheimer}, K., {Tristam}, K.R.W.: {Infrared
  interferometry and AGNs: Parsec-scale disks and dusty outflows}.
\newblock In: Optical and Infrared Interferometry and Imaging V,
  \emph{\procspie}, vol. 9907, p. 99070R (2016).
\newblock \doi{10.1117/12.2231077}

\bibitem{Buscher:2013}
{Buscher}, D.F., {Creech-Eakman}, M., {Farris}, A., {Haniff}, C.A., {Young},
  J.S.: {The Conceptual Design of the Magdalena Ridge Observatory
  Interferometer}.
\newblock Journal of Astronomical Instrumentation \textbf{2}, 1340001 (2013).
\newblock \doi{10.1142/S2251171713400011}

\bibitem{Buscher:1990}
{Buscher}, D.F., {Haniff}, C.A., {Baldwin}, J.E., {Warner}, P.J.: {Detection of
  a bright feature on the surface of Betelgeuse.}
\newblock MNRAS \textbf{245}, 7P (1990)

\bibitem{Butler:1996}
{Butler}, R.P., {Marcy}, G.W., {Williams}, E., {McCarthy}, C., {Dosanjh}, P.,
  {Vogt}, S.S.: {Attaining Doppler Precision of 3 M s-1}.
\newblock Publications of the Astronomical Society of the Pacific \textbf{108},
  500 (1996).
\newblock \doi{10.1086/133755}

\bibitem{Bonaccini:2004}
Calia, D.B., Hackenberg, W.K., Araujo, C., Guidolin, I., Alvarez, J.L.:
  {Laser-guide-star-related activities at ESO}.
\newblock In: D.B. Calia, B.L. Ellerbroek, R.~Ragazzoni (eds.) Advancements in
  Adaptive Optics, vol. 5490, pp. 974 -- 980. International Society for Optics
  and Photonics, SPIE (2004).
\newblock \doi{10.1117/12.552947}.
\newblock \urlprefix\url{https://doi.org/10.1117/12.552947}

\bibitem{Ceus:2012}
{Ceus}, D., {Reynaud}, F., {Woillez}, J., {Lai}, O., {Delage}, L., {Grossard},
  L., {Baudoin}, R., {Gomes}, J.T., {Bouyeron}, L., {Herrmann}, H., {Sohler},
  W.: {Application of frequency conversion of starlight to high-resolution
  imaging interferometry. On-sky sensitivity test of a single arm of the
  interferometer}.
\newblock Month. Not. R. Astron. Soc. \textbf{427}, L95--L98 (2012).
\newblock \doi{10.1111/j.1745-3933.2012.01352.x}

\bibitem{Ceus:2011}
{Ceus}, D., {Tonello}, A., {Grossard}, L., {Delage}, L., {Reynaud}, F.,
  {Herrmann}, H., {Sohler}, W.: {Phase closure retrieval in an
  infrared-to-visible upconversion interferometer for high resolution
  astronomical imaging}.
\newblock Optics Express \textbf{19}, 8616 (2011).
\newblock \doi{10.1364/OE.19.008616}

\bibitem{Charles:2012}
{Charles}, N., {Jovanovic}, N., {Gross}, S., {Stewart}, P., {Norris}, B.,
  {O'Byrne}, J., {Lawrence}, J.S., {Withford}, M.J., {Tuthill}, P.G.: {Design
  of optically path-length-matched, three-dimensional photonic circuits
  comprising uniquely routed waveguides}.
\newblock Appl. Opt. \textbf{51}, 6489 (2012).
\newblock \doi{10.1364/AO.51.006489}

\bibitem{Chesneau:2008}
{Chesneau}, O., {Banerjee}, D.P.K., {Millour}, F., {Nardetto}, N., {Sacuto},
  S., {Spang}, A., {Wittkowski}, M., {Ashok}, N.M., {Das}, R.K., {Hummel}, C.,
  {Kraus}, S., {Lagadec}, E., {Morel}, S., {Petr-Gotzens}, M., {Rantakyro}, F.,
  {Sch{\"o}ller}, M.: {VLTI monitoring of the dust formation event of the Nova
  V1280 Scorpii}.
\newblock A\&A \textbf{487}, 223--235 (2008).
\newblock \doi{10.1051/0004-6361:200809485}

\bibitem{Chilcote:2012}
{Chilcote}, J.K., {Larkin}, J.E., {Maire}, J., {Perrin}, M.D., {Fitzgerald},
  M.P., {Doyon}, R., {Thibault}, S., {Bauman}, B., {Macintosh}, B.A., {Graham},
  J.R., {Saddlemyer}, L.: {Performance of the integral field spectrograph for
  the Gemini Planet Imager}.
\newblock In: Ground-based and Airborne Instrumentation for Astronomy IV,
  \emph{Proc. SPIE}, vol. 8446, p. 84468W (2012).
\newblock \doi{10.1117/12.925790}

\bibitem{Ciliegi:2018}
Ciliegi, P., Diolaiti, E., Abicca, R., Agapito, G., Aliverti, M., Arcidiacono,
  C., Auricchio, N., Balestra, A., Baruffolo, A., Bellazzini, M., Bonaglia, M.,
  Bregoli, G., Brissaud, O., Busoni, L., Carlotti, A., Cascone, E., Correia,
  J.J., Cortecchia, F., Cosentino, G., D'Orazi, V., Dall'Ora, M., Caprio, V.D.,
  Rosa, A.D., Delboulbé, A., Antonio, I.D., Rico, G.D., Dolci, M., Esposito,
  S., Fantinel, D., Feautrier, P., Fiorentino, G., Foppiani, I., Giro, E.,
  Gluck, L., Grani, P., Greggio, D., Hénault, F., Jocou, L., Penna, P.L.,
  Lafrasse, S., Lauria, M., Coarer, E.L., Louarn, M.L., Lombini, M., Magnard,
  Y., Magrin, D., Maiorano, E., Mannucci, F., Marchetti, E., Maurel, D.,
  Michaud, L., Moraux, E., Morgante, G., Moulin, T., Oberti, S., Pariani, G.,
  Patti, M., Plantet, C., Podio, L., Puglisi, A., Rabou, P., Ragazzoni, R.,
  Redaelli, E., Riva, M., Rochat, S., Roussel, F., Roux, A., Salasnich, B.,
  Saracco, P., Schreiber, L., Spavone, M., Stadler, E., Sztefek, M.H., Terenzi,
  L., Valentini, A., Ventura, N., Vérinaud, C., Zaggia, S.: {MAORY for ELT:
  preliminary design overview}.
\newblock In: L.M. Close, L.~Schreiber, D.~Schmidt (eds.) Adaptive Optics
  Systems VI, vol. 10703, pp. 336 -- 345. International Society for Optics and
  Photonics, SPIE (2018).
\newblock \doi{10.1117/12.2313672}.
\newblock \urlprefix\url{https://doi.org/10.1117/12.2313672}

\bibitem{Claudi:2008}
{Claudi}, R.U., {Turatto}, M., {Gratton}, R.G., {Antichi}, J., {Bonavita}, M.,
  {Bruno}, P., {Cascone}, E., {De Caprio}, V., {Desidera}, S., {Giro}, E.,
  {Mesa}, D., {Scuderi}, S., {Dohlen}, K., {Beuzit}, J.L., {Puget}, P.: {SPHERE
  IFS: the spectro differential imager of the VLT for exoplanets search}.
\newblock In: Ground-based and Airborne Instrumentation for Astronomy II,
  \emph{Proc. SPIE}, vol. 7014, p. 70143E (2008).
\newblock \doi{10.1117/12.788366}

\bibitem{Colavita:1999}
Colavita, M.M., Wallace, J.K., Hines, B.E., Gursel, Y., Malbet, F., Palmer,
  D.L., Pan, X.P., Shao, M., Yu, J.W., Boden, A.F., Dumont, P.J., Gubler, J.,
  Koresko, C.D., Kulkarni, S.R., Lane, B.F., Mobley, D.W., van Belle, G.T.: The
  palomar testbed interferometer.
\newblock The Astrophysical Journal \textbf{510}(1), 505--521 (1999).
\newblock \doi{10.1086/306579}.
\newblock \urlprefix\url{https://doi.org/10.1086}

\bibitem{comte:1994}
Comte, G., Surace, C.: Slitless spectroscopy by photographic and ccd detectors
  across large fields.
\newblock In: Symposium-International Astronomical Union, vol. 161, pp.
  709--714. Cambridge University Press (1994)

\bibitem{Connes:1985}
{Connes}, P., {Froehly}, C., {Facq}, P.: {A fiber-linked version of project
  TRIO}.
\newblock In: N.~{Longdon}, O.~{Melita} (eds.) Kilometric Optical Arrays in
  Space, \emph{ESA Special Publication}, vol. 226 (1985)

\bibitem{Connes:1987}
{Connes}, P., {Shaklan}, S., {Roddier}, F.: {A Fiber-Linked Groundbased Array}.
\newblock In: J.W. {Goad} (ed.) Interferometric Imaging in Astronomy (1987)

\bibitem{Correia2006}
{Correia}, S., {Zinnecker}, H., {Ratzka}, T., {Sterzik}, M.F.: {A VLT/NACO
  survey for triple and quadruple systems among visual pre-main sequence
  binaries}.
\newblock \aap \textbf{459}, 909--926 (2006).
\newblock \doi{10.1051/0004-6361:20065545}

\bibitem{Corrigan:2016}
Corrigan, M., Harris, R.J., Thomson, R.R., MacLachlan, D.G., Allington-Smith,
  J., Myers, R., Morris, T.: Wavefront sensing using a photonic lantern.
\newblock In: SPIE Astronomical Telescopes+ Instrumentation, pp.
  990,969--990,969. International Society for Optics and Photonics (2016)

\bibitem{CoudeDuForesto:2003}
{Coud{\'e} du Foresto}, V., {Borde}, P.J., {Merand}, A., {Baudouin}, C.,
  {Remond}, A., {Perrin}, G.S., {Ridgway}, S.T., {ten Brummelaar}, T.A.,
  {McAlister}, H.A.: {FLUOR fibered beam combiner at the CHARA array}.
\newblock In: W.A. {Traub} (ed.) Interferometry for Optical Astronomy II,
  \emph{Proc. SPIE}, vol. 4838, pp. 280--285 (2003).
\newblock \doi{10.1117/12.459942}

\bibitem{CoudeDuForesto:1997b}
{Coud{\'e} du Foresto}, V., {Perrin}, G., {Mariotti}, J.M., {Lacasse}, M.,
  {Traub}, W.: {The FLUOR/IOTA fiber stellar interferometer}, p. 115.
\newblock EDP Sciences (1997)

\bibitem{CoudeDuForesto:1997}
{Coud{\'e} du Foresto}, V., {Ridgway}, S., {Mariotti}, J.M.: {Deriving object
  visibilities from interferograms obtained with a fiber stellar
  interferometer}.
\newblock A\&AS \textbf{121} (1997).
\newblock \doi{10.1051/aas:1997290}

\bibitem{CoudeDuForesto:1992}
{Coud{\'e} du Foresto}, V., {Ridgway}, S.T.: {Fluor - a Stellar Interferometer
  Using Single-Mode Fibers}.
\newblock In: J.M. {Beckers}, F.~{Merkle} (eds.) European Southern Observatory
  Conference and Workshop Proceedings, \emph{European Southern Observatory
  Conference and Workshop Proceedings}, vol.~39, p. 731 (1992)

\bibitem{crepp:2016}
Crepp, J.R., Crass, J., King, D., Bechter, A., Bechter, E., Ketterer, R.,
  Reynolds, R., Hinz, P., Kopon, D., Cavalieri, D., et~al.: ilocater: a
  diffraction-limited doppler spectrometer for the large binocular telescope.
\newblock In: Ground-based and Airborne Instrumentation for Astronomy VI, vol.
  9908, p. 990819. International Society for Optics and Photonics (2016)

\bibitem{croom2012sydney}
Croom, S.M., Lawrence, J.S., Bland-Hawthorn, J., Bryant, J.J., Fogarty, L.,
  Richards, S., Goodwin, M., Farrell, T., Miziarski, S., Heald, R., et~al.: The
  sydney-aao multi-object integral field spectrograph.
\newblock Monthly Notices of the Royal Astronomical Society \textbf{421}(1),
  872--893 (2012)

\bibitem{Crouzier2016}
{Crouzier}, A., {Malbet}, F., {Henault}, F., {L{\'e}ger}, A., {Cara}, C.,
  {LeDuigou}, J.M., {Preis}, O., {Kern}, P., {Delboulbe}, A., {Martin}, G.,
  {Feautrier}, P., {Stadler}, E., {Lafrasse}, S., {Rochat}, S., {Ketchazo}, C.,
  {Donati}, M., {Doumayrou}, E., {Lagage}, P.O., {Shao}, M., {Goullioud}, R.,
  {Nemati}, B., {Zhai}, C., {Behar}, E., {Potin}, S., {Saint-Pe}, M., {Dupont},
  J.: {A detector interferometric calibration experiment for high precision
  astrometry}.
\newblock \aap \textbf{595}, A108 (2016).
\newblock \doi{10.1051/0004-6361/201526321}

\bibitem{Cvetojevic:2017}
Cvetojevic, N., Jovanovic, N., Gross, S., Norris, B., Spaleniak, I., Schwab,
  C., Withford, M.J., Ireland, M., Tuthill, P., Guyon, O., Martinache, F.,
  Lawrence, J.S.: Modal noise in an integrated photonic lantern fed
  diffraction-limited spectrograph.
\newblock Opt. Express \textbf{25}(21), 25,546--25,565 (2017).
\newblock \doi{10.1364/OE.25.025546}.
\newblock
  \urlprefix\url{http://www.opticsexpress.org/abstract.cfm?URI=oe-25-21-25546}

\bibitem{Cvetojevic:2012}
Cvetojevic, N., Jovanovic, N., Lawrence, J., Withford, M., Bland-Hawthorn, J.:
  Developing arrayed waveguide grating spectrographs for multi-object
  astronomical spectroscopy.
\newblock Opt. Express \textbf{20}(3), 2062--2072 (2012).
\newblock \doi{10.1364/OE.20.002062}.
\newblock
  \urlprefix\url{http://www.opticsexpress.org/abstract.cfm?URI=oe-20-3-2062}

\bibitem{cvetojevic:2009}
Cvetojevic, N., Lawrence, J., Ellis, S., Bland-Hawthorn, J., Haynes, R.,
  Horton, A.: Characterization and on-sky demonstration of an integrated
  photonic spectrograph for astronomy.
\newblock Optics Express \textbf{17}(21), 18,643--18,650 (2009)

\bibitem{Defrere:2016}
{Defr{\`e}re}, D., {Hinz}, P.M., {Mennesson}, B., {Hoffmann}, W.F.,
  {Millan-Gabet}, R., {Skemer}, A.J., {Bailey}, V., {Danchi}, W.C., {Downey},
  E.C., {Durney}, O., {Grenz}, P., {Hill}, J.M., {McMahon}, T.J., {Montoya},
  M., {Spalding}, E., {Vaz}, A., {Absil}, O., {Arbo}, P., {Bailey}, H.,
  {Brusa}, G., {Bryden}, G., {Esposito}, S., {Gaspar}, A., {Haniff}, C.A.,
  {Kennedy}, G.M., {Leisenring}, J.M., {Marion}, L., {Nowak}, M., {Pinna}, E.,
  {Powell}, K., {Puglisi}, A., {Rieke}, G., {Roberge}, A., {Serabyn}, E.,
  {Sosa}, R., {Stapeldfeldt}, K., {Su}, K., {Weinberger}, A.J., {Wyatt}, M.C.:
  {Nulling Data Reduction and On-sky Performance of the Large Binocular
  Telescope Interferometer}.
\newblock Astroph. J. \textbf{824}, 66 (2016).
\newblock \doi{10.3847/0004-637X/824/2/66}

\bibitem{Defrere:2015}
{Defr{\`e}re}, D., {Hinz}, P.M., {Skemer}, A.J., {Kennedy}, G.M., {Bailey},
  V.P., {Hoffmann}, W.F., {Mennesson}, B., {Millan-Gabet}, R., {Danchi}, W.C.,
  {Absil}, O., {Arbo}, P., {Beichman}, C., {Brusa}, G., {Bryden}, G., {Downey},
  E.C., {Durney}, O., {Esposito}, S., {Gaspar}, A., {Grenz}, P., {Haniff}, C.,
  {Hill}, J.M., {Lebreton}, J., {Leisenring}, J.M., {Males}, J.R., {Marion},
  L., {McMahon}, T.J., {Montoya}, M., {Morzinski}, K.M., {Pinna}, E.,
  {Puglisi}, A., {Rieke}, G., {Roberge}, A., {Serabyn}, E., {Sosa}, R.,
  {Stapeldfeldt}, K., {Su}, K., {Vaitheeswaran}, V., {Vaz}, A., {Weinberger},
  A.J., {Wyatt}, M.C.: {First-light LBT Nulling Interferometric Observations:
  Warm Exozodiacal Dust Resolved within a Few AU of {$\eta$} Crv}.
\newblock Astroph. J. \textbf{799}, 42 (2015).
\newblock \doi{10.1088/0004-637X/799/1/42}

\bibitem{Dekany:2013}
Dekany, R., Roberts, J., Burruss, R., Bouchez, A., Truong, T., Baranec, C.,
  Guiwits, S., Hale, D., Angione, J., Trinh, T., Zolkower, J., Shelton, J.C.,
  Palmer, D., Henning, J., Croner, E., Troy, M., McKenna, D., Tesch, J.,
  Hildebrandt, S., Milburn, J.: Palm-3000: Exoplanet adaptive optics for the 5m
  hale telescope.
\newblock The Astrophysical Journal \textbf{776}(2), 130 (2013).
\newblock \urlprefix\url{http://stacks.iop.org/0004-637X/776/i=2/a=130}

\bibitem{Delacroix:2013}
Delacroix, C., Absil, O., Forsberg, P., Mawet, D., Christiaens, V., Karlsson,
  M., Boccaletti, A., Baudoz, P., Kuittinen, M., Vartiainen, I., Surdej, J.,
  Habraken, S.: Laboratory demonstration of a mid-infrared agpm vector vortex
  coronagraph.
\newblock A\&A \textbf{553}, A98 (2013)

\bibitem{Delacroix:2012}
{Delacroix}, C., {Forsberg}, P., {Karlsson}, M., {Mawet}, D., {Absil}, O.,
  {Hanot}, C., {Surdej}, J., {Habraken}, S.: {Design, manufacturing, and
  performance analysis of mid-infrared achromatic half-wave plates with diamond
  subwavelength gratings}.
\newblock Appl. Opt. \textbf{51}, 5897 (2012).
\newblock \doi{10.1364/AO.51.005897}

\bibitem{Delplancke2006}
{Delplancke}, F., {Derie}, F., {L{\'e}v{\^e}que}, S., {M{\'e}nardi}, S.,
  {Abuter}, R., {Andolfato}, L., {Ballester}, P., {de Jong}, J., {Di Lieto},
  N., {Duhoux}, P., {Frahm}, R., {Gitton}, P., {Glindemann}, A., {Palsa}, R.,
  {Puech}, F., {Sahlmann}, J., {Schuhler}, N., {Duc}, T.P., {Valat}, B.,
  {Wallander}, A.: {PRIMA for the VLTI: a status report}.
\newblock In: Society of Photo-Optical Instrumentation Engineers (SPIE)
  Conference Series, \emph{\procspie}, vol. 6268, p. 62680U (2006).
\newblock \doi{10.1117/12.660395}

\bibitem{Diab:2018}
{Diab}, M., {Minardi}, S.: {On the modal throughput of photonic lanterns in the
  presence of partial adaptive optic correction}.
\newblock In: Advances in Optical and Mechanical Technologies for Telescopes
  and Instrumentation III, \emph{Society of Photo-Optical Instrumentation
  Engineers (SPIE) Conference Series}, vol. 10706, p. 107064T (2018).
\newblock \doi{10.1117/12.2309657}

\bibitem{Diener:2017}
Diener, R., Tepper, J., Labadie, L., Pertsch, T., Nolte, S., Minardi, S.:
  Towards 3d-photonic, multi-telescope beam combiners for mid-infrared
  astrointerferometry.
\newblock Opt. Express \textbf{25}(16), 19,262--19,274 (2017).
\newblock \doi{10.1364/OE.25.019262}.
\newblock
  \urlprefix\url{http://www.opticsexpress.org/abstract.cfm?URI=oe-25-16-19262}

\bibitem{Dietrich:2017}
Dietrich, P.I., Harris, R.J., Blaicher, M., Corrigan, M.K., Morris, T.J.,
  Freude, W., Quirrenbach, A., Koos, C.: Printed freeform lens arrays on
  multi-core fibers for highly efficient coupling in astrophotonic systems.
\newblock Opt. Express \textbf{25}(15), 18,288--18,295 (2017).
\newblock \doi{10.1364/OE.25.018288}.
\newblock
  \urlprefix\url{http://www.opticsexpress.org/abstract.cfm?URI=oe-25-15-18288}

\bibitem{DosSantos:2015}
Dos~Santos, D., Rativa, D., Vohnsen, B.: Wavefront sensing using a
  liquid-filled photonic crystal fiber.
\newblock Optics Express \textbf{23}, 13,005--13,014 (2015).
\newblock \doi{10.1364/OE.23.013005}

\bibitem{Douglass:2018}
Douglass, G., Dreisow, F., Gross, S., Withford, M.J.: Femtosecond laser written
  arrayed waveguide gratings with integrated photonic lanterns.
\newblock Opt. Express \textbf{26}(2), 1497--1505 (2018).
\newblock \doi{10.1364/OE.26.001497}.
\newblock
  \urlprefix\url{http://www.opticsexpress.org/abstract.cfm?URI=oe-26-2-1497}

\bibitem{Downes:1996}
Downes, R.A., Wallace, D.: Spectroscopy of suspected variable stars.
\newblock Publications of the Astronomical Society of the Pacific
  \textbf{108}(720), 134 (1996).
\newblock \urlprefix\url{http://stacks.iop.org/1538-3873/108/i=720/a=134}

\bibitem{drew2010report}
Drew, J., Bergeron, J., Bouvier, J., et~al.: Report by the european telescope
  strategic review committee on europeÕs 2-4m telescopes over the decade to
  2020 (2010)

\bibitem{Dubbeldam:2000}
Dubbeldam, C.M., Allington-Smith, J.R., Pokrovski, S., Robertson, D.J.:
  Integral field unit for the gemini near-infrared spectrograph.
\newblock In: Astronomical Telescopes and Instrumentation, pp. 1181--1192.
  International Society for Optics and Photonics (2000)

\bibitem{Duchene2010}
{Duch{\^e}ne}, G., {McCabe}, C., {Pinte}, C., {Stapelfeldt}, K.R.,
  {M{\'e}nard}, F., {Duvert}, G., {Ghez}, A.M., {Maness}, H.L., {Bouy}, H.,
  {Barrado y Navascu{\'e}s}, D., {Morales-Calder{\'o}n}, M., {Wolf}, S.,
  {Padgett}, D.L., {Brooke}, T.Y., {Noriega-Crespo}, A.: {Panchromatic
  Observations and Modeling of the HV Tau C Edge-on Disk}.
\newblock \apj \textbf{712}, 112--129 (2010).
\newblock \doi{10.1088/0004-637X/712/1/112}

\bibitem{Duquennoy1991}
{Duquennoy}, A., {Mayor}, M.: {Multiplicity among solar-type stars in the solar
  neighbourhood. II - Distribution of the orbital elements in an unbiased
  sample}.
\newblock \aap \textbf{248}, 485--524 (1991)

\bibitem{ellis:1988}
Ellis, R.S., Parry, I.R.: Multiple object spectroscopy.
\newblock In: Instrumentation for Ground-Based Optical Astronomy, pp. 192--208.
  Springer (1988)

\bibitem{ellis:2011}
Ellis, S., Crouzier, A., Bland-Hawthorn, J., Lawrence, J.: Potential
  applications of ring resonators for astronomical instrumentation.
\newblock In: Conference on Lasers and Electro-Optics/Pacific Rim, p. C677.
  Optical Society of America (2011)

\bibitem{Ellis:2016}
{Ellis}, S.C., {Bauer}, S., {Bland-Hawthorn}, J., {Case}, S., {Content}, R.,
  {Fechner}, T., {Giannone}, D., {Haynes}, R., {Hernandez}, E., {Horton}, A.J.,
  {Klauser}, U., {Lawrence}, J.S., {Leon-Saval}, S.G., {Lindley}, E.,
  {L{\"o}hmannsr{\"o}ben}, H.G., {Min}, S.S., {Pai}, N., {Roth}, M.,
  {Shortridge}, K., {Staszak}, N.F., {Tims}, J., {Xavier}, P., {Zhelem}, R.:
  {PRAXIS: a near infrared spectrograph optimised for OH suppression}.
\newblock In: Ground-based and Airborne Instrumentation for Astronomy VI,
  \emph{Proc. SPIE}, vol. 9908, p. 99084A (2016).
\newblock \doi{10.1117/12.2232115}

\bibitem{Ellis:2017}
Ellis, S.C., Kuhlmann, S., Kuehn, K., Spinka, H., Underwood, D., Gupta, R.R.,
  Ocola, L.E., Liu, P., Wei, G., Stern, N.P., Bland-Hawthorn, J., Tuthill, P.:
  Photonic ring resonator filters for astronomical oh suppression.
\newblock Opt. Express \textbf{25}(14), 15,868--15,889 (2017).
\newblock \doi{10.1364/OE.25.015868}.
\newblock
  \urlprefix\url{http://www.opticsexpress.org/abstract.cfm?URI=oe-25-14-15868}

\bibitem{Errmann:2015}
{Errmann}, R., {Minardi}, S., {Labadie}, L., {Muthusubramanian}, B., {Dreisow},
  F., {Nolte}, S., {Pertsch}, T.: {Interferometric nulling of four channels
  with integrated optics}.
\newblock \ao \textbf{54}, 7449 (2015).
\newblock \doi{10.1364/AO.54.007449}

\bibitem{Errmann:2013}
{Errmann}, R., {Minardi}, S., {Pertsch}, T.: {A broad-band scalar vortex
  coronagraph}.
\newblock MNRAS \textbf{435}, 565--569 (2013).
\newblock \doi{10.1093/mnras/stt1317}

\bibitem{Ertel2014}
{Ertel}, S., {Absil}, O., {Defr{\`e}re}, D., {Le Bouquin}, J.B., {Augereau},
  J.C., {Marion}, L., {Blind}, N., {Bonsor}, A., {Bryden}, G., {Lebreton}, J.,
  {Milli}, J.: {A near-infrared interferometric survey of debris-disk stars.
  IV. An unbiased sample of 92 southern stars observed in H band with
  VLTI/PIONIER}.
\newblock \aap \textbf{570}, A128 (2014).
\newblock \doi{10.1051/0004-6361/201424438}

\bibitem{Evans:2016}
{Evans}, C.J., {Puech}, M., {Rodrigues}, M., {Barbuy}, B., {Cuby}, J.G.,
  {Dalton}, G., {Fitzsimons}, E., {Hammer}, F., {Jagourel}, P., {Kaper}, L.,
  {Morris}, S.L., {Morris}, T.J.: {Science requirements and trade-offs for the
  MOSAIC instrument for the European ELT}.
\newblock In: Ground-based and Airborne Instrumentation for Astronomy VI,
  \emph{Proc. SPIE}, vol. 9908, p. 99089J (2016).
\newblock \doi{10.1117/12.2231675}

\bibitem{Feger2014}
Feger, T., Bacigalupo, C., Bedding, T.R., Bento, J., Coutts, D.W., Ireland,
  M.J., Parker, Q.A., Rizzuto, A., Spaleniak, I.: Rhea: the ultra-compact
  replicable high-resolution exoplanet and asteroseismology spectrograph.
\newblock In: Ground-based and Airborne Instrumentation for Astronomy V, vol.
  9147, p. 91477I. International Society for Optics and Photonics (2014)

\bibitem{Feng:2008}
Feng, Y., Taylor, L., Calia, D.B.: Multiwatts narrow line width fiber raman
  amplifiers.
\newblock Opt. Exp. \textbf{16}, 10,927--10,932 (2008)

\bibitem{Feng:2009}
Feng, Y., Taylor, L., Calia, D.B.: 25 w raman-fiber-amplifier-based 589 nm
  laser for laser guide star.
\newblock Opt. Exp. \textbf{17}, 19,021--19,026 (2009)

\bibitem{Foo:2005}
Foo, G., Palacios, D., Swartzlander, G.: Optical vortex coronagraph.
\newblock Opt. Lett. \textbf{30}, 3308--3310 (2005)

\bibitem{Foy:1985}
{Foy}, R., {Labeyrie}, A.: {Feasibility of adaptive telescope with laser
  probe}.
\newblock A\&A \textbf{152}, L29--L31 (1985)

\bibitem{Freedman:2001}
{Freedman}, W.L., {Madore}, B.F., {Gibson}, B.K., {Ferrarese}, L., {Kelson},
  D.D., {Sakai}, S., {Mould}, J.R., {Kennicutt} Jr., R.C., {Ford}, H.C.,
  {Graham}, J.A., {Huchra}, J.P., {Hughes}, S.M.G., {Illingworth}, G.D.,
  {Macri}, L.M., {Stetson}, P.B.: {Final Results from the Hubble Space
  Telescope Key Project to Measure the Hubble Constant}.
\newblock \apj \textbf{553}, 47--72 (2001).
\newblock \doi{10.1086/320638}

\bibitem{Fridlund2004}
{Fridlund}, C.V.M.: {The Darwin mission}.
\newblock Advances in Space Research \textbf{34}, 613--617 (2004).
\newblock \doi{10.1016/j.asr.2003.05.031}

\bibitem{Fridlund:2004}
{Fridlund}, C.V.M.: {The Darwin mission}.
\newblock Advances in Space Research \textbf{34}, 613--617 (2004).
\newblock \doi{10.1016/j.asr.2003.05.031}

\bibitem{Fried:1966}
{Fried}, D.L.: {Limiting Resolution Looking Down Through the Atmosphere}.
\newblock Journal of the Optical Society of America (1917-1983) \textbf{56},
  1380 (1966)

\bibitem{Froehly:1981}
{Froehly}, C.: {Coherence and interferometry through optical fibers}.
\newblock In: M.H. {Ulrich}, K.~{Kjaer} (eds.) Scientific Importance of High
  Angular Resolution at Infrared and Optical Wavelengths, pp. 285--293 (1981)

\bibitem{Garufi2017}
{Garufi}, A., {Benisty}, M., {Stolker}, T., {Avenhaus}, H., {de Boer}, J..,
  {Pohl}, A., {Quanz}, S.P., {Dominik}, C., {Ginski}, C., {Thalmann}, C., {van
  Boekel}, R., {Boccaletti}, A., {Henning}, T., {Janson}, M., {Salter}, G.,
  {Schmid}, H.M., {Sissa}, E., {Langlois}, M., {Beuzit}, J.L., {Chauvin}, G.,
  {Mouillet}, D., {Augereau}, J.C., {Bazzon}, A., {Biller}, B., {Bonnefoy}, M.,
  {Buenzli}, E., {Cheetham}, A., {Daemgen}, S., {Desidera}, S., {Engler}, N.,
  {Feldt}, M., {Girard}, J., {Gratton}, R., {Hagelberg}, J., {Keller}, C.,
  {Keppler}, M., {Kenworthy}, M., {Kral}, Q., {Lopez}, B., {Maire}, A.L.,
  {Menard}, F., {Mesa}, D., {Messina}, S., {Meyer}, M.R., {Milli}, J., {Min},
  M., {Muller}, A., {Olofsson}, J., {Pawellek}, N., {Pinte}, C., {Szulagyi},
  J., {Vigan}, A., {Wahhaj}, Z., {Waters}, R., {Zurlo}, A.: {Three Years of
  SPHERE: The Latest View of the Morphology and Evolution of Protoplanetary
  Discs}.
\newblock The Messenger \textbf{169}, 32--37 (2017).
\newblock \doi{10.18727/0722-6691/5036}

\bibitem{Gatkine:2017}
Gatkine, P., Veilleux, S., Hu, Y., Bland-Hawthorn, J., Dagenais, M.: Arrayed
  waveguide grating spectrometers for astronomical applications: new results.
\newblock Optics Express \textbf{25}(15), 17,918--17,935 (2017)

\bibitem{Gatkine:2016}
Gatkine, P., Veilleux, S., Hu, Y., Zhu, T., Meng, Y., Bland-Hawthorn, J.,
  Dagenais, M.: Development of high-resolution arrayed waveguide grating
  spectrometers for astronomical applications: first results.
\newblock In: SPIE Astronomical Telescopes+ Instrumentation, pp.
  991,271--991,271. International Society for Optics and Photonics (2016)

\bibitem{Gattass:2006}
Gattass, R.R., Cerami, L.R., Mazur, E.: Micromachining of bulk glass with
  bursts of femtosecond laser pulses at variable repetition rates.
\newblock Optics Express \textbf{14}(12), 5279--5284 (2006)

\bibitem{Ghez1993}
{Ghez}, A.M., {Neugebauer}, G., {Matthews}, K.: {The multiplicity of T Tauri
  stars in the star forming regions Taurus-Auriga and Ophiuchus-Scorpius: A 2.2
  micron speckle imaging survey}.
\newblock \aj \textbf{106}, 2005--2023 (1993).
\newblock \doi{10.1086/116782}

\bibitem{gizis:1999}
Gizis, J.E., Reid, I.N., Monet, D.G.: A 2mass survey for brown dwarfs toward
  the hyades.
\newblock The Astronomical Journal \textbf{118}(2), 997 (1999)

\bibitem{Glenday:2015}
Glenday, A.G., Li, C.H., Langellier, N., Chang, G., Chen, L.J., Furesz, G.,
  Zibrov, A.A., K\"{a}rtner, F., Phillips, D.F., Sasselov, D., Szentgyorgyi,
  A., Walsworth, R.L.: Operation of a broadband visible-wavelength astro-comb
  with a high-resolution astrophysical spectrograph.
\newblock Optica \textbf{2}(3), 250--254 (2015).
\newblock \doi{10.1364/OPTICA.2.000250}.
\newblock
  \urlprefix\url{http://www.osapublishing.org/optica/abstract.cfm?URI=optica-2-3-250}

\bibitem{Goodman}
Goodman, J.W.: Introduction to Fourier optics.
\newblock Roberts (2003)

\bibitem{goodman:1981}
Goodman, J.W., Rawson, E.G.: Statistics of modal noise in fibers: a case of
  constrained speckle.
\newblock Optics letters \textbf{6}(7), 324--326 (1981)

\bibitem{Goto2006}
{Goto}, M., {Usuda}, T., {Dullemond}, C.P., {Henning}, T., {Linz}, H.,
  {Stecklum}, B., {Suto}, H.: {Inner Rim of a Molecular Disk Spatially Resolved
  in Infrared CO Emission Lines}.
\newblock \apj \textbf{652}, 758--762 (2006).
\newblock \doi{10.1086/506582}

\bibitem{Abuter:2017}
{Gravity Collaboration}, {Abuter}, R., {Accardo}, M., {Amorim}, A., {Anugu},
  N., {{\'A}vila}, G., {Azouaoui}, N., {Benisty}, M., {Berger}, J.P., {Blind},
  N., {Bonnet}, H., {Bourget}, P., {Brandner}, W., {Brast}, R., {Buron}, A.,
  {Burtscher}, L., {Cassaing}, F., {Chapron}, F., {Choquet}, {\'E}.,
  {Cl{\'e}net}, Y., {Collin}, C., {Coud{\'e} Du Foresto}, V., {de Wit}, W., {de
  Zeeuw}, P.T., {Deen}, C., {Delplancke-Str{\"o}bele}, F., {Dembet}, R.,
  {Derie}, F., {Dexter}, J., {Duvert}, G., {Ebert}, M., {Eckart}, A.,
  {Eisenhauer}, F., {Esselborn}, M., {F{\'e}dou}, P., {Finger}, G., {Garcia},
  P., {Garcia Dabo}, C.E., {Garcia Lopez}, R., {Gendron}, E., {Genzel}, R.,
  {Gillessen}, S., {Gonte}, F., {Gordo}, P., {Grould}, M., {Gr{\"o}zinger}, U.,
  {Guieu}, S., {Haguenauer}, P., {Hans}, O., {Haubois}, X., {Haug}, M.,
  {Haussmann}, F., {Henning}, T., {Hippler}, S., {Horrobin}, M., {Huber}, A.,
  {Hubert}, Z., {Hubin}, N., {Hummel}, C.A., {Jakob}, G., {Janssen}, A.,
  {Jochum}, L., {Jocou}, L., {Kaufer}, A., {Kellner}, S., {Kendrew}, S.,
  {Kern}, L., {Kervella}, P., {Kiekebusch}, M., {Klein}, R., {Kok}, Y., {Kolb},
  J., {Kulas}, M., {Lacour}, S., {Lapeyr{\`e}re}, V., {Lazareff}, B., {Le
  Bouquin}, J.B., {L{\`e}na}, P., {Lenzen}, R., {L{\'e}v{\^e}que}, S., {Lippa},
  M., {Magnard}, Y., {Mehrgan}, L., {Mellein}, M., {M{\'e}rand}, A.,
  {Moreno-Ventas}, J., {Moulin}, T., {M{\"u}ller}, E., {M{\"u}ller}, F.,
  {Neumann}, U., {Oberti}, S., {Ott}, T., {Pallanca}, L., {Panduro}, J.,
  {Pasquini}, L., {Paumard}, T., {Percheron}, I., {Perraut}, K., {Perrin}, G.,
  {Pfl{\"u}ger}, A., {Pfuhl}, O., {Phan Duc}, T., {Plewa}, P.M., {Popovic}, D.,
  {Rabien}, S., {Ram{\'{\i}}rez}, A., {Ramos}, J., {Rau}, C., {Riquelme}, M.,
  {Rohloff}, R.R., {Rousset}, G., {Sanchez-Bermudez}, J., {Scheithauer}, S.,
  {Sch{\"o}ller}, M., {Schuhler}, N., {Spyromilio}, J., {Straubmeier}, C.,
  {Sturm}, E., {Suarez}, M., {Tristram}, K.R.W., {Ventura}, N., {Vincent}, F.,
  {Waisberg}, I., {Wank}, I., {Weber}, J., {Wieprecht}, E., {Wiest}, M.,
  {Wiezorrek}, E., {Wittkowski}, M., {Woillez}, J., {Wolff}, B., {Yazici}, S.,
  {Ziegler}, D., {Zins}, G.: {First light for GRAVITY: Phase referencing
  optical interferometry for the Very Large Telescope Interferometer}.
\newblock A\&A \textbf{602}, A94 (2017).
\newblock \doi{10.1051/0004-6361/201730838}

\bibitem{gris-sanchez2018}
{Gris-S{\'a}nchez}, I., {Haynes}, D.M., {Ehrlich}, K., {Haynes}, R., {Birks},
  T.A.: {Multicore fibre photonic lanterns for precision radial velocity
  Science}.
\newblock \mnras \textbf{475}(3), 3065--3075 (2018).
\newblock \doi{10.1093/mnras/stx3278}

\bibitem{Gross:2015}
{Gross}, S., {Jovanovic}, N., {Sharp}, A., {Ireland}, M., {Lawrence}, J.,
  {Withford}, M.J.: {Low loss mid-infrared ZBLAN waveguides for future
  astronomical applications}.
\newblock Optics Express \textbf{23}, 7946 (2015).
\newblock \doi{10.1364/OE.23.007946}

\bibitem{Gurevich:2014}
{Gurevich}, Y.V., {St{\"u}rmer}, J., {Schwab}, C., {F{\"u}hrer}, T.,
  {Lamoreaux}, S.K., {Quirrenbach}, A., {Walther}, T.: {A laser locked
  Fabry-Perot etalon with 3 cm/s stability for spectrograph calibration}.
\newblock In: Ground-based and Airborne Instrumentation for Astronomy V,
  \emph{Proc. SPIE}, vol. 9147, p. 91477M (2014).
\newblock \doi{10.1117/12.2057008}

\bibitem{Gustafsson2008}
{Gustafsson}, M., {Labadie}, L., {Herbst}, T.M., {Kasper}, M.: {Spatially
  resolved H$_{2}$ emission from the disk around T Tau N}.
\newblock \aap \textbf{488}, 235--244 (2008).
\newblock \doi{10.1051/0004-6361:200809770}

\bibitem{Guyon:2003}
{Guyon}, O.: {Phase-induced amplitude apodization of telescope pupils for
  extrasolar terrestrial planet imaging}.
\newblock Astron. Astroph. \textbf{404}, 379--387 (2003).
\newblock \doi{10.1051/0004-6361:20030457}

\bibitem{Guyon:2009}
{Guyon}, O., {Matsuo}, T., {Angel}, R.: {Coronagraphic Low-Order Wave-Front
  Sensor: Principle and Application to a Phase-Induced Amplitude Coronagraph}.
\newblock Astroph. J. \textbf{693}, 75--84 (2009).
\newblock \doi{10.1088/0004-637X/693/1/75}

\bibitem{Guyon:2006}
{Guyon}, O., {Pluzhnik}, E.A., {Kuchner}, M.J., {Collins}, B., {Ridgway}, S.T.:
  {Theoretical Limits on Extrasolar Terrestrial Planet Detection with
  Coronagraphs}.
\newblock Astrophys. J. Suppl. Ser. \textbf{167}, 81--99 (2006).
\newblock \doi{10.1086/507630}

\bibitem{Guyon:1999}
Guyon, O., Roddier, C., Graves, J., Roddier, F., Cuevas, S., Espejo, C.,
  Gonzales, S.: The nulling stellar coronagaph: laboratory test and performance
  evaluation.
\newblock PASP \textbf{111}, 1321--1330 (1999)

\bibitem{Haguenauer:2006}
{Haguenauer}, P., {Serabyn}, E.: {Deep nulling of laser light with a
  single-mode-fiber beam combiner}.
\newblock \ao \textbf{45}, 2749--2754 (2006).
\newblock \doi{10.1364/AO.45.002749}

\bibitem{Halverson:2012}
Halverson, S., Mahadevan, S., Ramsey, L.W., Redman, S., Nave, G., Wilson, J.C.,
  Hearty, F., Holtzman, J.: {Development of a new, precise near-infrared
  Doppler wavelength reference: a fiber Fabry-Perot interferometer}.
\newblock In: I.S. McLean, S.K. Ramsay, H.~Takami (eds.) Ground-based and
  Airborne Instrumentation for Astronomy IV, vol. 8446, pp. 1198 -- 1213.
  International Society for Optics and Photonics, SPIE (2013).
\newblock \doi{10.1117/12.925716}.
\newblock \urlprefix\url{https://doi.org/10.1117/12.925716}

\bibitem{Halverson:2015}
Halverson, S., Roy, A., Mahadevan, S., Schwab, C.: {\textquotedblleft}{MODAL}
  {NOISE}{\textquotedblright} {IN} {SINGLE}-{MODE} {FIBERS}: A {CAUTIONARY}
  {NOTE} {FOR} {HIGH} {PRECISION} {RADIAL} {VELOCITY} {INSTRUMENTS}.
\newblock The Astrophysical Journal \textbf{814}(2), L22 (2015).
\newblock \doi{10.1088/2041-8205/814/2/l22}.
\newblock
  \urlprefix\url{https://doi.org/10.1088\%2F2041-8205\%2F814\%2F2\%2Fl22}

\bibitem{Haniff:1987}
{Haniff}, C.A., {Mackay}, C.D., {Titterington}, D.J., {Sivia}, D., {Baldwin},
  J.E.: {The first images from optical aperture synthesis}.
\newblock Nature \textbf{328}, 694--696 (1987).
\newblock \doi{10.1038/328694a0}

\bibitem{Hankla:2006}
Hankla, A.K., Bartholomew, J., Groff, K., Lee, I., McKinnie, I.T., Moule, G.,
  Rogers, N., Tiemann, B., Tracy, A.J., VanHoudt, P., Adkins, S.M.,
  d'Orgeville, C.: 20 w and 50 w solid-state sodium beacon guidestar laser
  systems for the keck i and gemini south telescopes.
\newblock In: Proc. SPIE, vol. 6272, p.~1G (2006)

\bibitem{Hanot:2011}
{Hanot}, C., {Mennesson}, B., {Martin}, S., {Liewer}, K., {Loya}, F., {Mawet},
  D., {Riaud}, P., {Absil}, O., {Serabyn}, E.: {Improving Interferometric Null
  Depth Measurements using Statistical Distributions: Theory and First Results
  with the Palomar Fiber Nuller}.
\newblock Astroph. J. \textbf{729}, 110 (2011).
\newblock \doi{10.1088/0004-637X/729/2/110}

\bibitem{harris:2012}
Harris, R.J., Allington-Smith, J.: Applications of integrated photonic
  spectrographs in astronomy.
\newblock Monthly Notices of the Royal Astronomical Society \textbf{428}(4),
  3139--3150 (2012)

\bibitem{harris:2015}
Harris, R.J., MacLachlan, D.G., Choudhury, D., Morris, T.J., Gendron, E.,
  Basden, A.G., Brown, G., Allington-Smith, J.R., Thomson, R.R.: Photonic
  spatial reformatting of stellar light for diffraction-limited spectroscopy.
\newblock Monthly Notices of the Royal Astronomical Society \textbf{450}(1),
  428--434 (2015)

\bibitem{hartman:2011}
Hartman, J., Bakos, G., Noyes, R., Sip{\H{o}}cz, B., Kov{\'a}cs, G., Mazeh, T.,
  Shporer, A., P{\'a}l, A.: A photometric variability survey of field k and m
  dwarf stars with hatnet.
\newblock The Astronomical Journal \textbf{141}(5), 166 (2011)

\bibitem{Haynes2018}
{Haynes}, D.M., {Gris-Sanchez}, I., {Birks}, T.A., {Haynes}, R.: {Optical fiber
  modal noise suppression in the NIR region using multicore fiber and photonic
  lanterns}.
\newblock In: \procspie, \emph{Society of Photo-Optical Instrumentation
  Engineers (SPIE) Conference Series}, vol. 10706, p. 1070665 (2018).
\newblock \doi{10.1117/12.2314224}

\bibitem{Haynes2014}
{Haynes}, D.M., {Gris-Sanchez}, I., {Ehrlich}, K., {Birks}, T.A., {Giannone},
  D., {Haynes}, R.: {New multicore low mode noise scrambling fiber for
  applications in high-resolution spectroscopy}.
\newblock In: \procspie, \emph{Society of Photo-Optical Instrumentation
  Engineers (SPIE) Conference Series}, vol. 9151, p. 915155 (2014).
\newblock \doi{10.1117/12.2057170}

\bibitem{heacox1986application}
Heacox, W.D.: On the application of optical-fiber image scramblers to
  astronomical spectroscopy.
\newblock The Astronomical Journal \textbf{92}, 219--229 (1986)

\bibitem{Heacox1988}
Heacox, W.D.: Wavelength-precise slit spectroscopy with optical fiber image
  scramblers.
\newblock In: Fiber optics in astronomy, vol.~3, pp. 204--235 (1988)

\bibitem{Hecht:2012}
Hecht, E.: Optics.
\newblock Pearson (2012)

\bibitem{Henning2013}
{Henning}, T., {Semenov}, D.: {Chemistry in Protoplanetary Disks}.
\newblock Chemical Reviews \textbf{113}, 9016--9042 (2013).
\newblock \doi{10.1021/cr400128p}

\bibitem{Herr:2013}
Herr, T., Brasch, V., Jost, J.D., Wang, C.Y., Kondratiev, N.M., Gorodetsky,
  M.L., Kippenberg, T.J.: Temporal solitons in optical microresonators.
\newblock Nature Photonics \textbf{8}(2), 145--152 (2014).
\newblock \doi{10.1038/nphoton.2013.343}.
\newblock \urlprefix\url{https://doi.org/10.1038/nphoton.2013.343}

\bibitem{Herr:2012}
Herr, T., Hartinger, K., Riemensberger, J., Wang, C.Y., Gavartin, E.,
  Holzwarth, R., Gorodetsky, M.L., Kippenberg, T.J.: Universal formation
  dynamics and noise of kerr-frequency combs in microresonators.
\newblock Nature Photonics \textbf{6}(7), 480--487 (2012).
\newblock \doi{10.1038/nphoton.2012.127}.
\newblock \urlprefix\url{https://doi.org/10.1038/nphoton.2012.127}

\bibitem{Hill:2004}
Hill, G.J., MacQueen, P.J., Tejada, C., Cobos, F.J., Palunas, P., Gebhardt, K.,
  Drory, N.: Virus: a massively replicated ifu spectrograph for het.
\newblock In: Proc.SPIE, vol. 5492, pp. 5492 -- 5492 -- 11 (2004).
\newblock \doi{10.1117/12.552474}.
\newblock \urlprefix\url{http://dx.doi.org/10.1117/12.552474}

\bibitem{Hill:1980}
{Hill}, J.M., {Angel}, J.R.P., {Scott}, J.S., {Lindley}, D., {Hintzen}, P.:
  {Multiple object spectroscopy - The Medusa spectrograph}.
\newblock Astrophys. J. Lett. \textbf{242}, L69--L72 (1980).
\newblock \doi{10.1086/183405}

\bibitem{hoeijmakers2018mapping}
{Hoeijmakers}, H.J., {Schwarz}, H., {Snellen}, I.A.G., {de Kok}, R.J.,
  {Bonnefoy}, M., {Chauvin}, G., {Lagrange}, A.M., {Girard}, J.H.:
  {Medium-resolution integral-field spectroscopy for high-contrast exoplanet
  imaging. Molecule maps of the {\ensuremath{\beta}} Pictoris system with
  SINFONI}.
\newblock \aap \textbf{617}, A144 (2018).
\newblock \doi{10.1051/0004-6361/201832902}

\bibitem{Hoenig2014}
{H{\"o}nig}, S.F., {Watson}, D., {Kishimoto}, M., {Hjorth}, J.: {A
  dust-parallax distance of 19 megaparsecs to the supermassive black hole in
  NGC 4151}.
\newblock \nat \textbf{515}, 528--530 (2014).
\newblock \doi{10.1038/nature13914}

\bibitem{horton:2013}
Horton, A., Ellis, S., Lawrence, J., Bland-Hawthorn, J.: Praxis: a low
  background nir spectrograph for fibre bragg grating oh suppression.
\newblock arXiv preprint arXiv:1301.0680  (2013)

\bibitem{hottinger2018}
{Hottinger}, P., {Harris}, R.J., {Dietrich}, P.I., {Blaicher}, M., {Gl{\"u}ck},
  M., {Bechter}, A., {Crass}, J., {Pott}, J.U., {Koos}, C., {Sawodny}, O.,
  {Quirrenbach}, A.: {Micro-lens arrays as tip-tilt sensor for single mode
  fiber coupling}.
\newblock In: \procspie, \emph{Society of Photo-Optical Instrumentation
  Engineers (SPIE) Conference Series}, vol. 10706, p. 1070629 (2018).
\newblock \doi{10.1117/12.2312015}

\bibitem{Hsiao:2009}
{Hsiao}, H.K., {Winick}, K.A., {Monnier}, J.D., {Berger}, J.P.: {An infrared
  integrated optic astronomical beam combiner for stellar interferometry at 3-4
  {$\mu$}m}.
\newblock Opt. Exp. \textbf{17}, 18,489--18,500 (2009)

\bibitem{Hubbard1979}
Hubbard, E., Angel, J., Gresham, M.: Operation of a long fused silica fiber as
  a link between telescope and spectrograph.
\newblock The Astrophysical Journal \textbf{229}, 1074--1078 (1979)

\bibitem{Huby:2017}
{Huby}, E., {Bottom}, M., {Femenia}, B., {Ngo}, H., {Mawet}, D., {Serabyn}, E.,
  {Absil}, O.: {On-sky performance of the QACITS pointing control technique
  with the Keck/NIRC2 vortex coronagraph}.
\newblock Astron. Astroph. \textbf{600}, A46 (2017).
\newblock \doi{10.1051/0004-6361/201630232}

\bibitem{Huby:2013}
{Huby}, E., {Duch{\^e}ne}, G., {Marchis}, F., {Lacour}, S., {Perrin}, G.,
  {Kotani}, T., {Choquet}, {\'E}., {Gates}, E.L., {Lai}, O., {Allard}, F.:
  {FIRST, a fibered aperture masking instrument. II. Spectroscopy of the
  Capella binary system at the diffraction limit}.
\newblock A\&A \textbf{560}, A113 (2013).
\newblock \doi{10.1051/0004-6361/201321894}

\bibitem{Huby:2012}
{Huby}, E., {Perrin}, G., {Marchis}, F., {Lacour}, S., {Kotani}, T.,
  {Duch{\^e}ne}, G., {Choquet}, E., {Gates}, E.L., {Woillez}, J.M., {Lai}, O.,
  {F{\'e}dou}, P., {Collin}, C., {Chapron}, F., {Arslanyan}, V., {Burns}, K.J.:
  {FIRST, a fibered aperture masking instrument. I. First on-sky test results}.
\newblock A\&A \textbf{541}, A55 (2012).
\newblock \doi{10.1051/0004-6361/201118517}

\bibitem{Iuzzolino2014}
Iuzzolino, M., Tozzi, A., Sanna, N., Zangrilli, L., Oliva, E.: Preliminary
  results on the characterization and performances of zblan fiber for infrared
  spectrographs.
\newblock In: Ground-based and Airborne Instrumentation for Astronomy V, vol.
  9147, p. 914766. International Society for Optics and Photonics (2014)

\bibitem{Iwamuro:1994}
{Iwamuro}, F., {Maihara}, T., {Oya}, S., {Tsukamoto}, H., {Hall}, D.N.B.,
  {Cowie}, L.L., {Tokunaga}, A.T., {Pickles}, A.J.: {Development of an
  OH-airglow suppressor spectrograph}.
\newblock \pasj \textbf{46}, 515--521 (1994)

\bibitem{Janson2010}
{Janson}, M., {Bergfors}, C., {Goto}, M., {Brandner}, W., {Lafreni{\`e}re}, D.:
  {Spatially Resolved Spectroscopy of the Exoplanet HR 8799 c}.
\newblock \apjl \textbf{710}, L35--L38 (2010).
\newblock \doi{10.1088/2041-8205/710/1/L35}

\bibitem{Jenkins:2008}
{Jenkins}, C.: {Optical vortex coronagraphs on ground-based telescopes}.
\newblock \mnras \textbf{384}, 515--524 (2008).
\newblock \doi{10.1111/j.1365-2966.2007.12744.x}

\bibitem{Jeys:1989}
Jeys, T.H., Brailove, A.A., Mooradian, A.: Sum frequency generation of sodium
  resonance radiation.
\newblock Appl. Opt. \textbf{28}, 2588--2591 (1989)

\bibitem{Ge:1998}
Jian~Ge James Roger P.~Angel, J.C.S.: Optical spectroscopy with a
  near-single-mode fiber-feed and adaptive optics.
\newblock In: Proc.SPIE, vol. 3355, pp. 3355 -- 3355 -- 11 (1998).
\newblock \doi{10.1117/12.316767}.
\newblock \urlprefix\url{http://dx.doi.org/10.1117/12.316767}

\bibitem{Jovanovic:2017b}
Jovanovic, N., Cvetojevic, N., Norris, B., Betters, C., Schwab, C., Lozi, J.,
  Guyon, O., Gross, S., Martinache, F., Tuthill, P., Doughty, D., Minowa, Y.,
  Takato, N., Lawrence, J.: Demonstration of an efficient, photonic-based
  astronomical spectrograph on an 8-m telescope.
\newblock Opt. Express \textbf{25}(15), 17,753--17,766 (2017).
\newblock \doi{10.1364/OE.25.017753}.
\newblock
  \urlprefix\url{http://www.opticsexpress.org/abstract.cfm?URI=oe-25-15-17753}

\bibitem{Jovanovic:2014}
{Jovanovic}, N., {Guyon}, O., {Martinache}, F., {Clergeon}, C., {Singh}, G.,
  {Kudo}, T., {Newman}, K., {Kuhn}, J., {Serabyn}, E., {Norris}, B., {Tuthill},
  P., {Stewart}, P., {Huby}, E., {Perrin}, G., {Lacour}, S., {Vievard}, S.,
  {Murakami}, N., {Fumika}, O., {Minowa}, Y., {Hayano}, Y., {White}, J., {Lai},
  O., {Marchis}, F., {Duchene}, G., {Kotani}, T., {Woillez}, J.: {Development
  and recent results from the Subaru coronagraphic extreme adaptive optics
  system}.
\newblock In: Ground-based and Airborne Instrumentation for Astronomy V,
  \emph{Proc. SPIE}, vol. 9147, p. 91471Q (2014).
\newblock \doi{10.1117/12.2057249}

\bibitem{Jovanovic:2015}
{Jovanovic}, N., {Martinache}, F., {Guyon}, O., {Clergeon}, C., {Singh}, G.,
  {Kudo}, T., {Garrel}, V., {Newman}, K., {Doughty}, D., {Lozi}, J., {Males},
  J., {Minowa}, Y., {Hayano}, Y., {Takato}, N., {Morino}, J., {Kuhn}, J.,
  {Serabyn}, E., {Norris}, B., {Tuthill}, P., {Schworer}, G., {Stewart}, P.,
  {Close}, L., {Huby}, E., {Perrin}, G., {Lacour}, S., {Gauchet}, L.,
  {Vievard}, S., {Murakami}, N., {Oshiyama}, F., {Baba}, N., {Matsuo}, T.,
  {Nishikawa}, J., {Tamura}, M., {Lai}, O., {Marchis}, F., {Duchene}, G.,
  {Kotani}, T., {Woillez}, J.: {The Subaru Coronagraphic Extreme Adaptive
  Optics System: Enabling High-Contrast Imaging on Solar-System Scales}.
\newblock Publications of the Astronomical Society of the Pacific \textbf{127},
  890 (2015).
\newblock \doi{10.1086/682989}

\bibitem{jovanovic:2017}
Jovanovic, N., Schwab, C., Guyon, O., Lozi, J., Cvetojevic, N., Martinache, F.,
  Leon-Saval, S., Norris, B., Gross, S., Doughty, D., et~al.: Efficient
  injection from large telescopes into single-mode fibres: Enabling the era of
  ultra-precision astronomy.
\newblock Astronomy \& Astrophysics \textbf{604}, A122 (2017)

\bibitem{Jovanovic:2012}
{Jovanovic}, N., {Tuthill}, P.G., {Norris}, B., {Gross}, S., {Stewart}, P.,
  {Charles}, N., {Lacour}, S., {Ams}, M., {Lawrence}, J.S., {Lehmann}, A.,
  {Niel}, C., {Robertson}, J.G., {Marshall}, G.D., {Ireland}, M., {Fuerbach},
  A., {Withford}, M.J.: {Starlight demonstration of the Dragonfly instrument:
  an integrated photonic pupil-remapping interferometer for high-contrast
  imaging}.
\newblock Month. Not. R. Astron. Soc. \textbf{427}, 806--815 (2012).
\newblock \doi{10.1111/j.1365-2966.2012.21997.x}

\bibitem{jurgenson2016expres}
Jurgenson, C., Fischer, D., McCracken, T., Sawyer, D., Szymkowiak, A., Davis,
  A., Muller, G., Santoro, F.: Expres: a next generation rv spectrograph in the
  search for earth-like worlds.
\newblock In: Ground-based and Airborne Instrumentation for Astronomy VI, vol.
  9908, p. 99086T. International Society for Optics and Photonics (2016)

\bibitem{Kalas2007}
{Kalas}, P., {Duchene}, G., {Fitzgerald}, M.P., {Graham}, J.R.: {Discovery of
  an Extended Debris Disk around the F2 V Star HD 15745}.
\newblock \apjl \textbf{671}, L161--L164 (2007).
\newblock \doi{10.1086/525252}

\bibitem{Kalas2008}
{Kalas}, P., {Graham}, J.R., {Chiang}, E., {Fitzgerald}, M.P., {Clampin}, M.,
  {Kite}, E.S., {Stapelfeldt}, K., {Marois}, C., {Krist}, J.: {Optical Images
  of an Exosolar Planet 25 Light-Years from Earth}.
\newblock Science \textbf{322}, 1345 (2008).
\newblock \doi{10.1126/science.1166609}

\bibitem{Kalas2005}
{Kalas}, P., {Graham}, J.R., {Clampin}, M.: {A planetary system as the origin
  of structure in Fomalhaut's dust belt}.
\newblock \nat \textbf{435}, 1067--1070 (2005).
\newblock \doi{10.1038/nature03601}

\bibitem{Kammerer2018}
{Kammerer}, J., {Quanz}, S.P.: {Simulating the exoplanet yield of a space-based
  mid-infrared interferometer based on Kepler statistics}.
\newblock \aap \textbf{609}, A4 (2018).
\newblock \doi{10.1051/0004-6361/201731254}

\bibitem{KenchingtonGoldsmith2016}
{Kenchington Goldsmith}, H.D., {Cvetojevic}, N., {Ireland}, M., {Ma}, P.,
  {Tuthill}, P., {Eggleton}, B., {Lawrence}, J.S., {Debbarma}, S.,
  {Luther-Davies}, B., {Madden}, S.J.: {Chalcogenide glass planar MIR couplers
  for future chip based Bracewell interferometers}.
\newblock In: Optical and Infrared Interferometry and Imaging V, \emph{Proc.
  SPIE}, vol. 9907, p. 990730 (2016).
\newblock \doi{10.1117/12.2232199}

\bibitem{KenchingtonGoldsmith2017a}
{Kenchington Goldsmith}, H.D., {Cvetojevic}, N., {Ireland}, M., {Madden}, S.:
  {Fabrication tolerant chalcogenide mid-infrared multimode interference
  coupler design with applications for Bracewell nulling interferometry}.
\newblock Optics Express \textbf{25}, 3038 (2017).
\newblock \doi{10.1364/OE.25.003038}

\bibitem{KenchingtonGoldsmith2017b}
{Kenchington Goldsmith}, H.D., {Ireland}, M., {Ma}, P., {Cevetojevic}, N.,
  {Madden}, S.: {Improving the extinction bandwidth of MMI chalcogenide
  photonic chip based MIR nulling interferometers}.
\newblock Optics Express \textbf{25}, 16,813 (2017).
\newblock \doi{10.1364/OE.25.016813}

\bibitem{Kern:2009}
{Kern}, P., {Le Co{\"a}rer}, E., {Benech}, P.: {On-chip spectro-detection for
  fully integrated coherent beam combiners}.
\newblock Opt. Expr. \textbf{17}, 1976--1987 (2009).
\newblock \doi{10.1364/OE.17.001976}

\bibitem{Kern:1997}
{Kern}, P., {Malbet}, F., {Schanen-Duport}, I., {Benech}, P.: {Integrated
  optics single-mode interferometric beam combiner for near infrared
  astronomy}, p. 115.
\newblock EDP Sciences (1997)

\bibitem{Kervella:2000}
{Kervella}, P., {Coud{\'e} du Foresto}, V., {Glindemann}, A., {Hofmann}, R.:
  {VINCI: the VLT Interferometer commissioning instrument}.
\newblock In: P.~{L{\'e}na}, A.~{Quirrenbach} (eds.) Interferometry in Optical
  Astronomy, \emph{Proc. SPIE}, vol. 4006, pp. 31--42 (2000).
\newblock \doi{10.1117/12.390227}

\bibitem{Kimura:2010}
Kimura, M., Maihara, T., Iwamuro, F., Akiyama, M., Tamura, N., Dalton, G.B.,
  Takato, N., Tait, P., Ohta, K., Eto, S., et~al.: Fibre multi-object
  spectrograph (fmos) for the subaru telescope.
\newblock Publications of the Astronomical Society of Japan \textbf{62}(5),
  1135--1147 (2010)

\bibitem{Kloppenborg:2010}
Kloppenborg, B., Stencel, R., Monnier, J.D., Schaefer, G., Zhao, M., Baron, F.,
  McAlister, H., ten Brummelaar, T., Che, X., Farrington, C., Pedretti, E.,
  Sallave-Goldfinger, P.J., Sturmann, J., Sturmann, L., Thureau, N., Turner,
  N., Carroll, S.M.: Infrared images of the transiting disk in the $\epsilon$
  aurigae system.
\newblock Nature \textbf{464}, 870--872 (2010)

\bibitem{Kotani:2005}
Kotani, T., Perrin, G., Vergnole, S., Woillez, J., Guerin, J.: Characterization
  of fluoride fibers for the optical hawaiian array for nanoradian astronomy
  project.
\newblock Appl. Opt. \textbf{44}, 5029--5035 (2005)

\bibitem{Kraus:2012}
{Kraus}, A.L., {Ireland}, M.J.: {LkCa 15: A Young Exoplanet Caught at
  Formation?}
\newblock Astroph. J. \textbf{745}, 5 (2012).
\newblock \doi{10.1088/0004-637X/745/1/5}

\bibitem{Kraus:2016}
{Kraus}, S., {Monnier}, J.D., {Ireland}, M.J., {Duch{\^e}ne}, G., {Espaillat},
  C., {H{\"o}nig}, S., {Juhasz}, A., {Mordasini}, C., {Olofsson}, J.,
  {Paladini}, C., {Stassun}, K., {Turner}, N., {Vasisht}, G., {Harries}, T.J.,
  {Bate}, M.R., {Gonzalez}, J.F., {Matter}, A., {Zhu}, Z., {Panic}, O.,
  {Regaly}, Z., {Morbidelli}, A., {Meru}, F., {Wolf}, S., {Ilee}, J., {Berger},
  J.P., {Zhao}, M., {Kral}, Q., {Morlok}, A., {Bonsor}, A., {Ciardi}, D.,
  {Kane}, S.R., {Kratter}, K., {Laughlin}, G., {Pepper}, J., {Raymond}, S.,
  {Labadie}, L., {Nelson}, R.P., {Weigelt}, G., {ten Brummelaar}, T.,
  {Pierens}, A., {Oudmaijer}, R., {Kley}, W., {Pope}, B., {Jensen}, E.L.N.,
  {Bayo}, A., {Smith}, M., {Boyajian}, T., {Quiroga-Nu{\~n}ez}, L.H.,
  {Millan-Gabet}, R., {Chiavassa}, A., {Gallenne}, A., {Reynolds}, M., {de
  Wit}, W.J., {Wittkowski}, M., {Millour}, F., {Gandhi}, P., {Ramos Almeida},
  C., {Alonso Herrero}, A., {Packham}, C., {Kishimoto}, M., {Tristram}, K.R.W.,
  {Pott}, J.U., {Surdej}, J., {Buscher}, D., {Haniff}, C., {Lacour}, S.,
  {Petrov}, R., {Ridgway}, S., {Tuthill}, P., {van Belle}, G., {Armitage}, P.,
  {Baruteau}, C., {Benisty}, M., {Bitsch}, B., {Paardekooper}, S.J., {Pinte},
  C., {Masset}, F., {Rosotti}, G.: {Planet Formation Imager (PFI): science
  vision and key requirements}.
\newblock In: Optical and Infrared Interferometry and Imaging V, \emph{Proc.
  SPIE}, vol. 9907, p. 99071K (2016).
\newblock \doi{10.1117/12.2231067}

\bibitem{Kuhn:1962}
Kuhn, T.S.: The structure of scientific revolutions.
\newblock The University of Chicago (1962)

\bibitem{Labadie:2008}
{Labadie}, L., {Kern}, P., {Labeye}, P., {Lecoarer}, E., {Vigreux-Bercovici},
  C., {Pradel}, A., {Broquin}, J.E., {Kirschner}, V.: {Technology challenges
  for space interferometry: The option of mid-infrared integrated optics}.
\newblock Advances in Space Research \textbf{41}, 1975--1982 (2008).
\newblock \doi{10.1016/j.asr.2007.07.013}

\bibitem{Labadie:2007}
{Labadie}, L., {Le Coarer}, E., {Maurand}, R., {Labeye}, P., {Kern}, P.,
  {Arezki}, B., {Broquin}, J.E.: {Mid-infrared laser light nulling experiment
  using single-mode conductive waveguides}.
\newblock A\&A \textbf{471}, 355--360 (2007).
\newblock \doi{10.1051/0004-6361:20067005}

\bibitem{Labadie:2011}
Labadie, L., Martin, G., Anheier, N.C., Arezki, B., Qiao, H.A., Bernacki, B.,
  Kern, P.: First fringes with an integrated-optics beam combiner at 10 $\mu$m.
  a new step towards instrument miniaturization for mid-infrared
  interferometry.
\newblock A\&A \textbf{531}, A48--A54 (2011)

\bibitem{Labadie:2018}
Labadie, L., Minardi, S., Tepper, J., Diener, R., Muthusubramanian, B., Pott,
  J.U., Gross, S., Arriola, A., Withford, M.: {Photonics-based mid-infrared
  interferometry I: 4-year results of the ALSI project and future prospects}.
\newblock In: Optical and Infrared Interferometry and Imaging VI, \emph{Proc.
  SPIE}, vol. 10701, pp. 10,701--46 (2018)

\bibitem{Labadie:2009}
{Labadie}, L., {Wallner}, O.: {Mid-infrared guided optics: a perspective for
  astronomical instruments}.
\newblock Optics Express \textbf{17}, 1947--1962 (2009).
\newblock \doi{10.1364/OE.17.001947}

\bibitem{Lacour:2014}
{Lacour}, S., {Lapeyr{\`e}re}, V., {Gauchet}, L., {Arroud}, S., {Gourgues}, R.,
  {Martin}, G., {Heidmann}, S., {Haubois}, X., {Perrin}, G.: {CubeSats as
  pathfinders for planetary detection: the FIRST-S satellite}.
\newblock In: Space Telescopes and Instrumentation 2014: Optical, Infrared, and
  Millimeter Wave, \emph{Proc. SPIE}, vol. 9143, p. 91432N (2014).
\newblock \doi{10.1117/12.2057381}

\bibitem{Lacour:2007}
{Lacour}, S., {Thi{\'e}baut}, E., {Perrin}, G.: {High dynamic range imaging
  with a single-mode pupil remapping system: a self-calibration algorithm for
  redundant interferometric arrays}.
\newblock Month. Not. R. Astron. Soc. \textbf{374}, 832--846 (2007).
\newblock \doi{10.1111/j.1365-2966.2006.11198.x}

\bibitem{Lagrange2009}
{Lagrange}, A.M., {Gratadour}, D., {Chauvin}, G., {Fusco}, T., {Ehrenreich},
  D., {Mouillet}, D., {Rousset}, G., {Rouan}, D., {Allard}, F., {Gendron},
  {\'E}., {Charton}, J., {Mugnier}, L., {Rabou}, P., {Montri}, J., {Lacombe},
  F.: {A probable giant planet imaged in the {$\beta$} Pictoris disk. VLT/NaCo
  deep L'-band imaging}.
\newblock \aap \textbf{493}, L21--L25 (2009).
\newblock \doi{10.1051/0004-6361:200811325}

\bibitem{Langrock:2005}
{Langrock}, C., {Diamanti}, E., {Roussev}, R.V., {Yamamoto}, Y., {Fejer}, M.M.,
  {Takesue}, H.: {Highly efficient single-photon detection at communication
  wavelengths by use of upconversion in reverse-proton-exchanged periodically
  poled LiNbO$_{3}$ waveguides}.
\newblock Optics Letters \textbf{30}, 1725--1727 (2005).
\newblock \doi{10.1364/OL.30.001725}

\bibitem{Larkin:2010}
Larkin, J.E., Moore, A.M., Barton, E.J., Bauman, B., Bui, K., Canfield, J.,
  Crampton, D., Delacroix, A., Fletcher, M., Hale, D., et~al.: The infrared
  imaging spectrograph (iris) for tmt: instrument overview.
\newblock In: SPIE Astronomical Telescopes+ Instrumentation, pp.
  773,529--773,529. International Society for Optics and Photonics (2010)

\bibitem{Laurent:2010}
Laurent, F., Adjali, L., Arns, J., Bacon, R., Boudon, D., Caillier, P.,
  Daguisé, E., Delabre, B., Dubois, J.P., Godefroy, P., Jarno, A., Jorden, P.,
  Kosmalski, J., Lap\'ere, V., Lizon, J.L., Loupias, M., Pecontal, A., Reiss,
  R., Remillieux, A., Renault, E., Rupprecht, G., Salaun, Y.: Muse integral
  field unit: Test results on the first out of 24.
\newblock In: Proc. SPIE, vol. 7739, p.~95 (2010)

\bibitem{Lawrence:2012}
Lawrence, J., Bland-Hawthorn, J., Bryant, J., Brzeski, J., Colless, M., Croom,
  S., Gers, L., Gilbert, J., Gillingham, P., Goodwin, M., Heijmans, J., Horton,
  A., Ireland, M., Miziarski, S., Saunders, W., Smith, G.: Hector: a
  high-multiplex survey instrument for spatially resolved galaxy spectroscopy.
\newblock In: Proc.SPIE, vol. 8446, pp. 8446 -- 8446 -- 11 (2012).
\newblock \doi{10.1117/12.925260}.
\newblock \urlprefix\url{http://dx.doi.org/10.1117/12.925260}

\bibitem{Lazareff2017}
{Lazareff}, B., {Berger}, J.P., {Kluska}, J., {Le Bouquin}, J.B., {Benisty},
  M., {Malbet}, F., {Koen}, C., {Pinte}, C., {Thi}, W.F., {Absil}, O., {Baron},
  F., {Delboulb{\'e}}, A., {Duvert}, G., {Isella}, A., {Jocou}, L., {Juhasz},
  A., {Kraus}, S., {Lachaume}, R., {M{\'e}nard}, F., {Millan-Gabet}, R.,
  {Monnier}, J.D., {Moulin}, T., {Perraut}, K., {Rochat}, S., {Soulez}, F.,
  {Tallon}, M., {Thi{\'e}baut}, E., {Traub}, W., {Zins}, G.: {Structure of
  Herbig AeBe disks at the milliarcsecond scale . A statistical survey in the H
  band using PIONIER-VLTI}.
\newblock \aap \textbf{599}, A85 (2017).
\newblock \doi{10.1051/0004-6361/201629305}

\bibitem{LeCoarer:2007}
{Le Coarer}, E., {Blaize}, S., {Benech}, P., {Stefanon}, I., {Morand}, A.,
  {L{\'e}rondel}, G., {Leblond}, G., {Kern}, P., {Fedeli}, J.M., {Royer}, P.:
  {Wavelength-scale stationary-wave integrated Fourier-transform spectrometry}.
\newblock Nature Photonics \textbf{1}, 473--478 (2007).
\newblock \doi{10.1038/nphoton.2007.138}

\bibitem{leFevre:2003}
Le~F{\`e}vre, O., Sa{\"\i}sse, M., Mancini, D., Brau-Nogu{\'e}, S., Caputi, O.,
  Castinel, L., d’Odorico, S., Garilli, B., Kissler, M., Lucuix, C., et~al.:
  Commissioning and performances of the vlt-vimos.
\newblock In: Proc. of SPIE Vol, vol. 4841, p. 1671 (2003)

\bibitem{Lee:2012}
{Lee}, C., {Chu}, S.T., {Little}, B.E., {Bland-Hawthorn}, J., {Leon-Saval}, S.:
  {Portable frequency combs for optical frequency metrology}.
\newblock Opt. Exp. \textbf{20}, 16,671 (2012).
\newblock \doi{10.1364/OE.20.016671}

\bibitem{Lee:2006}
Lee, J.H., Foo, G., Johnson, E.G., Swartzlander, G.A.J.: Experimental
  verification of an optical vortex coronagraph.
\newblock Phys. Rev. Lett. \textbf{97}, 053,901 (2006)

\bibitem{Lehmann:2018b}
{Lehmann}, L., {Darr{\'e}}, P., {Boulogne}, H., {Delage}, L., {Grossard}, L.,
  {Reynaud}, F.: {Multichannel spectral mode of the ALOHA up-conversion
  interferometer}.
\newblock \mnras \textbf{477}(1), 190--194 (2018).
\newblock \doi{10.1093/mnras/sty648}

\bibitem{Lehmann:2018}
{Lehmann}, L., {Darr{\'e}}, P., {Szemendera}, L., {Gomes}, J.T., {Baudoin}, R.,
  {Ceus}, D., {Brustlein}, S., {Delage}, L., {Grossard}, L., {Reynaud}, F.:
  {ALOHA{\textemdash}Astronomical Light Optical Hybrid Analysis. From
  experimental demonstrations to a MIR instrument proposal}.
\newblock Experimental Astronomy \textbf{46}(3), 447--456 (2018).
\newblock \doi{10.1007/s10686-018-9585-2}

\bibitem{Lehmann:2019}
{Lehmann}, L., {Delage}, L., {Grossard}, L., {Reynaud}, F., {Golden}, S.,
  {Woods}, C., {Webster}, L., {Sturmann}, J., {Brummelaar}, T.t.:
  {Environmental characterisation and stabilisation of a 2{\texttimes}200-meter
  outdoor fibre interferometer at the CHARA Array}.
\newblock Experimental Astronomy \textbf{47}(3), 303--312 (2019).
\newblock \doi{10.1007/s10686-019-09627-x}

\bibitem{Leinert1993}
{Leinert}, C., {Zinnecker}, H., {Weitzel}, N., {Christou}, J., {Ridgway}, S.T.,
  {Jameson}, R., {Haas}, M., {Lenzen}, R.: {A systematic approach for young
  binaries in Taurus}.
\newblock \aap \textbf{278}, 129--149 (1993)

\bibitem{lemke:2011}
Lemke, U., Corbett, J., Allington-Smith, J., Murray, G.: Modal noise prediction
  in fibre spectroscopy--i. visibility and the coherent model.
\newblock Monthly Notices of the Royal Astronomical Society \textbf{417}(1),
  689--697 (2011)

\bibitem{LeonSaval:2005}
Leon-Saval, S., Birks, T., Bland-Hawthorn, J., Englund, M.: Multimode fiber
  devices with single-mode performance.
\newblock Optics letters \textbf{30}(19), 2545--2547 (2005)

\bibitem{LeonSaval:2012}
Leon-Saval, S.G., Betters, C.H., Bland-Hawthorn, J.: The photonic tiger: a
  multicore fiber-fed spectrograph.
\newblock In: Proc. of SPIE Vol, vol. 8450, pp. 84,501K--1 (2012)

\bibitem{LeonSaval:2017}
Leon-Saval, S.G., Betters, C.H., Salazar-Gil, J.R., Min, S.S., Gris-Sanchez,
  I., Birks, T.A., Lawrence, J., Haynes, R., Haynes, D., Roth, M., Veilleux,
  S., Bland-Hawthorn, J.: Divide and conquer: an efficient solution to highly
  multimoded photonic lanterns from multicore fibres.
\newblock Opt. Express \textbf{25}(15), 17,530--17,540 (2017).
\newblock \doi{10.1364/OE.25.017530}.
\newblock
  \urlprefix\url{http://www.opticsexpress.org/abstract.cfm?URI=oe-25-15-17530}

\bibitem{Lindley2014}
Lindley, E., Min, S.S., Leon-Saval, S., Cvetojevic, N., Lawrence, J., Ellis,
  S., Bland-Hawthorn, J.: Demonstration of uniform multicore fiber bragg
  gratings.
\newblock Optics express \textbf{22}(25), 31,575--31,581 (2014)

\bibitem{Lippa2016}
{Lippa}, M., {Gillessen}, S., {Blind}, N., {Kok}, Y., {Yaz{\i}c{\i}}, {\c S}.,
  {Weber}, J., {Pfuhl}, O., {Haug}, M., {Kellner}, S., {Wieprecht}, E.,
  {Eisenhauer}, F., {Genzel}, R., {Hans}, O., {Hau{\ss}mann}, F., {Huber}, D.,
  {Kratschmann}, T., {Ott}, T., {Plattner}, M., {Rau}, C., {Sturm}, E.,
  {Waisberg}, I., {Wiezorrek}, E., {Perrin}, G., {Perraut}, K., {Brandner}, W.,
  {Straubmeier}, C., {Amorim}, A.: {The metrology system of the VLTI instrument
  GRAVITY}.
\newblock In: Optical and Infrared Interferometry and Imaging V,
  \emph{\procspie}, vol. 9907, p. 990722 (2016).
\newblock \doi{10.1117/12.2232272}

\bibitem{Lozi:2009}
{Lozi}, J., {Martinache}, F., {Guyon}, O.: {Phase-Induced Amplitude Apodization
  on Centrally Obscured Pupils: Design and First Laboratory Demonstration for
  the Subaru Telescope Pupil}.
\newblock \pasp \textbf{121}, 1232 (2009).
\newblock \doi{10.1086/648392}

\bibitem{Ma:2013}
{Ma}, P., {Choi}, D.Y., {Yu}, Y., {Gai}, X., {Yang}, Z., {Debbarma}, S.,
  {Madden}, S., {Luther-Davies}, B.: {Low-loss chalcogenide waveguides for
  chemical sensing in the mid-infrared}.
\newblock Optics Express \textbf{21}, 29,927 (2013).
\newblock \doi{10.1364/OE.21.029927}

\bibitem{maclachlan:2016}
MacLachlan, D.G., Harris, R.J., Choudhury, D., Simmonds, R.D., Salter, P.S.,
  Booth, M.J., Allington-Smith, J.R., Thomson, R.R.: Development of integrated
  mode reformatting components for diffraction-limited spectroscopy.
\newblock Optics letters \textbf{41}(1), 76--79 (2016)

\bibitem{maclachlan:2016_hybrid}
MacLachlan, D.G., Harris, R.J., Gris-S{\'a}nchez, I., Morris, T.J., Choudhury,
  D., Gendron, E., Basden, A.G., Spaleniak, I., Arriola, A., Birks, T.A.,
  et~al.: Efficient photonic reformatting of celestial light for
  diffraction-limited spectroscopy.
\newblock Monthly Notices of the Royal Astronomical Society \textbf{464}(4),
  4950--4957 (2016)

\bibitem{Malbet2012}
{Malbet}, F., {L{\'e}ger}, A., {Shao}, M., {Goullioud}, R., {Lagage}, P.O.,
  {Brown}, A.G.A., {Cara}, C., {Durand}, G., {Eiroa}, C., {Feautrier}, P.,
  {Jakobsson}, B., {Hinglais}, E., {Kaltenegger}, L., {Labadie}, L.,
  {Lagrange}, A.M., {Laskar}, J., {Liseau}, R., {Lunine}, J., {Maldonado}, J.,
  {Mercier}, M., {Mordasini}, C., {Queloz}, D., {Quirrenbach}, A., {Sozzetti},
  A., {Traub}, W., {Absil}, O., {Alibert}, Y., {Andrei}, A.H., {Arenou}, F.,
  {Beichman}, C., {Chelli}, A., {Cockell}, C.S., {Duvert}, G., {Forveille}, T.,
  {Garcia}, P.J.V., {Hobbs}, D., {Krone-Martins}, A., {Lammer}, H., {Meunier},
  N., {Minardi}, S., {Moitinho de Almeida}, A., {Rambaux}, N., {Raymond}, S.,
  {R{\"o}ttgering}, H.J.A., {Sahlmann}, J., {Schuller}, P.A., {S{\'e}gransan},
  D., {Selsis}, F., {Surdej}, J., {Villaver}, E., {White}, G.J., {Zinnecker},
  H.: {High precision astrometry mission for the detection and characterization
  of nearby habitable planetary systems with the Nearby Earth Astrometric
  Telescope (NEAT)}.
\newblock Experimental Astronomy \textbf{34}, 385--413 (2012).
\newblock \doi{10.1007/s10686-011-9246-1}

\bibitem{Mariotti:1996}
{Mariotti}, J.M., {Coud{\'e} du Foresto}, V., {Perrin}, G., {Zhao}, P., {Lena},
  P.: {Interferometric connection of large ground-based telescopes.}
\newblock A\&AS \textbf{116}, 381--393 (1996)

\bibitem{Marois2008}
{Marois}, C., {Macintosh}, B., {Barman}, T., {Zuckerman}, B., {Song}, I.,
  {Patience}, J., {Lafreni{\`e}re}, D., {Doyon}, R.: {Direct Imaging of
  Multiple Planets Orbiting the Star HR 8799}.
\newblock Science \textbf{322}, 1348 (2008).
\newblock \doi{10.1126/science.1166585}

\bibitem{Martin:2017}
{Martin}, G., {Bhuyan}, M., {Troles}, J., {D'Amico}, C., {Stoian}, R., {Le
  Coarer}, E.: {Near infrared spectro-interferometer using femtosecond laser
  written GLS embedded waveguides and nano-scatterers}.
\newblock Opt. Expr. \textbf{25}, 8386 (2017).
\newblock \doi{10.1364/OE.25.008386}

\bibitem{Martin:2016}
Martin, G., Pugnat, T., Gardillou, F., Cassagnettes, C., Barbier, D., Guyot,
  C., Hauden, J., Huby, E., Lacour, S.: {Novel multi-telescopes beam combiners
  for next generation instruments (FIRST/SUBARU)}.
\newblock In: F.~Malbet, M.J. Creech-Eakman, P.G. Tuthill (eds.) Optical and
  Infrared Interferometry and Imaging V, vol. 9907, pp. 764 -- 771.
  International Society for Optics and Photonics, SPIE (2016).
\newblock \doi{10.1117/12.2233105}.
\newblock \urlprefix\url{https://doi.org/10.1117/12.2233105}

\bibitem{Martin:2008}
{Martin}, S., {Serabyn}, E., {Liewer}, K., {Loya}, F., {Mennesson}, B.,
  {Hanot}, C., {Mawet}, D.: {The development and applications of a ground-based
  fiber nulling coronagraph}.
\newblock In: Optical and Infrared Interferometry, \emph{Proc. SPIE}, vol.
  7013, p. 70131Y (2008).
\newblock \doi{10.1117/12.789484}

\bibitem{Mathieu1992}
{Mathieu}, R.D.: {The Short-Period Frequency Among Low-Mass PreMain Sequence
  Stars}.
\newblock In: H.A. {McAlister}, W.I. {Hartkopf} (eds.) Complementary Approaches
  to Double and Multiple Star Research, \emph{IAU Colloquium}, vol.~32, pp.
  30--40 (1992).
\newblock \doi{10.1017/S1743921307009672}

\bibitem{Mawet:2013a}
Mawet, D., Absil, O., Delacroix, C., Girard, J.H., Milli, J., O'Neal, J.,
  Baudoz, P., Boccaletti, A., Bourget, P., Christiaens, V., Forsberg, P.,
  Gonte, F., Habraken, S., Hanot, C., Karlsson, M., Kasper, M., Lizon, J.L.,
  Muzic, K., Olivier, R., na, E.P., Slusarenko, N., Tacconi-Garman, L.E.,
  Surdej, J.: L'-band agpm vector vortex coronagraph's first light on vlt/naco
  - discovery of a late-type companion at two beam widths from an f0v star.
\newblock A\&A \textbf{553}, L13 (2013)

\bibitem{Mawet:2013b}
{Mawet}, D., {Pueyo}, L., {Carlotti}, A., {Mennesson}, B., {Serabyn}, E.,
  {Wallace}, J.K.: {Ring-apodized Vortex Coronagraphs for Obscured Telescopes.
  I. Transmissive Ring Apodizers}.
\newblock Astrophys. J. Suppl. Ser. \textbf{209}, 7 (2013).
\newblock \doi{10.1088/0067-0049/209/1/7}

\bibitem{Mawet2012}
{Mawet}, D., {Pueyo}, L., {Lawson}, P., {Mugnier}, L., {Traub}, W.,
  {Boccaletti}, A., {Trauger}, J.T., {Gladysz}, S., {Serabyn}, E., {Milli}, J.,
  {Belikov}, R., {Kasper}, M., {Baudoz}, P., {Macintosh}, B., {Marois}, C.,
  {Oppenheimer}, B., {Barrett}, H., {Beuzit}, J.L., {Devaney}, N., {Girard},
  J., {Guyon}, O., {Krist}, J., {Mennesson}, B., {Mouillet}, D., {Murakami},
  N., {Poyneer}, L., {Savransky}, D., {V{\'e}rinaud}, C., {Wallace}, J.K.:
  {Review of small-angle coronagraphic techniques in the wake of ground-based
  second-generation adaptive optics systems}.
\newblock In: Space Telescopes and Instrumentation 2012: Optical, Infrared, and
  Millimeter Wave, \emph{\procspie}, vol. 8442, p. 844204 (2012).
\newblock \doi{10.1117/12.927245}

\bibitem{Mawet:2005}
Mawet, D., Riaud, P., Absil, O., Surdej, J.: Annular groove phase mask
  coronagraph.
\newblock Astron. J. \textbf{633}, 1191--1200 (2005)

\bibitem{Mawet:2010}
Mawet, D., Serabyn, E., Liewer, K., Burruss, R., Hickey, J., Shemo, D.: The
  vector vortex coronagraph: laboratory results and first light at palomar
  observatory.
\newblock Astroph. J. \textbf{709}, 53--57 (2010)

\bibitem{Mawet:2009}
Mawet, D., Serabyn, E., Liewer, K., Hanot, C., McEldowney, S., Shemo, D.,
  O'Brien, N.: Optical vectorial vortex coronagraphs using liquid crystal
  polymers: theory, manufacturing and laboratory demonstration.
\newblock Opt. Exp. \textbf{17}, 1902--1918 (2009)

\bibitem{Mayor:2014}
Mayor, M., Lovis, C., Santos, N.C.: Doppler spectroscopy as a path to the
  detection of earth-like planets.
\newblock Nature \textbf{513}(7518), 328 (2014)

\bibitem{Mayor:1995}
{Mayor}, M., {Queloz}, D.: {A Jupiter-mass companion to a solar-type star}.
\newblock Nature \textbf{378}, 355--359 (1995).
\newblock \doi{10.1038/378355a0}

\bibitem{Mayor:2009}
{Mayor}, M., {Udry}, S., {Lovis}, C., {Pepe}, F., {Queloz}, D., {Benz}, W.,
  {Bertaux}, J.L., {Bouchy}, F., {Mordasini}, C., {Segransan}, D.: {The HARPS
  search for southern extra-solar planets. XIII. A planetary system with 3
  super-Earths (4.2, 6.9, and 9.2 M)}.
\newblock A\&A \textbf{493}, 639--644 (2009).
\newblock \doi{10.1051/0004-6361:200810451}

\bibitem{McCabe2006}
{McCabe}, C., {Ghez}, A.M., {Prato}, L., {Duch{\^e}ne}, G., {Fisher}, R.S.,
  {Telesco}, C.: {Investigating Disk Evolution: A High Spatial Resolution
  Mid-Infrared Survey of T Tauri Stars}.
\newblock \apj \textbf{636}, 932--951 (2006).
\newblock \doi{10.1086/498207}

\bibitem{McGregor:2003}
McGregor, P.J., Hart, J., Conroy, P.G., Pfitzner, M.L., Bloxham, G.J., Jones,
  D.J., Downing, M.D., Dawson, M., Young, P., Jarnyk, M., et~al.: Gemini
  near-infrared integral field spectrograph (nifs).
\newblock In: Instrument Design and Performance for Optical/Infrared
  Ground-based Telescopes, vol. 4841, pp. 1581--1591 (2003)

\bibitem{McMahon:1975}
{McMahon}, D.H.: {Efficiency limitations imposed by thermodynamics on optical
  coupling in fiber-optic data links}.
\newblock Journal of the Optical Society of America (1917-1983) \textbf{65},
  1479 (1975)

\bibitem{Mennesson2014}
{Mennesson}, B., {Millan-Gabet}, R., {Serabyn}, E., {Colavita}, M.M., {Absil},
  O., {Bryden}, G., {Wyatt}, M., {Danchi}, W., {Defr{\`e}re}, D., {Dor{\'e}},
  O., {Hinz}, P., {Kuchner}, M., {Ragland}, S., {Scott}, N., {Stapelfeldt}, K.,
  {Traub}, W., {Woillez}, J.: {Constraining the Exozodiacal Luminosity Function
  of Main-sequence Stars: Complete Results from the Keck Nuller Mid-infrared
  Surveys}.
\newblock \apj \textbf{797}, 119 (2014).
\newblock \doi{10.1088/0004-637X/797/2/119}

\bibitem{Mennesson:2014}
{Mennesson}, B., {Millan-Gabet}, R., {Serabyn}, E., {Colavita}, M.M., {Absil},
  O., {Bryden}, G., {Wyatt}, M., {Danchi}, W., {Defr{\`e}re}, D., {Dor{\'e}},
  O., {Hinz}, P., {Kuchner}, M., {Ragland}, S., {Scott}, N., {Stapelfeldt}, K.,
  {Traub}, W., {Woillez}, J.: {Constraining the Exozodiacal Luminosity Function
  of Main-sequence Stars: Complete Results from the Keck Nuller Mid-infrared
  Surveys}.
\newblock Astroph. J. \textbf{797}, 119 (2014).
\newblock \doi{10.1088/0004-637X/797/2/119}

\bibitem{MillanGabet:2011}
{Millan-Gabet}, R., {Serabyn}, E., {Mennesson}, B., {Traub}, W.A., {Barry},
  R.K., {Danchi}, W.C., {Kuchner}, M., {Stark}, C.C., {Ragland}, S.,
  {Hrynevych}, M., {Woillez}, J., {Stapelfeldt}, K., {Bryden}, G., {Colavita},
  M.M., {Booth}, A.J.: {Exozodiacal Dust Levels for Nearby Main-sequence Stars:
  A Survey with the Keck Interferometer Nuller}.
\newblock Astroph. J. \textbf{734}, 67 (2011).
\newblock \doi{10.1088/0004-637X/734/1/67}

\bibitem{Miller:2013}
Miller, D.A.B.: Self-aligning universal beam coupler.
\newblock Opt. Express \textbf{21}(5), 6360--6370 (2013).
\newblock \doi{10.1364/OE.21.006360}.
\newblock
  \urlprefix\url{http://www.opticsexpress.org/abstract.cfm?URI=oe-21-5-6360}

\bibitem{Millour2011}
{Millour}, F., {Meilland}, A., {Chesneau}, O., {Stee}, P., {Kanaan}, S.,
  {Petrov}, R., {Mourard}, D., {Kraus}, S.: {Imaging the spinning gas and dust
  in the disc around the supergiant A[e] star HD 62623}.
\newblock \aap \textbf{526}, A107 (2011).
\newblock \doi{10.1051/0004-6361/201016193}

\bibitem{Millour:2011}
{Millour}, F., {Meilland}, A., {Chesneau}, O., {Stee}, P., {Kanaan}, S.,
  {Petrov}, R., {Mourard}, D., {Kraus}, S.: {Imaging the spinning gas and dust
  in the disc around the supergiant A[e] star HD 62623}.
\newblock A\&A \textbf{526}, A107 (2011).
\newblock \doi{10.1051/0004-6361/201016193}

\bibitem{Minardi:2012}
Minardi, S., Dreisow, F., Gr\"afe, M., Nolte, S., Pertsch, T.:
  Three-dimensional photonic component for multichannel coherence measurements.
\newblock Opt. Lett. \textbf{37}, 3030--3032 (2012)

\bibitem{Minardi:2016}
{Minardi}, S., {Lacour}, S., {Berger}, J.P., {Labadie}, L., {Thomson}, R.R.,
  {Haniff}, C., {Ireland}, M.: {Beam combination schemes and technologies for
  the Planet Formation Imager}.
\newblock In: Optical and Infrared Interferometry and Imaging V, \emph{Proc.
  SPIE}, vol. 9907, p. 99071N (2016).
\newblock \doi{10.1117/12.2232656}

\bibitem{Minardi:2010}
Minardi, S., Pertsch, T.: Interferometric beam combination with discrete
  optics.
\newblock Opt. Lett. \textbf{35}, 3009--3011 (2010)

\bibitem{Molaro:2013}
{Molaro}, P., {Esposito}, M., {Monai}, S., {Lo Curto}, G., {Gonz{\'a}lez
  Hern{\'a}ndez}, J.I., {H{\"a}nsch}, T.W., {Holzwarth}, R., {Manescau}, A.,
  {Pasquini}, L., {Probst}, R.A., {Rebolo}, R., {Steinmetz}, T., {Udem}, T.,
  {Wilken}, T.: {A frequency comb calibrated solar atlas}.
\newblock A\&A \textbf{560}, A61 (2013).
\newblock \doi{10.1051/0004-6361/201322324}

\bibitem{Monnier:2004}
{Monnier}, J.D., {Berger}, J.P., {Millan-Gabet}, R., {ten Brummelaar}, T.A.:
  {The Michigan Infrared Combiner (MIRC): IR imaging with the CHARA Array}.
\newblock In: W.A. {Traub} (ed.) New Frontiers in Stellar Interferometry,
  \emph{Proc. SPIE}, vol. 5491, p. 1370 (2004).
\newblock \doi{10.1117/12.550804}

\bibitem{Monnier2002}
{Monnier}, J.D., {Millan-Gabet}, R.: {On the Interferometric Sizes of Young
  Stellar Objects}.
\newblock \apj \textbf{579}, 694--698 (2002).
\newblock \doi{10.1086/342917}

\bibitem{Monnier:2007}
{Monnier}, J.D., {Zhao}, M., {Pedretti}, E., {Thureau}, N., {Ireland}, M.,
  {Muirhead}, P., {Berger}, J.P., {Millan-Gabet}, R., {Van Belle}, G., {ten
  Brummelaar}, T., {McAlister}, H., {Ridgway}, S., {Turner}, N., {Sturmann},
  L., {Sturmann}, J., {Berger}, D.: {Imaging the Surface of Altair}.
\newblock Science \textbf{317}(5836), 342 (2007).
\newblock \doi{10.1126/science.1143205}

\bibitem{Murakami:2013}
{Murakami}, N., {Hamaguchi}, S., {Sakamoto}, M., {Fukumoto}, R., {Ise}, A.,
  {Oka}, K., {Baba}, N., {Tamura}, M.: {Design and laboratory demonstration of
  an achromatic vector vortex coronagraph}.
\newblock Opt. Expr. \textbf{21}, 7400 (2013).
\newblock \doi{10.1364/OE.21.007400}

\bibitem{Murray:2009}
Murray, G., Allington-Smith, J.: Strategies for spectroscopy on extremely large
  telescopes--ii. diverse-field spectroscopy.
\newblock Monthly Notices of the Royal Astronomical Society \textbf{399}(1),
  209--218 (2009)

\bibitem{Norris:2014}
Norris, B., Cvetojevic, N., Gross, S., Jovanovic, N., Stewart, P.N., Charles,
  N., Lawrence, J.S., Withford, M.J., Tuthill, P.: High-performance 3d
  waveguide architecture for astronomical pupil-remapping interferometry.
\newblock Opt. Express \textbf{22}(15), 18,335--18,353 (2014).
\newblock \doi{10.1364/OE.22.018335}.
\newblock
  \urlprefix\url{http://www.opticsexpress.org/abstract.cfm?URI=oe-22-15-18335}

\bibitem{norris2020all}
Norris, B., Wei, J., Betters, C., Wong, A., Leon-Saval, S.: All-fibre
  focal-plane wavefront sensor.
\newblock In: Conference on Lasers and Electro-Optics/Pacific Rim, p. PDP\_1.
  Optical Society of America (2020)

\bibitem{Norris:2020}
{Norris}, B.R.M., {Cvetojevic}, N., {Lagadec}, T., {Jovanovic}, N., {Gross},
  S., {Arriola}, A., {Gretzinger}, T., {Martinod}, M.A., {Guyon}, O., {Lozi},
  J., {Withford}, M.J., {Lawrence}, J.S., {Tuthill}, P.: {First on-sky
  demonstration of an integrated-photonic nulling interferometer: the GLINT
  instrument}.
\newblock MNRAS \textbf{491}(3), 4180--4193 (2020).
\newblock \doi{10.1093/mnras/stz3277}

\bibitem{Offer:1998}
{Offer}, A.R., {Bland-Hawthorn}, J.: {Rugate filters for OH-suppressed imaging
  at near-infrared wavelengths}.
\newblock MNRAS \textbf{299}, 176--188 (1998).
\newblock \doi{10.1046/j.1365-8711.1998.01760.x}

\bibitem{Ohnaka2017}
{Ohnaka}, K., {Weigelt}, G., {Hofmann}, K.H.: {Vigorous atmospheric motion in
  the red supergiant star Antares}.
\newblock \nat \textbf{548}, 310--312 (2017).
\newblock \doi{10.1038/nature23445}

\bibitem{Oppenheimer2009}
{Oppenheimer}, B.R., {Hinkley}, S.: {High-Contrast Observations in Optical and
  Infrared Astronomy}.
\newblock \araa \textbf{47}, 253--289 (2009).
\newblock \doi{10.1146/annurev-astro-082708-101717}

\bibitem{oshagh:2016}
Oshagh, M., Dreizler, S., Santos, N., Figueira, P., Reiners, A.: Can stellar
  activity make a planet seem misaligned?
\newblock Astronomy \& Astrophysics \textbf{593}, A25 (2016)

\bibitem{Padgett1999}
{Padgett}, D.L., {Brandner}, W., {Stapelfeldt}, K.R., {Strom}, S.E., {Terebey},
  S., {Koerner}, D.: {HUBBLE SPACE TELESCOPE/NICMOS Imaging of Disks and
  Envelopes around Very Young Stars}.
\newblock \aj \textbf{117}, 1490--1504 (1999).
\newblock \doi{10.1086/300781}

\bibitem{Paladini2018}
{Paladini}, C., {Baron}, F., {Jorissen}, A., {Le Bouquin}, J.B., {Freytag}, B.,
  {van Eck}, S., {Wittkowski}, M., {Hron}, J., {Chiavassa}, A., {Berger}, J.P.,
  {Siopis}, C., {Mayer}, A., {Sadowski}, G., {Kravchenko}, K., {Shetye}, S.,
  {Kerschbaum}, F., {Kluska}, J., {Ramstedt}, S.: {Large granulation cells on
  the surface of the giant star {$\pi$}$^{1}$ Gruis}.
\newblock \nat \textbf{553}, 310--312 (2018).
\newblock \doi{10.1038/nature25001}

\bibitem{Pasquini:2002}
{Pasquini}, L., {Avila}, G., {Blecha}, A., {Cacciari}, C., {Cayatte}, V.,
  {Colless}, M., {Damiani}, F., {de Propris}, R., {Dekker}, H., {di
  Marcantonio}, P., {Farrell}, T., {Gillingham}, P., {Guinouard}, I., {Hammer},
  F., {Kaufer}, A., {Hill}, V., {Marteaud}, M., {Modigliani}, A., {Mulas}, G.,
  {North}, P., {Popovic}, D., {Rossetti}, E., {Royer}, F., {Santin}, P.,
  {Schmutzer}, R., {Simond}, G., {Vola}, P., {Waller}, L., {Zoccali}, M.:
  {Installation and commissioning of FLAMES, the VLT Multifibre Facility}.
\newblock The Messenger \textbf{110}, 1--9 (2002)

\bibitem{Paufique:2012}
Paufique, J., Argomedo, J., Arsenault, R., Conzelmann, R., Donaldson, R.,
  Hubin, N., Jochum, L., Jost, A., Kiekebusch, M., Kolb, J., Kuntschner, H.,
  Louarn, M.L., Madec, P.Y., Siebenmorgen, R., Tordo, S.: Status of the graal
  system development: very wide-field correction with 4 laser guide-stars.
\newblock In: Proc. SPIE, vol. 8447, p. 116 (2012)

\bibitem{peacock:2001}
Peacock, J.A., Cole, S., Norberg, P., Baugh, C.M., Bland-Hawthorn, J., Bridges,
  T., Cannon, R.D., Colless, M., Collins, C., Couch, W., et~al.: A measurement
  of the cosmological mass density from clustering in the 2df galaxy redshift
  survey.
\newblock Nature \textbf{410}(6825), 169 (2001)

\bibitem{Pedichini:2017}
{Pedichini}, F., {Stangalini}, M., {Ambrosino}, F., {Puglisi}, A., {Pinna}, E.,
  {Bailey}, V., {Carbonaro}, L., {Centrone}, M., {Christou}, J., {Esposito},
  S., {Farinato}, J., {Fiore}, F., {Giallongo}, E., {Hill}, J.M., {Hinz}, P.M.,
  {Sabatini}, L.: {High Contrast Imaging in the Visible: First Experimental
  Results at the Large Binocular Telescope}.
\newblock \aj \textbf{154}, 74 (2017).
\newblock \doi{10.3847/1538-3881/aa7ff3}

\bibitem{Pedretti:2018}
Pedretti, E., Piacentini, S., Corrielli, G., Osellame, R., Minardi, S.: A
  six-apertures discrete beam combiners for j-band interferometry.
\newblock In: Proc. SPIE, vol. 10701, pp. 10,701--25. International Society for
  Optics and Photonics (2018)

\bibitem{Pepe2010}
{Pepe}, F.A., {Cristiani}, S., {Rebolo Lopez}, R., {Santos}, N.C., {Amorim},
  A., {Avila}, G., {Benz}, W., {Bonifacio}, P., {Cabral}, A., {Carvas}, P.,
  {Cirami}, R., {Coelho}, J., {Comari}, M., {Coretti}, I., {De Caprio}, V.,
  {Dekker}, H., {Delabre}, B., {Di Marcantonio}, P., {D'Odorico}, V., {Fleury},
  M., {Garc{\'{\i}}a}, R., {Herreros Linares}, J.M., {Hughes}, I., {Iwert}, O.,
  {Lima}, J., {Lizon}, J.L., {Lo Curto}, G., {Lovis}, C., {Manescau}, A.,
  {Martins}, C., {M{\'e}gevand}, D., {Moitinho}, A., {Molaro}, P., {Monteiro},
  M., {Monteiro}, M., {Pasquini}, L., {Mordasini}, C., {Queloz}, D., {Rasilla},
  J.L., {Rebord{\~a}o}, J.M., {Santana Tschudi}, S., {Santin}, P., {Sosnowska},
  D., {Span{\`o}}, P., {Tenegi}, F., {Udry}, S., {Vanzella}, E., {Viel}, M.,
  {Zapatero Osorio}, M.R., {Zerbi}, F.: {ESPRESSO: the Echelle spectrograph for
  rocky exoplanets and stable spectroscopic observations}.
\newblock In: Ground-based and Airborne Instrumentation for Astronomy III,
  \emph{\procspie}, vol. 7735, p. 77350F (2010).
\newblock \doi{10.1117/12.857122}

\bibitem{Perrin:2000}
{Perrin}, G., {Lai}, O., {Lena}, P.J., {Coud{\'e} du Foresto}, V.: {Fibered
  large interferometer on top of Mauna Kea: OHANA, the optical Hawaiian array
  for nanoradian astronomy}.
\newblock In: P.~{L{\'e}na}, A.~{Quirrenbach} (eds.) Interferometry in Optical
  Astronomy, \emph{Proc. SPIE}, vol. 4006, pp. 708--714 (2000).
\newblock \doi{10.1117/12.390272}

\bibitem{Perrin:2006}
Perrin, G., Woillez, J., Lai, O., Gu\'erin, J., Kotani, T., Wizinowich, P.L.,
  Mignant, D.L., Hrynevych, M., Gathright, J., L\'ena, P., Chaffee, F.,
  Vergnole, S., Delage, L., Reynaud, F., Adamson, A.J., Berthod, C., Brient,
  B., Collin, C., Cr\'etenet, J., Dauny, F., Del\'eglise, C., F\'edou, P.,
  Goeltzenlichter, T., Guyon, O., Hulin, R., Marlot, C., Marteaud, M., Melse,
  B.T., Nishikawa, J., Reess, J.M., Ridgway, S.T., Rigaut, F., Roth, K.,
  Tokunaga, A.T., Zieglera, D.: Interferometric coupling of the keck telescopes
  with single-mode fibers.
\newblock Science \textbf{311}, 194 (2006)

\bibitem{Perrin2006}
{Perrin}, M.D., {Duch{\^e}ne}, G., {Kalas}, P., {Graham}, J.R.: {Discovery of
  an Optically Thick, Edge-on Disk around the Herbig Ae Star PDS 144N}.
\newblock \apj \textbf{645}, 1272--1282 (2006).
\newblock \doi{10.1086/504510}

\bibitem{Peters:2008}
Peters, M.A., Close, L.M., Rademacher, M., Stalcup, T., Swartzlander, G.A.,
  Ford, E., Abdul-Malik, R.S.: A high-strehl low-resolution optical imager
  (bessel): detection of a 0.7$\lambda/d$ separation binary from the ground.
\newblock New Astr. \textbf{13}, 359--369 (2008)

\bibitem{Peters:2012}
Peters, M.A., Groff, T., Kasdin, N.J., McElwain, M.W., Galvin, M., Carr, M.A.,
  Lupton, R., Gunn, J.E., Knapp, G., Gong, Q., et~al.: Conceptual design of the
  coronagraphic high angular resolution imaging spectrograph (charis) for the
  subaru telescope.
\newblock In: Ground-based and Airborne Instrumentation for Astronomy IV, vol.
  8446, p. 84467U. International Society for Optics and Photonics (2012)

\bibitem{Petrov:2007}
{Petrov}, R.G., {Malbet}, F., {Weigelt}, G., {Antonelli}, P., {Beckmann}, U.,
  {Bresson}, Y., {Chelli}, A., {Dugu{\'e}}, M., {Duvert}, G., {Gennari}, S.,
  {Gl{\"u}ck}, L., {Kern}, P., {Lagarde}, S., {Le Coarer}, E., {Lisi}, F.,
  {Millour}, F., {Perraut}, K., {Puget}, P., {Rantakyr{\"o}}, F.,
  {Robbe-Dubois}, S., {Roussel}, A., {Salinari}, P., {Tatulli}, E., {Zins}, G.,
  {Accardo}, M., {Acke}, B., {Agabi}, K., {Altariba}, E., {Arezki}, B.,
  {Aristidi}, E., {Baffa}, C., {Behrend}, J., {Bl{\"o}cker}, T., {Bonhomme},
  S., {Busoni}, S., {Cassaing}, F., {Clausse}, J.M., {Colin}, J., {Connot}, C.,
  {Delboulb{\'e}}, A., {Domiciano de Souza}, A., {Driebe}, T., {Feautrier}, P.,
  {Ferruzzi}, D., {Forveille}, T., {Fossat}, E., {Foy}, R., {Fraix-Burnet}, D.,
  {Gallardo}, A., {Giani}, E., {Gil}, C., {Glentzlin}, A., {Heiden}, M.,
  {Heininger}, M., {Hernandez Utrera}, O., {Hofmann}, K.H., {Kamm}, D.,
  {Kiekebusch}, M., {Kraus}, S., {Le Contel}, D., {Le Contel}, J.M., {Lesourd},
  T., {Lopez}, B., {Lopez}, M., {Magnard}, Y., {Marconi}, A., {Mars}, G.,
  {Martinot-Lagarde}, G., {Mathias}, P., {M{\`e}ge}, P., {Monin}, J.L.,
  {Mouillet}, D., {Mourard}, D., {Nussbaum}, E., {Ohnaka}, K., {Pacheco}, J.,
  {Perrier}, C., {Rabbia}, Y., {Rebattu}, S., {Reynaud}, F., {Richichi}, A.,
  {Robini}, A., {Sacchettini}, M., {Schertl}, D., {Sch{\"o}ller}, M.,
  {Solscheid}, W., {Spang}, A., {Stee}, P., {Stefanini}, P., {Tallon}, M.,
  {Tallon-Bosc}, I., {Tasso}, D., {Testi}, L., {Vakili}, F., {von der
  L{\"u}he}, O., {Valtier}, J.C., {Vannier}, M., {Ventura}, N.: {AMBER, the
  near-infrared spectro-interferometric three-telescope VLTI instrument}.
\newblock A\&A \textbf{464}, 1--12 (2007).
\newblock \doi{10.1051/0004-6361:20066496}

\bibitem{Phillips:2012}
{Phillips}, D.F., {Glenday}, A.G., {Li}, C.H., {Cramer}, C., {Furesz}, G.,
  {Chang}, G., {Benedick}, A.J., {Chen}, L.J., {K{\"a}rtner}, F.X.,
  {Korzennik}, S., {Sasselov}, D., {Szentgyorgyi}, A., {Walsworth}, R.L.:
  {Calibration of an astrophysical spectrograph below 1 m/s using a laser
  frequency comb}.
\newblock Opt. Exp. \textbf{20}, 13,711--13,726 (2012).
\newblock \doi{10.1364/OE.20.013711}

\bibitem{Pojmanski:1997}
{Pojmanski}, G.: {The All Sky Automated Survey}.
\newblock \actaa \textbf{47}, 467--481 (1997)

\bibitem{Pontoppidan2008}
{Pontoppidan}, K.M., {Blake}, G.A., {van Dishoeck}, E.F., {Smette}, A.,
  {Ireland}, M.J., {Brown}, J.: {Spectroastrometric Imaging of Molecular Gas
  within Protoplanetary Disk Gaps}.
\newblock \apj \textbf{684}, 1323-1329 (2008).
\newblock \doi{10.1086/590400}

\bibitem{Pontoppidan2005}
{Pontoppidan}, K.M., {Dullemond}, C.P., {van Dishoeck}, E.F., {Blake}, G.A.,
  {Boogert}, A.C.A., {Evans} II, N.J., {Kessler-Silacci}, J.E., {Lahuis}, F.:
  {Ices in the Edge-on Disk CRBR 2422.8-3423: Spitzer Spectroscopy and Monte
  Carlo Radiative Transfer Modeling}.
\newblock \apj \textbf{622}, 463--481 (2005).
\newblock \doi{10.1086/427688}

\bibitem{Primmerman:1991}
{Primmerman}, C.A., {Murphy}, D.V., {Page}, D.A., {Zollars}, B.G., {Barclay},
  H.T.: {Compensation of atmospheric optical distortion using a synthetic
  beacon}.
\newblock Nature \textbf{353}, 141--143 (1991).
\newblock \doi{10.1038/353141a0}

\bibitem{racine:1999}
Racine, R., Walker, G.A., Nadeau, D., Doyon, R., Marois, C.: Speckle noise and
  the detection of faint companions.
\newblock Publications of the Astronomical Society of the Pacific
  \textbf{111}(759), 587 (1999)

\bibitem{Rawson:1980}
Rawson, E.G., Goodman, J.W., Norton, R.E.: Frequency dependence of modal noise
  in multimode optical fibers.
\newblock J. Opt. Soc. Am. \textbf{70}(8), 968--976 (ts).
\newblock \doi{10.1364/JOSA.70.000968}.
\newblock
  \urlprefix\url{http://www.osapublishing.org/abstract.cfm?URI=josa-70-8-968}

\bibitem{rebolo:1995}
Rebolo, R., Osorio, M.Z., Martin, E.: Discovery of a brown dwarf in the
  pleiades star cluster.
\newblock Nature \textbf{377}(6545), 129 (1995)

\bibitem{Renard2010}
{Renard}, S., {Malbet}, F., {Benisty}, M., {Thi{\'e}baut}, E., {Berger}, J.P.:
  {Milli-arcsecond images of the Herbig Ae star HD 163296}.
\newblock \aap \textbf{519}, A26 (2010).
\newblock \doi{10.1051/0004-6361/201014910}

\bibitem{Reynaud:1992}
{Reynaud}, F., {Alleman}, J.J., {Connes}, P.: {Interferometric control of fiber
  lengths for a coherent telescope array}.
\newblock Appl. Opt. \textbf{31}, 3736--3743 (1992).
\newblock \doi{10.1364/AO.31.003736}

\bibitem{Rhodes:1973}
{Rhodes}, W.T., {Goodman}, J.W.: {Interferometric technique for recording and
  restoring images degraded by unknown aberrations.}
\newblock Journal of the Optical Society of America (1917-1983) \textbf{63},
  647--657 (1973)

\bibitem{Roddier:1997}
Roddier, F., Roddier, C.: Stellar coronagraph with phase mask.
\newblock PASP \textbf{109}, 815--820 (1997)

\bibitem{Rodenas:2012}
Rodenas, A., Martin, G., Arzeki, B., Psaila, N.D., Jose, G., Jha, A., Labadie,
  L., Kern, P., Kar, A.K., Thomson, R.R.: Three-dimensional mid-infrared
  photonic circuits in chalcogenide glass.
\newblock Opt. Lett. \textbf{37}, 392--394 (2012)

\bibitem{Roettenbacher2016}
{Roettenbacher}, R.M., {Monnier}, J.D., {Korhonen}, H., {Aarnio}, A.N.,
  {Baron}, F., {Che}, X., {Harmon}, R.O., {K{\H o}v{\'a}ri}, Z., {Kraus}, S.,
  {Schaefer}, G.H., {Torres}, G., {Zhao}, M., {Ten Brummelaar}, T.A.,
  {Sturmann}, J., {Sturmann}, L.: {No Sun-like dynamo on the active star
  {$\zeta$} Andromedae from starspot asymmetry}.
\newblock \nat \textbf{533}, 217--220 (2016).
\newblock \doi{10.1038/nature17444}

\bibitem{Rogstad:1968}
Rogstad, D.H.: A technique for measuring the visibility phase with an optical
  interferometer in the presence of atmospheric seeing.
\newblock Appl. Opt. \textbf{7}, 585 (1968)

\bibitem{Rohloff:1991}
{Rohloff}, R.R., {Leinert}, C.: {Properties of fiber optics for application in
  astronomical interferometry}.
\newblock Appl. Opt. \textbf{30}, 5031--5036 (1991).
\newblock \doi{10.1364/AO.30.005031}

\bibitem{Rouan:2000}
{Rouan}, D., {Riaud}, P., {Boccaletti}, A., {Cl{\'e}net}, Y., {Labeyrie}, A.:
  {The Four-Quadrant Phase-Mask Coronagraph. I. Principle}.
\newblock Pub. Astron. Soc. Pac. \textbf{112}, 1479--1486 (2000).
\newblock \doi{10.1086/317707}

\bibitem{Ruane:2014}
{Ruane}, G.J., {Kanburapa}, P., {Han}, J., {Swartzlander}, G.A.: {Vortex-phase
  filtering technique for extracting spatial information from unresolved
  sources}.
\newblock Appl. Opt. \textbf{53}, 4503 (2014).
\newblock \doi{10.1364/AO.53.004503}

\bibitem{Robinson:2014}
S.~Robinson, B., Boroson, D., Burianek, D., Murphy, D., Khatri, F., Biswas, A.,
  Sodnik, Z., Burnside, J., Kansky, J., M.~Cornwell, D.: The nasa lunar laser
  communication demonstration—successful high-rate laser communications to
  and from the moon.
\newblock In: 13th International Conference on Space Operations, SpaceOps 2014
  (2014)

\bibitem{Saito:2007}
Saito, N., Akagawa, K., Ito, M., Takazawa, A., Saito, Y.H.Y., Ito, M., Takami,
  H., Iye, M., Wada, S.: Sodium $d_2$ resonance radiation single pass
  sum-frequency generation with actively mode-locked nd:yag lasers.
\newblock Opt. Lett. \textbf{32}, 1965--1967 (2007)

\bibitem{Saito:2010}
Saito, Y., Hayano, Y., Ito, M., Minowa, Y., Egner, S., Oya, S., Watanabe, M.,
  Hattori, M., Garrel, V., Akagawa, K., Guyon, O., Colley, S., Golota, T.,
  Saito, N., Takazawa, A., Ito, M., Takami, H., Wada, S., Iye, M.: The
  performance of the laser guide star system for the subaru telescope.
\newblock In: Proc. SPIE, vol. 7736, p.~53 (2010)

\bibitem{Sana2014}
{Sana}, H., {Le Bouquin}, J.B., {Lacour}, S., {Berger}, J.P., {Duvert}, G.,
  {Gauchet}, L., {Norris}, B., {Olofsson}, J., {Pickel}, D., {Zins}, G.,
  {Absil}, O., {de Koter}, A., {Kratter}, K., {Schnurr}, O., {Zinnecker}, H.:
  {Southern Massive Stars at High Angular Resolution: Observational Campaign
  and Companion Detection}.
\newblock \apjs \textbf{215}, 15 (2014).
\newblock \doi{10.1088/0067-0049/215/1/15}

\bibitem{Saviauk:2013}
{Saviauk}, A., {Minardi}, S., {Dreisow}, F., {Nolte}, S., {Pertsch}, T.:
  {3D-integrated optics component for astronomical spectro-interferometry}.
\newblock Appl. Opt. \textbf{52}, 4556 (2013).
\newblock \doi{10.1364/AO.52.004556}

\bibitem{Schoedel2002}
{Sch{\"o}del}, R., {Ott}, T., {Genzel}, R., {Hofmann}, R., {Lehnert}, M.,
  {Eckart}, A., {Mouawad}, N., {Alexander}, T., {Reid}, M.J., {Lenzen}, R.,
  {Hartung}, M., {Lacombe}, F., {Rouan}, D., {Gendron}, E., {Rousset}, G.,
  {Lagrange}, A.M., {Brandner}, W., {Ageorges}, N., {Lidman}, C., {Moorwood},
  A.F.M., {Spyromilio}, J., {Hubin}, N., {Menten}, K.M.: {A star in a 15.2-year
  orbit around the supermassive black hole at the centre of the Milky Way}.
\newblock Nature \textbf{419}, 694--696 (2002).
\newblock \doi{10.1038/nature01121}

\bibitem{Schroeder1987}
{Schroeder}, D.J.: {Astronomical optics}.
\newblock Elsevier (1987)

\bibitem{Seemann:2014}
{Seemann}, U., {Anglada-Escude}, G., {Baade}, D., {Bristow}, P., {Dorn}, R.J.,
  {Follert}, R., {Gojak}, D., {Grunhut}, J., {Hatzes}, A.P., {Heiter}, U.,
  {Ives}, D.J., {Jeep}, P., {Jung}, Y., {K{\"a}ufl}, H.U., {Kerber}, F.,
  {Klein}, B., {Lizon}, J.L., {Lockhart}, M., {L{\"o}winger}, T., {Marquart},
  T., {Oliva}, E., {Paufique}, J., {Piskunov}, N., {Pozna}, E., {Reiners}, A.,
  {Smette}, A., {Smoker}, J., {Stempels}, E., {Valenti}, E.: {Wavelength
  calibration from 1-5{$\mu$}m for the CRIRES+ high-resolution spectrograph at
  the VLT}.
\newblock In: Ground-based and Airborne Instrumentation for Astronomy V,
  \emph{Proc. SPIE}, vol. 9147, p. 91475G (2014).
\newblock \doi{10.1117/12.2056668}

\bibitem{Serabyn:2017}
{Serabyn}, E., {Huby}, E., {Matthews}, K., {Mawet}, D., {Absil}, O., {Femenia},
  B., {Wizinowich}, P., {Karlsson}, M., {Bottom}, M., {Campbell}, R.,
  {Carlomagno}, B., {Defr{\`e}re}, D., {Delacroix}, C., {Forsberg}, P., {Gomez
  Gonzalez}, C., {Habraken}, S., {Jolivet}, A., {Liewer}, K., {Lilley}, S.,
  {Piron}, P., {Reggiani}, M., {Surdej}, J., {Tran}, H., {Vargas Catal{\'a}n},
  E., {Wertz}, O.: {The W. M. Keck Observatory Infrared Vortex Coronagraph and
  a First Image of HIP 79124 B}.
\newblock Astron. J. \textbf{153}, 43 (2017).
\newblock \doi{10.3847/1538-3881/153/1/43}

\bibitem{Serabyn:2010}
Serabyn, E., Mawet, D., Burruss, R.: An image of an exoplanet separated by two
  diffraction beamwidths from a star.
\newblock Nature \textbf{464}, 1018--1020 (2010)

\bibitem{Serabyn:2012}
{Serabyn}, E., {Mennesson}, B., {Colavita}, M.M., {Koresko}, C., {Kuchner},
  M.J.: {The Keck Interferometer Nuller}.
\newblock Astroph. J. \textbf{748}, 55 (2012).
\newblock \doi{10.1088/0004-637X/748/1/55}

\bibitem{Serabyn:2007}
Serabyn, E., Wallace, K., Troy, M., Mennesson, B., Haguenauer, P., Gappinger,
  R., Burruss, R.: Extreme adaptive optics imaging with a clear and
  well-corrected off-axis telescope subaperture.
\newblock Astroph. J. \textbf{658}, 1386--1391 (2007)

\bibitem{Shaklan:1992}
{Shaklan}, S., {Reynaud}, F., {Froehly}, C.: {Multimode fiber-optic broad
  spectral band interferometer}.
\newblock Appl. Opt. \textbf{31}, 749--756 (1992).
\newblock \doi{10.1364/AO.31.000749}

\bibitem{Shaklan:1987}
Shaklan, S., Roddier, F.: Single-mode fiber optics in long-baseline
  interferometer.
\newblock Appl. Opt. \textbf{26}, 2159--2163 (1987)

\bibitem{Shaklan:1988}
Shaklan, S., Roddier, F.: Coupling starlight into single-mode optical fiber
  optics.
\newblock Appl. Opt. \textbf{27}, 2334--2338 (1988)

\bibitem{Sharp:2006}
Sharp, R., Saunders, W., Smith, G., Churilov, V., Correll, D., Dawson, J.,
  Farrel, T., Frost, G., Haynes, R., Heald, R., et~al.: Performance of aaomega:
  the aat multi-purpose fiber-fed spectrograph.
\newblock In: Society of Photo-Optical Instrumentation Engineers (SPIE)
  Conference Series, vol. 6269 (2006)

\bibitem{Sharples:2004}
Sharples, R.M., Bender, R., Lehnert, M.D., Ramsay~Howat, S.K., Bremer, M.N.,
  Davies, R.L., Genzel, R., Hofmann, R., Ivison, R.J., Saglia, R., Thatte,
  N.A.: Kmos: an infrared multiple-object integral field spectrograph for the
  eso vlt.
\newblock In: Society of Photo-Optical Instrumentation Engineers (SPIE)
  Conference Series, vol. 5492, pp. 5492 -- 5492 -- 8 (2004).
\newblock \doi{10.1117/12.550495}.
\newblock \urlprefix\url{http://dx.doi.org/10.1117/12.550495}

\bibitem{smee:2013}
Smee, S.A., Gunn, J.E., Uomoto, A., Roe, N., Schlegel, D., Rockosi, C.M., Carr,
  M.A., Leger, F., Dawson, K.S., Olmstead, M.D., et~al.: The multi-object,
  fiber-fed spectrographs for the sloan digital sky survey and the baryon
  oscillation spectroscopic survey.
\newblock The Astronomical Journal \textbf{146}(2), 32 (2013)

\bibitem{Smith:1984}
{Smith}, B.A., {Terrile}, R.J.: {A circumstellar disk around Beta Pictoris}.
\newblock Science \textbf{226}, 1421--1424 (1984).
\newblock \doi{10.1126/science.226.4681.1421}

\bibitem{Smith:2006}
{Smith}, D., {Zuber}, M., {Sun}, X., {Neumann}, G., {Cavanaugh}, J., {McGarry},
  J., {Zagwodzki}, T.: {Two-way Laser Link over Interplanetary Distance}.
\newblock Science \textbf{311}, 53--53 (2006).
\newblock \doi{10.1126/science.1120091}

\bibitem{Snellen2015}
{Snellen}, I., {de Kok}, R., {Birkby}, J.L., {Brandl}, B., {Brogi}, M.,
  {Keller}, C., {Kenworthy}, M., {Schwarz}, H., {Stuik}, R.: {Combining
  high-dispersion spectroscopy with high contrast imaging: Probing rocky
  planets around our nearest neighbors}.
\newblock \aap \textbf{576}, A59 (2015).
\newblock \doi{10.1051/0004-6361/201425018}

\bibitem{SnyderLove}
Snyder, A.W., Love, J.D.: Optical Waveguide Theory.
\newblock Kluwer, London (2007)

\bibitem{Sodnik:2014}
Sodnik, Z., Smit, H., Sans, M., Giggenbach, D., Becker, P., Mata~Calvo, R.,
  Fuchs, C., Zayer, I., Lanucara, M., Schulz, K.J., Widmer, J., Arnold, F.,
  Alonso, A., Montilla, I.: Results from a lunar laser communication experiment
  between nasa's ladee satellite and esa's optical ground station.
\newblock In: Proc. ICSOS, pp. S2--1 (2014)

\bibitem{spaleniak:2014}
Spaleniak, I., Gross, S., Jovanovic, N., Williams, R.J., Lawrence, J.S.,
  Ireland, M.J., Withford, M.J.: Multiband processing of multimode light:
  combining 3d photonic lanterns with waveguide bragg gratings.
\newblock Laser \& Photonics Reviews \textbf{8}(1) (2014)

\bibitem{spaleniak:2013}
Spaleniak, I., Jovanovic, N., Gross, S., Ireland, M.J., Lawrence, J.S.,
  Withford, M.J.: Integrated photonic building blocks for next-generation
  astronomical instrumentation ii: the multimode to single mode transition.
\newblock Optics express \textbf{21}(22), 27,197--27,208 (2013)

\bibitem{Steinmetz:2008}
Steinmetz, T., Wilken, T., Araujo-Hauck, C., Holzwarth, R., H{\"a}nsch, T.W.,
  Pasquini, L., Manescau, A., D'odorico, S., Murphy, M.T., Kentischer, T.,
  et~al.: Laser frequency combs for astronomical observations.
\newblock Science \textbf{321}(5894), 1335--1337 (2008)

\bibitem{Stoll:2017}
Stoll, A., Zhang, Z., Haynes, R., Roth, M.: High-resolution
  arrayed-waveguide-gratings in astronomy: Design and fabrication challenges.
\newblock In: Photonics, vol.~4, p.~30. Multidisciplinary Digital Publishing
  Institute (2017)

\bibitem{Stroebele:2012}
Str\"'obele, S., Penna, P.L., Arsenault, R., Conzelmann, R., Delabre, B.,
  Duchateau, M., Dorn, R., Fedrigo, E., Hubin, N., Quentin, J., Jolley, P.,
  Kiekebusch, M., Kirchbauer, J., Klein, B., Kolb, J., Kuntschner, H., Louarn,
  M.L., Lizon, J., Madec, P.Y., Pettazzi, L., Soenke, C., Tordo, S., Vernet,
  J., Muradore, R.: Galacsi system design and analysis.
\newblock In: Proc. SPIE, vol. 8447, p.~37 (2012)

\bibitem{Su:2017}
{Su}, T., {Scott}, R.P., {Ogden}, C., {Thurman}, S.T., {Kendrick}, R.L.,
  {Duncan}, A., {Yu}, R., {Yoo}, S.J.B.: {Experimental demonstration of
  interferometric imaging using photonic integrated circuits}.
\newblock Opt. Expr. \textbf{25}, 12,653 (2017).
\newblock \doi{10.1364/OE.25.012653}

\bibitem{Shu:2019}
Suh, M.G., Yi, X., Lai, Y.H., Leifer, S., Grudinin, I.S., Vasisht, G., Martin,
  E.C., Fitzgerald, M.P., Doppmann, G., Wang, J., Mawet, D., Papp, S.B.,
  Diddams, S.A., Beichman, C., Vahala, K.: Searching for exoplanets using a
  microresonator astrocomb.
\newblock Nature Photonics \textbf{13}(1), 25--30 (2019).
\newblock \doi{10.1038/s41566-018-0312-3}.
\newblock \urlprefix\url{https://doi.org/10.1038/s41566-018-0312-3}

\bibitem{Swain2003}
{Swain}, M., {Vasisht}, G., {Akeson}, R., {Monnier}, J., {Millan-Gabet}, R.,
  {Serabyn}, E., {Creech-Eakman}, M., {van Belle}, G., {Beletic}, J.,
  {Beichman}, C., {Boden}, A., {Booth}, A., {Colavita}, M., {Gathright}, J.,
  {Hrynevych}, M., {Koresko}, C., {Le Mignant}, D., {Ligon}, R., {Mennesson},
  B., {Neyman}, C., {Sargent}, A., {Shao}, M., {Thompson}, R., {Unwin}, S.,
  {Wizinowich}, P.: {Interferometer Observations of Subparsec-Scale Infrared
  Emission in the Nucleus of NGC 4151}.
\newblock \apjl \textbf{596}, L163--L166 (2003).
\newblock \doi{10.1086/379235}

\bibitem{Swartzlander:2005}
Swartzlander, G.A.J.: Broadband nulling of a vortex phase mask.
\newblock Opt. Lett. \textbf{30}, 2876--2878 (2005)

\bibitem{Swartzlander:2006}
Swartzlander, G.A.J.: Achromatic optical vortex lens.
\newblock Opt. Lett. \textbf{31}, 2042--2044 (2006)

\bibitem{Taha2017}
{Taha}, A.S., {Labadie}, L., {Pantin}, E., {Matter}, A., {Alvarez}, C.,
  {Esquej}, P., {Grellmann}, R., {Rebolo}, R., {Telesco}, C., {Wolf}, S.: {The
  spatial extent of Polycyclic Aromatic Hydrocarbons emission in the Herbig
  star HD 179218}.
\newblock ArXiv e-prints  (2017)

\bibitem{Takami2001}
{Takami}, M., {Bailey}, J., {Gledhill}, T.M., {Chrysostomou}, A., {Hough},
  J.H.: {Circumstellar structure of RU Lupi down to au scales}.
\newblock \mnras \textbf{323}, 177--187 (2001).
\newblock \doi{10.1046/j.1365-8711.2001.04172.x}

\bibitem{Tatulli:2007b}
Tatulli, E., Duvert, G.: Amber data reduction.
\newblock New Astronomy Reviews \textbf{51}(8), 682 -- 696 (2007).
\newblock \doi{https://doi.org/10.1016/j.newar.2007.06.010}.
\newblock
  \urlprefix\url{http://www.sciencedirect.com/science/article/pii/S1387647307000693}.
\newblock Observation and Data Reduction with the VLT Interferometer

\bibitem{Taylor:2010}
Taylor, L., Fen, Y., Calia, D.B.: 50 w visible laser source at 589 nm obtained
  via frequency doubling of three coherently combined narrow-band raman fiber
  amplifiers.
\newblock Opt. Exp. \textbf{18}, 8540--8555 (2010)

\bibitem{Bummelaar:2016}
{ten Brummelaar}, T.A., {Gies}, D.G., {McAlister}, H.A., {Ridgway}, S.T.,
  {Sturmann}, J., {Sturmann}, L., {Schaefer}, G.H., {Turner}, N.H.,
  {Farrington}, C.D., {Scott}, N.J., {Monnier}, J.D., {Ireland}, M.J.: {An
  update on the CHARA array}.
\newblock In: Optical and Infrared Interferometry and Imaging V, \emph{Proc.
  SPIE}, vol. 9907, p. 990703 (2016).
\newblock \doi{10.1117/12.2232125}

\bibitem{Tepper:2017b}
Tepper, J., Labadie, L., Gross, S., Arriola, A., Minardi, S., Diener, R.,
  Withford, M.J.: Ultrafast laser inscription in zblan integrated optics chips
  for mid-ir beam combination in astronomical interferometry.
\newblock Opt. Express \textbf{25}(17), 20,642--20,653 (2017).
\newblock \doi{10.1364/OE.25.020642}.
\newblock
  \urlprefix\url{http://www.opticsexpress.org/abstract.cfm?URI=oe-25-17-20642}

\bibitem{Tepper:2017a}
{Tepper, J.}, {Labadie, L.}, {Diener, R.}, {Minardi, S.}, {Pott, J.-U.},
  {Thomson, R.}, {Nolte, S.}: Integrated optics prototype beam combiner for
  long baseline interferometry in the l and m bands.
\newblock A\&A \textbf{602}, A66 (2017).
\newblock \doi{10.1051/0004-6361/201630138}.
\newblock \urlprefix\url{https://doi.org/10.1051/0004-6361/201630138}

\bibitem{Boehm2017}
{The Theia Collaboration}, {Boehm}, C., {Krone-Martins}, A., {Amorim}, A.,
  {Anglada-Escude}, G., {Brandeker}, A., {Courbin}, F., {Ensslin}, T.,
  {Falcao}, A., {Freese}, K., {Holl}, B., {Labadie}, L., {Leger}, A., {Malbet},
  F., {Mamon}, G., {McArthur}, B., {Mora}, A., {Shao}, M., {Sozzetti}, A.,
  {Spolyar}, D., {Villaver}, E., {Albertus}, C., {Bertone}, S., {Bouy}, H.,
  {Boylan-Kolchin}, M., {Brown}, A., {Brown}, W., {Cardoso}, V., {Chemin}, L.,
  {Claudi}, R., {Correia}, A.C.M., {Crosta}, M., {Crouzier}, A., {Cyr-Racine},
  F.Y., {Damasso}, M., {da Silva}, A., {Davies}, M., {Das}, P., {Dayal}, P.,
  {de Val-Borro}, M., {Diaferio}, A., {Erickcek}, A., {Fairbairn}, M.,
  {Fortin}, M., {Fridlund}, M., {Garcia}, P., {Gnedin}, O., {Goobar}, A.,
  {Gordo}, P., {Goullioud}, R., {Hambly}, N., {Hara}, N., {Hobbs}, D., {Hog},
  E., {Holland}, A., {Ibata}, R., {Jordi}, C., {Klioner}, S., {Kopeikin}, S.,
  {Lacroix}, T., {Laskar}, J., {Le Poncin-Lafitte}, C., {Luri}, X., {Majumdar},
  S., {Makarov}, V., {Massey}, R., {Mennesson}, B., {Michalik}, D., {Moitinho
  de Almeida}, A., {Mourao}, A., {Moustakas}, L., {Murray}, N., {Muterspaugh},
  M., {Oertel}, M., {Ostorero}, L., {Perez-Garcia}, A., {Platais}, I., {de
  Mora}, J.P.i., {Quirrenbach}, A., {Randall}, L., {Read}, J., {Regos}, E.,
  {Rory}, B., {Rybicki}, K., {Scott}, P., {Schneider}, J., {Scholtz}, J.,
  {Siebert}, A., {Tereno}, I., {Tomsick}, J., {Traub}, W., {Valluri}, M.,
  {Walker}, M., {Walton}, N., {Watkins}, L., {White}, G., {Evans}, D.W.,
  {Wyrzykowski}, L., {Wyse}, R.: {Theia: Faint objects in motion or the new
  astrometry frontier}.
\newblock ArXiv e-prints  (2017)

\bibitem{Thomson:2012}
Thomson, R., Harris, R., Birks, T., Brown, G., Allington-Smith, J.,
  Bland-Hawthorn, J.: Ultrafast laser inscription of a 121-waveguide fan-out
  for astrophotonics.
\newblock Optics letters \textbf{37}(12), 2331--2333 (2012)

\bibitem{Tinney:2004}
Tinney, C., Ryder, S., Ellis, S., Churilov, V., Dawson, J., Smith, G., Waller,
  L., Whittard, J., Haynes, R., Lankshear, A., et~al.: Iris2: a working
  infrared multi-object spectrograph \& camera.
\newblock In: Proc. of SPIE Vol, vol. 5492, p. 999 (2004)

\bibitem{traub2010direct}
Traub, W.A., Oppenheimer, B.R.: Direct imaging of exoplanets.
\newblock University of Arizona Press, Tucson (2010)

\bibitem{Trinh:2013}
Trinh, C.Q., Ellis, S.C., Bland-Hawthorn, J., Lawrence, J.S., Horton, A.J.,
  Leon-Saval, S.G., Shortridge, K., Bryant, J., Case, S., Colless, M., et~al.:
  Gnosis: the first instrument to use fiber bragg gratings for oh suppression.
\newblock The Astronomical Journal \textbf{145}(2), 51 (2013)

\bibitem{Tuthill:2010}
{Tuthill}, P., {Jovanovic}, N., {Lacour}, S., {Lehmann}, A., {Ams}, M.,
  {Marshall}, G., {Lawrence}, J., {Withford}, M., {Robertson}, G., {Ireland},
  M., {Pope}, B., {Stewart}, P.: {Photonic technologies for a pupil remapping
  interferometer}.
\newblock In: Optical and Infrared Interferometry II, \emph{Proc. SPIE}, vol.
  7734, p. 77341P (2010).
\newblock \doi{10.1117/12.856770}

\bibitem{udalski:1994}
Udalski, A., Szymanski, M., Kaluzny, J., Kubiak, M., Mateo, M., Krzeminski, W.:
  The optical gravitational lensing experiment: The discovery of three further
  microlensing events in the direction of the galactic bulge.
\newblock The Astrophysical Journal \textbf{426}, 69--72 (1994)

\bibitem{Udem:2002}
{Udem}, T., {Holzwarth}, R., {H{\"a}nsch}, T.W.: {Optical frequency metrology}.
\newblock Nature \textbf{416}, 233--237 (2002).
\newblock \doi{10.1038/416233a}

\bibitem{Unwin2008}
{Unwin}, S.C., {Shao}, M., {Tanner}, A.M., {Allen}, R.J., {Beichman}, C.A.,
  {Boboltz}, D., {Catanzarite}, J.H., {Chaboyer}, B.C., {Ciardi}, D.R.,
  {Edberg}, S.J., {Fey}, A.L., {Fischer}, D.A., {Gelino}, C.R., {Gould}, A.P.,
  {Grillmair}, C., {Henry}, T.J., {Johnston}, K.V., {Johnston}, K.J., {Jones},
  D.L., {Kulkarni}, S.R., {Law}, N.M., {Majewski}, S.R., {Makarov}, V.V.,
  {Marcy}, G.W., {Meier}, D.L., {Olling}, R.P., {Pan}, X., {Patterson}, R.J.,
  {Pitesky}, J.E., {Quirrenbach}, A., {Shaklan}, S.B., {Shaya}, E.J.,
  {Strigari}, L.E., {Tomsick}, J.A., {Wehrle}, A.E., {Worthey}, G.: {Taking the
  Measure of the Universe: Precision Astrometry with SIM PlanetQuest}.
\newblock \pasp \textbf{120}, 38 (2008).
\newblock \doi{10.1086/525059}

\bibitem{VanBoekel2004}
{van Boekel}, R., {Min}, M., {Leinert}, C., {Waters}, L.B.F.M., {Richichi}, A.,
  {Chesneau}, O., {Dominik}, C., {Jaffe}, W., {Dutrey}, A., {Graser}, U.,
  {Henning}, T., {de Jong}, J., {K{\"o}hler}, R., {de Koter}, A., {Lopez}, B.,
  {Malbet}, F., {Morel}, S., {Paresce}, F., {Perrin}, G., {Preibisch}, T.,
  {Przygodda}, F., {Sch{\"o}ller}, M., {Wittkowski}, M.: {The building blocks
  of planets within the `terrestrial' region of protoplanetary disks}.
\newblock \nat \textbf{432}, 479--482 (2004).
\newblock \doi{10.1038/nature03088}

\bibitem{VanDenAncker1999}
{van den Ancker}, M.E., {Wesselius}, P.R., {Tielens}, A.G.G.M., {van Dishoeck},
  E.F., {Spinoglio}, L.: {ISO spectroscopy of shocked gas in the vicinity of T
  Tauri}.
\newblock \aap \textbf{348}, 877--887 (1999)

\bibitem{VanDishoek2004}
{van Dishoeck}, E.F.: {ISO Spectroscopy of Gas and Dust: From Molecular Clouds
  to Protoplanetary Disks}.
\newblock \araa \textbf{42}, 119--167 (2004).
\newblock \doi{10.1146/annurev.astro.42.053102.134010}

\bibitem{VanKerckhoven2002}
{Van Kerckhoven}, C., {Tielens}, A.G.G.M., {Waelkens}, C.: {Nanodiamonds around
  HD 97048 and Elias 1}.
\newblock \aap \textbf{384}, 568--584 (2002).
\newblock \doi{10.1051/0004-6361:20011814}

\bibitem{Vanderriest:1980}
{Vanderriest}, C.: {A fiber-optics dissector for spectroscopy of nebulosities
  around quasars and similar objects}.
\newblock Publ. Astron. Soc. Pac. \textbf{92}, 858--862 (1980).
\newblock \doi{10.1086/130764}

\bibitem{VargasCatalan:2016}
{Vargas Catal{\'a}n}, E., {Huby}, E., {Forsberg}, P., {Jolivet}, A., {Baudoz},
  P., {Carlomagno}, B., {Delacroix}, C., {Habraken}, S., {Mawet}, D., {Surdej},
  J., {Absil}, O., {Karlsson}, M.: {Optimizing the subwavelength grating of
  L-band annular groove phase masks for high coronagraphic performance}.
\newblock Astron. Astroph. \textbf{595}, A127 (2016).
\newblock \doi{10.1051/0004-6361/201628739}

\bibitem{Vergnole:2005}
Vergnole, S., Kotani, T., Perrin, G., Delage, L., Reynaud, F.: Calibration of
  silica fibers for the optical hawaiian array for nanoradian astronomy
  ('ohana): temperature dependence of differential chromatic dispersion.
\newblock Opt. Comm. \textbf{251}, 115--123 (2005)

\bibitem{Veron-Cetty2000}
{V{\'e}ron-Cetty}, M.P., {V{\'e}ron}, P.: {The emission line spectrum of active
  galactic nuclei and the unifying scheme}.
\newblock \aapr \textbf{10}, 81--133 (2000).
\newblock \doi{10.1007/s001590000006}

\bibitem{Vilas:1987}
Vilas, F., Smith, B.A.: Coronagraph for astronomical imaging and
  spectrophotometry.
\newblock Appl. Opt. \textbf{26}, 664--668 (1987)

\bibitem{Watson:1995}
Watson, F.G.: Multifiber waveguide spectrograph for astronomy?
\newblock In: Proc.SPIE, vol. 2476, pp. 2476 -- 2476 -- 7 (1995).
\newblock \doi{10.1117/12.211840}.
\newblock \urlprefix\url{http://dx.doi.org/10.1117/12.211840}

\bibitem{Watson:1997}
Watson, F.G.: Waveguide spectrographs for astronomy?
\newblock In: Proc.SPIE, vol. 2871, pp. 2871 -- 2871 -- 6 (1997).
\newblock \doi{10.1117/12.269030}.
\newblock \urlprefix\url{http://dx.doi.org/10.1117/12.269030}

\bibitem{Weber:2004}
{Weber}, V., {Barillot}, M., {Haguenauer}, P., {Kern}, P.Y., {Schanen-Duport},
  I., {Labeye}, P.R., {Pujol}, L., {Sodnik}, Z.: {Nulling interferometer based
  on an integrated optics combiner}.
\newblock In: W.A. {Traub} (ed.) New Frontiers in Stellar Interferometry,
  \emph{Proc. SPIE}, vol. 5491, p. 842 (2004).
\newblock \doi{10.1117/12.550581}

\bibitem{Weinberger:2015}
{Weinberger}, A.J., {Bryden}, G., {Kennedy}, G.M., {Roberge}, A.,
  {Defr{\`e}re}, D., {Hinz}, P.M., {Millan-Gabet}, R., {Rieke}, G., {Bailey},
  V.P., {Danchi}, W.C., {Haniff}, C., {Mennesson}, B., {Serabyn}, E., {Skemer},
  A.J., {Stapelfeldt}, K.R., {Wyatt}, M.C.: {Target Selection for the LBTI
  Exozodi Key Science Program}.
\newblock Astroph. J. Suppl. S. \textbf{216}, 24 (2015).
\newblock \doi{10.1088/0067-0049/216/2/24}

\bibitem{Welford:1982}
{Welford}, W.T., {Winston}, R.: {Nonconventional optical systems and brightness
  theroem}.
\newblock Appl. Opt. \textbf{21}, 1531--1533 (1982).
\newblock \doi{10.1364/AO.21.001531}

\bibitem{Wildi:2010}
{Wildi}, F., {Pepe}, F., {Chazelas}, B., {Lo Curto}, G., {Lovis}, C.: {A
  Fabry-Perot calibrator of the HARPS radial velocity spectrograph: performance
  report}.
\newblock In: Ground-based and Airborne Instrumentation for Astronomy III,
  \emph{Proc. SPIE}, vol. 7735, p. 77354X (2010).
\newblock \doi{10.1117/12.857951}

\bibitem{Wildi:2011}
{Wildi}, F., {Pepe}, F., {Chazelas}, B., {Lo Curto}, G., {Lovis}, C.: {The
  performance of the new Fabry-Perot calibration system of the radial velocity
  spectrograph HARPS}.
\newblock In: Techniques and Instrumentation for Detection of Exoplanets V,
  \emph{Proc. SPIE}, vol. 8151, p. 81511F (2011).
\newblock \doi{10.1117/12.901550}

\bibitem{Wilken:2012}
{Wilken}, T., {Curto}, G.L., {Probst}, R.A., {Steinmetz}, T., {Manescau}, A.,
  {Pasquini}, L., {Gonz{\'a}lez Hern{\'a}ndez}, J.I., {Rebolo}, R.,
  {H{\"a}nsch}, T.W., {Udem}, T., {Holzwarth}, R.: {A spectrograph for
  exoplanet observations calibrated at the centimetre-per-second level}.
\newblock Nature \textbf{485}, 611--614 (2012).
\newblock \doi{10.1038/nature11092}

\bibitem{Woillez:2017}
{Woillez}, J., {Lai}, O., {Perrin}, G., {Reynaud}, F., {Baril}, M., {Dong}, Y.,
  {F{\'e}dou}, P.: {AGILIS: Agile Guided Interferometer for Longbaseline
  Imaging Synthesis. Demonstration and concepts of reconfigurable optical
  imaging interferometers}.
\newblock A\&A \textbf{602}, A116 (2017).
\newblock \doi{10.1051/0004-6361/201730500}

\bibitem{yerolatsitis:2017}
Yerolatsitis, S., Harrington, K., Birks, T.: All-fibre pseudo-slit
  reformatters.
\newblock Optics Express \textbf{25}(16), 18,713--18,721 (2017)

\bibitem{Yi:2016}
{Yi}, X., {Vahala}, K., {Li}, J., {Diddams}, S., {Ycas}, G., {Plavchan}, P.,
  {Leifer}, S., {Sandhu}, J., {Vasisht}, G., {Chen}, P., {Gao}, P., {Gagne},
  J., {Furlan}, E., {Bottom}, M., {Martin}, E.C., {Fitzgerald}, M.P.,
  {Doppmann}, G., {Beichman}, C.: {Demonstration of a near-IR line-referenced
  electro-optical laser frequency comb for precision radial velocity
  measurements in astronomy}.
\newblock Nature Communications \textbf{7}, 10436 (2016).
\newblock \doi{10.1038/ncomms10436}

\bibitem{york:2000}
York, D.G., Adelman, J., Anderson~Jr, J.E., Anderson, S.F., Annis, J., Bahcall,
  N.A., Bakken, J., Barkhouser, R., Bastian, S., Berman, E., et~al.: The sloan
  digital sky survey: Technical summary.
\newblock The Astronomical Journal \textbf{120}(3), 1579 (2000)

\bibitem{Zajnulina:2015}
{Zajnulina}, M., {Boggio}, J.M.C., {B{\"o}hm}, M., {Rieznik}, A.A., {Fremberg},
  T., {Haynes}, R., {Roth}, M.M.: {Generation of optical frequency combs via
  four-wave mixing processes for low- and medium-resolution astronomy}.
\newblock Applied Physics B: Lasers and Optics \textbf{120}, 171--184 (2015).
\newblock \doi{10.1007/s00340-015-6121-1}

\end{thebibliography}

\end{document}